\documentclass[a4paper,11pt]{article}
\pdfoutput=1 
\usepackage{jheppub}
\usepackage[utf8]{inputenc}
\usepackage[english]{babel}
\usepackage{amsmath, amssymb,graphicx}
\newcommand{\be}{\begin{equation}}
\newcommand{\ee}{\end{equation}}
\newcommand{\bea}{\begin{eqnarray}}
\newcommand{\eea}{\end{eqnarray}}
\newcommand{\nn}{\nonumber}

\title{Analytic black branes in Lifshitz-like backgrounds and thermalization}

\author[a]{Irina Ya. Aref'eva,}
\author[b,c,d]{Anastasia A. Golubtsova}
\author[e]{and Eric Gourgoulhon}

\affiliation[a]{Steklov Mathematical Institute, Russian Academy of Sciences,\\Gubkina str. 8, 119991, Moscow, Russia}
\affiliation[b]{Bogoliubov Laboratory of Theoretical Physics, Joint Institute for Nuclear research,\\Joliot-Curie str. 6, Dubna, 141980, Russia}
\affiliation[c]{Dubna State University,\\Universitetskaya str. 19, Dubna, 141980, Russia}
\affiliation[d]{Peoples' Friendship University of Russia,\\Miklukho-Maklaya str. 6, Moscow, 117198, Russia}
\affiliation[e]{Laboratoire Univers et Th\'{e}ories, Observatoire de Paris,
CNRS, Universit\'e Paris Diderot \\5 place Jules Janssen, 92190 Meudon, France}

\emailAdd{arefeva@mi.ras.ru}
\emailAdd{siedhe@gmail.com}
\emailAdd{eric.gourgoulhon@obspm.fr}

\abstract{
Using black brane solutions in 5d Lifshitz-like backgrounds with arbitrary dynamical exponent $\nu$, we construct the Vaidya geometry, 
asymptoting to the Lifshitz-like spacetime, which represents a thin shell infalling at the speed of light.  
We apply the new Lifshitz-Vaidya background to study the thermalization process of the quark-gluon plasma via the thin shell approach previously successfully used in several backgrounds. We find that the thermalization  depends on the chosen direction because of the spatial anisotropy. 
The plasma thermalizes thus faster in the transversal direction than in the longitudinal one.
To probe the system described by the Lifshitz-like backgrounds, we also calculate the holographic 
entanglement entropy for the subsystems delineated along both transversal and longitudinal directions.  
We show that the entropy has some universality in the behavior for both subsystems. 
At the same time, we find that certain characteristics strongly depend on the critical exponent $\nu$. \\
}


\begin{document}
\maketitle
\flushbottom
\newpage
\section{Introduction}
\label{sec:intro}

 The gravity/gauge duality provide an alternative tool for understanding dynamics of the strong coupling system, where standard methods lacks. One such system is the quark-gluon plasma (QGP), which can be produced in heavy-ion collisions and  represents  a strongly coupled fluid with a small viscosity \cite{Solana}.  The QGP goes through several stages of evolution. It is believed that the QGP is created after a very short time after the collision $\tau_{\rm therm} \approx $ few $0.1 \, {\rm fm}/c$
 and the holographic approach, in particular, is aimed to describe  this and a short nearest  period  of evolution \cite{IA,DeWolf}.
 There are indications that in this time the QGP is anisotropic. Since at time
 scales of $\tau \approx$ few $0.1\, {\rm fm}/c$ it is in thermal equilibrium, one can try to apply  anisotropic  holographic hydrodynamics  to  describe its isotropization.  The anisotropic stage of the QGP takes place for $0.1\, {\rm fm}/c \lesssim \tau \lesssim 0.3 - 2\, {\rm fm}/c$ \cite{1312.2285} and can be studied also holographically \cite{Giataganas,Giataganas:2013lga}.

Through the gauge/gravity duality, the  thermalization of the field theory
in the boundary corresponds to the process of black hole formation in the bulk.  According to the holographic dictionary, the scenario of a heavy-ion collision can be represented as a shock wave collision in which trapped surface is formed \cite{GPY}-\cite{APP}.   After the collision the shocks slowly decay, leaving the plasma described by hydrodynamics in the middle. The creation of the black hole is  also described by the Vaidya metric of an infalling shell with a horizon corresponding to the location of the trapped surface \cite{BBBC}-\cite{AV}.

By now both standard existing models and the holographic approach with AdS backgrounds, as well as its conformally equivalent deformations for  bulk geometries, have failed to reproduce the
particle multiplicity at high energies. However, if one performs the holographic estimations of multiplicities in Lifshitz-like spacetimes \cite{TAYLOR, Taylor16, ALT}, one can fit the experimental data for certain values of the critical exponents \cite{AG}. In this paper,
we consider following the 5-dimensional Lifshitz-like metric:
\begin{equation}\label{Intr.1}
ds^{2} =L^{2}\left[\frac{\left(-dt^{2} + dx^{2}\right)}{z^{2}} + \frac{\left(dy^{2}_{1} + dy^{2}_{2}\right)}{z^{2/\nu}} +  \frac{d z^{2}}{z^{2}}\right].
\end{equation}
The choice of the geometry (\ref{Intr.1})  is motivated by studies of  the anisotropic phase of the QGP. As it is known, the QGP in the 4d gauge theory can be characterized by the energy-momentum tensor $\langle T_{\mu \nu}\rangle = \textrm{diag}(\varepsilon, p_{L}, p_{T}, p_{T})$, with the particle momenta $\langle p^{2}_{L} \rangle < \langle p^{2}_{T} \rangle$  at early times of the QGP formation. To reproduce this anisotropy from the gravity side, one of the possible backgrounds is the Lifshitz-like metric (\ref{Intr.1}).
It has been shown in \cite{AG} that for the wall-on-wall collision in the 5d Lifshitz-like background with the critical exponent $\nu=4$, the dependence of multiplicity on the energy
is desirable, i.e. behaves as $E^{1/3}$.

Another possible implementation of the 5d Lifshitz-like spacetime is
\begin{equation}\label{Intr.2}
ds^{2} =L^{2}\left[\frac{-dt^{2}}{z^{2}} + \frac{dx^{2}}{z^{2/\nu}} + \frac{\left(dy^{2}_{1} + dy^{2}_{2}\right)}{z^{2}} +  \frac{d z^{2}}{z^{2}}\right],
\end{equation}
which differs from (\ref{Intr.1}) by  the anisotropic scaling taking place only for one spatial direction. The embedding  of this background and its non-zero temperature generalization into supergravity IIB was done in \cite{ALT} for $\nu = 3/2$. Solutions interpolating between Lifshitz-like (\ref{Intr.2}) and AdS geometries  were intensively studied in \cite{MT2}-\cite{CGS}  within the context of applications to the anisotropic QGP. However,  the results for multiplicities calculated using the background
(\ref{Intr.2})  in \cite{AG} do not fit the experimental data unlike the case of the metric (\ref{Intr.1}).

Since  after the shock wave collision the trapped surface argument supports black hole formation, it is natural to construct  the corresponding Vaidya-type solution. 
In the present paper we start from the generalization of  (\ref{Intr.1})  to the non-zero temperature case for an arbitrary critical exponent.  
Further, we construct a Vaidya-type geometry asymptoting to the Lifshitz-like solution to model a gravitation collapse in order to study the holographic thermalization. The Vaidya metric with Lifshitz scaling was used for the  examination of  the holographic thermalization in \cite{KKVT, IYaA}. There, it has been shown that for the metric with anisotropy between time and spatial directions the propagation of thermalization represents a similar ''horizon''
behavior as that seen in the AdS case. The Vaidya metric was also generalized to the Lifshitz spacetimes with a hyperscaling violating factor  \cite{AAM,FFKKTT,VKLT}.\\
As another application of our solutions, we consider the time evolution of the holographic entanglement entropy during the process of thermalization.  The behavior of the entanglement entropy modeling the thermalization and ``quench'' processes is a subject of intensive studies during last years, see \cite{Hubeny:2012ry,Liu:2013qca,FFKKTT,Kol:2014nqa,Fischetti:2014zja,Fonda:2015nma} and refs. therein. For Lifshitz metrics the time evolution of the entanglement entropy turns out to have a linear regime \cite{KKVT}. In this work, we examine the influence of spatial anisotropy on the behavior of the entanglement entropy.

The paper is organized as follows. Sect.~\ref{Sect:2} is devoted to constructing the exact solutions which asymptotes to the Lifshitz-like metric (\ref{Intr.1}). In Sect.~\ref{Sect:2.1} we present the 5d black brane background. In Sect.~\ref{Sect:2.2}  we generalize it to the Vaidya type solution, which describes a thin shell collapsing to a black hole in the Lifshitz-like background.  In Sect.~\ref{Sect:3} we numerically calculate thermalization times using our Lifshtiz-Vaidya  type solution. Sect.~\ref{Sect:4} is devoted to studies of the holographic entanglement entropy at equilibrium as well as its out-of-equilibrium behavior. We conclude in Sect.~\ref{Sect:6} with a discussion of our results. Appendix ~\ref{App:A} collects some technical details used for constructing analytic solutions. In Appendix ~\ref{App:B} we present some details concerning numerical solutions to EOM related the functional of the entanglement entropy.

\section{Gravity backgrounds}\label{Sect:2}
\subsection{Black branes in Lifshitz-like backgrounds}\label{Sect:2.1}
In \cite{AG} we considered a collision of two domain walls in the five-dimensional Lifshitz-like background
\begin{eqnarray}\label{1.1}
ds^{2} = \tilde{r}^{2\nu}\left(-dt^{2} + dx^{2}\right) + \tilde{r}^{2}(dy^{2}_{1} + dy^{2}_{2}) + \frac{d\tilde{r}^{2}}{\tilde{r}^{2}},
\end{eqnarray}
where $\nu$ is the critical exponent.
Note that (\ref{1.1}) is equivalent to (\ref{Intr.1}) via the change of coordinate
$z = {\tilde r}^{-\nu}$ and the rescaling $(t,x,y_1,y_2)\mapsto \nu^{-1} (t,x,y_1,y_2)$.

In  \cite{ALT, AG}  the metric ansatz  (\ref{1.1}) was considered for a 5d model
governed by the action
\begin{eqnarray}\label{1.1a}
S = \frac{1}{16\pi G_{5}}\int d^{5}x\sqrt{|g|}\left(R + \Lambda   - \frac{1}{6}\left(H^{2}_{(3)} + m^{2}_{0}B^{2}_{(2)}\right)\right),
\end{eqnarray}
where $m_{0}$ and $\Lambda$ are constant and the 3-form $H_{(3)}$ and the 2-form
$B_{(2)}$ are related by
\begin{eqnarray}\label{1.1b}
H_{(3)} = dB_{(2)} .
\end{eqnarray}
 However, it seems difficult to find an analytic black brane (hole) solution for the model (\ref{1.1a}) due  the dependence (\ref{1.1b}) for the gauge fields.

In this paper we consider another bulk theory, possessing the metric (\ref{1.1}) as a solution to Einstein equations,  with the following action
\begin{eqnarray}\label{1.2a}
S = \int d^{5} x \sqrt{|g|}\left(R[g] + \Lambda - \frac{1}{2}(\partial\phi)^{2} - \frac{1}{4}e^{\lambda \phi}F^{2}_{(2)} \right),
\end{eqnarray}
where the 2-form $F_{(2)}$ is the gauge field with
\begin{equation}
F^{2}_{(2)} = F_{mn}F^{mn},
\end{equation}
 $\phi$ is the dilaton scalar field, $\lambda$ is a dilatonic coupling constant and $\Lambda$ is the cosmological constant\footnote{More precisely, $\Lambda=-2\bar\Lambda$, where
 $\bar\Lambda$ is the standard cosmological constant.}.
The model (\ref{1.2a}) can be considered as a truncated supergravity $IIA$ in the style of \cite{ALT}. Another possible underlying theory is the 5d $SO(6)$ gauged supergravity \cite{DGP}.

The Einstein equations of motion can be written as
\begin{eqnarray}\label{1.3}
R_{mn} = - \frac{\Lambda}{3}g_{mn} + \frac{1}{2}(\partial_{m} \phi)(\partial_{n} \phi)  + \frac{1}{2}e^{\lambda \phi} F_{mp}F_{n}^{\ p} - \frac{1}{12}e^{\lambda \phi}F^{2}_{(2)}g_{mn}.
\end{eqnarray}
The scalar field equation is
\begin{equation}\label{1.4}
\Box \phi = \frac{1}{4}\lambda e^{\lambda \phi} F^{2}_{(2)}, \quad\textrm{with} \quad \Box \phi
 = \frac{1}{\sqrt{|g|}}\partial_{m}(g^{mn}\sqrt{|g|}\partial_{n}\phi).
\end{equation}
Finally the equation of motion for the gauge field is
\begin{equation}\label{1.5}
D_{m}\left(e^{\lambda \phi}F^{mn}\right) = 0.
\end{equation}

Introducing the new variable
\begin{equation}\label{1.5a}
r =\ln{\tilde{r}},
\end{equation}
 one can rewrite  (\ref{1.1}) as
\begin{eqnarray}\label{1.2}
ds^{2} = e^{2\nu r}\left(-dt^{2} + dx^{2}\right) + e^{2r}(dy^{2}_{1} + dy^{2}_{2})  + dr^{2}.
\end{eqnarray}
We then select the following anzatz for the dilaton and Maxwell fields:
\begin{eqnarray}\label{1.2b}
\phi = \phi(r), \quad e^{\lambda \phi} = \mu e^{4r}, \\ \label{1.2c}
F_{(2)} = \frac{1}{2}q\,  dy_{1}\wedge dy_{2},
\end{eqnarray}
where $\mu$ and $q$ are two constants.
One can see that this ansatz has some features. Firstly, the dilaton has the linear dependence in the radial coordinate  (\ref{1.2b}). Black hole solutions  with a linear dilaton in the supergravity context were discussed in \cite{RKAP}. At the same time the similar ansatz for the gauge fields (\ref{1.2c}) emerges to support $AdS_{2}\times \mathbb{R}^{3}$, $AdS_{3}\times \mathbb{R}^{2}$, $AdS_{2}\times \mathbb{R}^{2}$ solutions and their non-zero temperature analogues of gauged supergravity in \cite{DGP}-\cite{DGP1}. The 6-dimensional  $Lif_{4}\times \mathbb{R}^{2}$ background with a constant two-form field was found in \cite{EDPK}.

The model (\ref{1.2a}) with the fields given by  (\ref{1.2b})-(\ref{1.2c}) can be generalized to the non-zero temperature case without changing the field ansatz. The metric of the black brane solution reads
\begin{equation}\label{2.2}
ds^{2} = e^{2\nu r}\left( - f(r)dt^{2} + dx^{2}\right) + e^{2r}\left(dy^{2}_{1}  + dy^{2}_{2}\right) + \frac{dr^{2}}{f(r)},
\end{equation}
with the blackening function given by
\begin{equation}\label{2.2a}
f(r) = 1- me^{-(2\nu+2)r}.
\end{equation}
For the particular case $\nu = 4$, the metric (\ref{2.2})-(\ref{2.2a}) along
with the ansatz (\ref{1.2b})-(\ref{1.2c})
solves the field equations (\ref{1.3})-(\ref{1.5}) provided that the constants
take the following values:
\begin{eqnarray}\label{2.2b}
\lambda = \pm \frac{2}{\sqrt{3}}, \quad \Lambda = 90, \quad \mu q^{2} =  240.
\end{eqnarray}
See Appendix~\ref{App:A1} for details.

If the dilaton is constant and the Maxwell field vanishes,
the metric (\ref{2.2}) with $\nu=1$ turns out to be the black brane solution
in the AdS background:
\begin{equation}\label{4.6}
ds^{2} = e^{2r}\left( - f(r)dt^{2} + dx^{2}\right) + e^{2r}\left(dy^{2}_{1}  + dy^{2}_{2}\right) + \frac{dr^{2}}{f(r)}
\end{equation}
with
\begin{equation}\label{4.6a}
f(r) = 1- me^{-4r}
\end{equation}
or in terms of variable $\tilde{r}$
\begin{equation}\label{4.7}
ds^{2} = \tilde{r}^{2}\left(-f(\tilde{r})dt^{2} + dx^{2}\right) + \tilde{r}^{2}(dy^{2}_{1} + dy^{2}_{2}) + \frac{d\tilde{r}^{2}}{f(\tilde{r})\tilde{r}^{2}},
\end{equation}
where
\begin{equation}\label{4.7a}
f(\tilde{r}) = 1- \frac{m}{\tilde{r}^{4}}.
\end{equation}
This corresponds to $r\rightarrow 0$ or the UV limit.

\subsection{The Vaidya-Lifshitz geometry}\label{Sect:2.2}

To study the thermalization process we need to use the infalling shell approach based on the Vaidya solution \cite{Vad}.
First, we introduce the coordinate $z=e^{-\nu r}$, which, after the
rescaling $(t,x,y_1,y_2)\mapsto \nu^{-1} (t,x,y_1,y_2)$, allows one to
rewrite the metric (\ref{2.2}) in the form
\begin{equation}\label{V3.2}
ds^{2} = z^{-2}\left(-f( z)dt^{2} + dx^{2}\right) + z^{-2/\nu}(dy^{2}_{1} + dy^{2}_{2})  + \frac{d z^{2}}{ z^2 f(z)},
\end{equation}
with the blackening function
\begin{equation} \label{V3.2c}
f = 1- m z^{2+ 2/\nu}.
\end{equation}

To write down the Vaidya-Lifshitz solution, one should consider the ingoing
null geodesics
\be
dt + \frac{dz}{f(z)} = 0
\ee
and introduce
the Eddington-Finkelstein coordinate system $(v,x,y_1,y_2,z)$ via
\be\label{v2}
dv= dt + \frac{dz}{f(z)}.
\ee

Owing to (\ref{v2}) we can represent  (\ref{V3.2}) in the following form

\begin{equation} \label{V3.2a}
ds^{2} =- z^{-2}f(v,z)dv^2  - 2z^{-2} dv d z + z^{-2}dx^{2}+ {z}^{-2/\nu}(dy^{2}_{1} + dy^{2}_{2}),
\end{equation}
with
\begin{equation}\label{V3.2b}
f(v,z) = 1- m(v) z^{2+ 2/\nu},
\end{equation}
where the mass function $m(v)$ determines the thickness of the shell falling along $v = 0$ and captures the information about the black hole formation. For the infinite thin shell $m(v)$ has the form
\begin{equation}
m(v) = M\theta(v),
\end{equation}
where $M$ is a constant and $\theta(v)$ is the Heaviside function. One can also consider a
smooth function $m(v)$ and get, for instance,
\be\label{f}
f(v,z) = 1- \frac{m}{2}\left(1+\tanh \frac{v}{\alpha}\right)z^{2+2/\nu},\ee
where $m$ and $\alpha$ are two constants.

The solution (\ref{V3.2a})-(\ref{V3.2b}) interpolates between the vacuum Lifshitz solution (\ref{1.1})
inside the shell ($v<0$) and the Lifshitz black brane geometry (\ref{V3.2})-(\ref{V3.2c}) outside the shell ($v>0$).

The check of the equations of motion for the background  (\ref{V3.2a})-(\ref{V3.2b}) is given in Appendix~\ref{App:A2}.

\section{The thermalization process}\label{Sect:3}
In \cite{AG} we have shown that there is a trapped surface, which forms in the collision of two shock waves in the background (\ref{1.1}), controlled by boundary points $z_{a}$ and $z_{b}$, with $z_{a}<z_{b}$. This trapped surface defines the location of the horizon for (\ref{V3.2a})-(\ref{V3.2b}).

Calculations of  the thermalization time $t_{\rm therm}$ at the scale $\ell$  is based on finding  geodesics with endpoints located at the distance  $\ell$ for a bulk particle. Then, the thermalization time $t_{\rm therm}$ is the time when this geodesic covered by the shell (\ref{V3.2a})-(\ref{V3.2b}).

The general case for the Lagrangian of the pointlike probe has the form
\begin{eqnarray}
\label{398}
\mathcal{L} = \sqrt{-z^{-2}f(z) \frac{dv}{d\tau}\frac{dv}{d\tau} -2z^{-2}\frac{dv}{d\tau}\frac{dz}{d\tau}+ z^{-2}\frac{dx}{d\tau}\frac{dx}{d\tau} + z^{-2/\nu}\left(\frac{d y_{1}}{d\tau}\frac{dy_{1}}{d\tau} + \frac{d y_{2}}{d\tau}\frac{dy_{2}}{d\tau}\right)},
\end{eqnarray}
where $\tau$ is a parameter. Here we have two possibilities for the choice of $\tau$ with respect to the transverse directions.

\subsection{Thermalization along the longitudinal direction }

Consider the first case taking  $\tau = x$, which can be interpreted as a longitudinal direction.
Now we obtain

\begin{eqnarray}
\label{Lagr2.a}\mathcal{L} = \frac{\sqrt{\mathcal{R}_x}}{z},
\end{eqnarray}
where we define
\begin{eqnarray}
\mathcal{R}&=&1 - f(z)(v^{'}_{x})^{2} - 2 v^{'}_{x} z^{'}_{x}.
\end{eqnarray}
The  integrals of motion corresponding to (\ref{Lagr2.a}) are
\begin{eqnarray}
\label{intJ}
\mathcal{J}& = &- \frac{1}{z\sqrt{\mathcal{R}_x}},  \\ \label{int1C1}
\mathcal{I} & = &\frac{f(z)v^{'}_{x} + z^{'}_{x}}{z\sqrt{\mathcal{R}}}.
\end{eqnarray}

From the relations (\ref{intJ}) and (\ref{int1C1}) we get
\begin{eqnarray}\label{diffzzx}
z^{'}_{x}& = &\pm \sqrt{f(z)\left(\frac{1}{z^{2}\mathcal{J}^{2}}  - 1\right)
+ \frac{\mathcal{I}^{2}}{\mathcal{J}^{2}}}.
\end{eqnarray}

\begin{figure}[tbp]
\centering
 \includegraphics[scale=0.45]{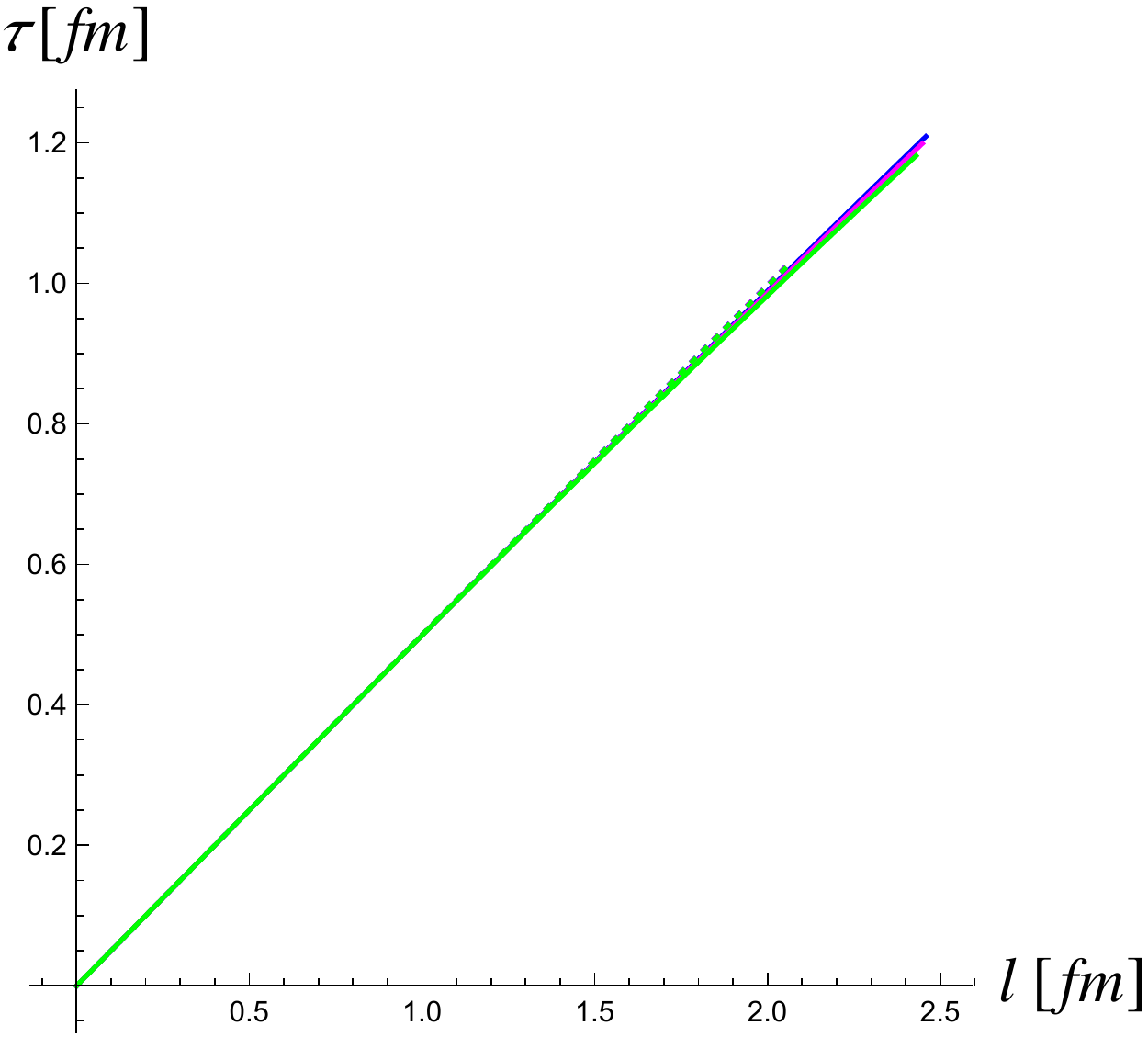}A$\,\,\,\,\,\,\,\,\,\,\,$ $\,\,\,\,\,\,\,\,\,\,\,$
  \includegraphics[scale=0.16]{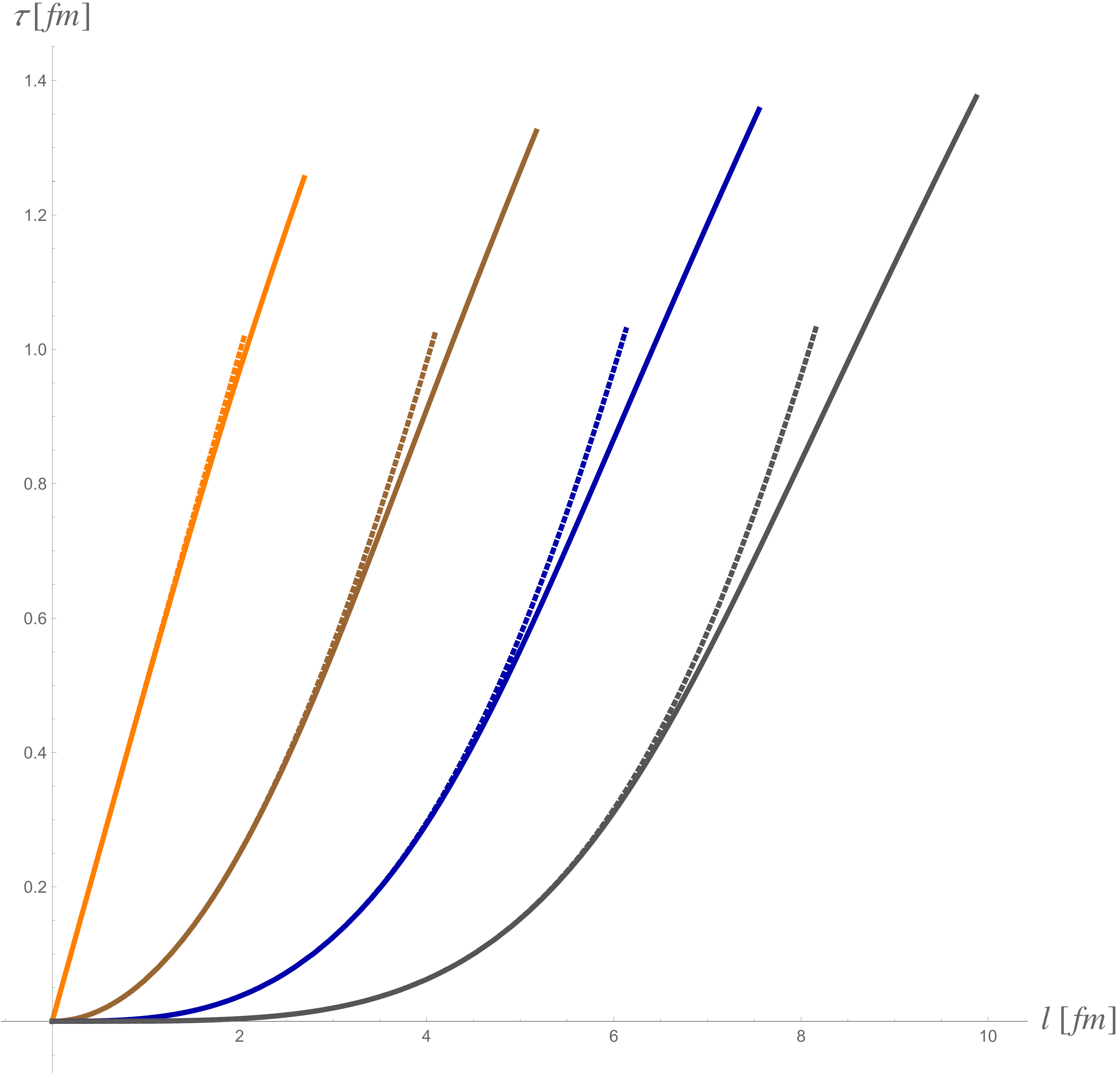}B
 \caption{The thermalization time $\tau$ as a function of $\ell$ for the 5-dimensional Lifshitz metric (\ref{V3.2})-(\ref{V3.2b}) for $\nu =2,3,4$.
 {\bf A:} Thermalization along the longitudinal direction with $m = 0.5$ and $m = 0.1$. All lines coincide.
 {\bf B:} Thermalization along the transversal direction, $\nu = 1$ (orange), $\nu = 2$ (brown), $\nu =3$ (blue) and $\nu=4$ (gray).  The solid and dotted curves correspond to $m = 0.5$ and $m = 0.1$, respectively.}\label{fig:1}
\end{figure}

The turning point $z_{*}$ can be found from the equation
\begin{eqnarray}\label{TP1}
f(z_{*})\left(\frac{1}{z^{2}_{*}} - \mathcal{J}^{2}\right) + \mathcal{I}^{2}= 0.
\end{eqnarray}
For simplicity, we put  $\mathcal{I}  = 0$  and we get from (\ref{TP1})
\begin{eqnarray}\label{TP1.a}
 \mathcal{J}^{2} = \frac{1}{z^{2}_{*}}.
\end{eqnarray}
For  the distance $\ell$ between the ends of the geodesic and the thermalization time
one gets
\begin{eqnarray}\label{ellt.1a}
\ell &=& 2z_{*}\int^{1}_{0} \frac{wdw}{\sqrt{f(z_{*}w)(1-w^2 )}},\\ \label{ellt.1b}
t_{\rm therm} &=&  z_{*}\int^{1}_{0}\frac{dw}{f(z_{*}w)}.
\end{eqnarray}
 Note that here we assume that the turning point lies above the horizon, i.e. $z_h>z_*$.
The behavior of the thermalization time as a function of the distances for (\ref{ellt.1a})-(\ref{ellt.1b}) is represented in Fig.\ref{fig:1}.{A}.
We see that the thermalization time behaves linearly with $\ell$.
The results match to those for modified AdS models from \cite{IYaA} and coincide for all values of the dynamical exponent $\nu$.

\subsection{Thermalization along the transversal direction}

Now turn to the second case when  $\tau = y_{1} $, that we interpret as the thermalization along a transversal direction. From (\ref{398}) we have
\begin{eqnarray}\label{Lagr.2}
\mathcal{L} = \frac{ \sqrt{ \mathcal{R}}}{z},
\end{eqnarray}
where we put
\begin{equation}
\mathcal{R} = z^{2 -2/\nu} - f(z) (\dot{v}_{y})^{2} -2\dot{v}_{y}\dot{z}_{y}.
\end{equation}

The integrals of motion corresponding to (\ref{Lagr.2}) read

\begin{eqnarray}\label{En2}
\mathcal{J} & =&-\frac{z^{1- 2/\nu}}{\sqrt{\mathcal{R}}},\\
\label{intI1}
\mathcal{I} & = &\frac{f(z)\dot{v}_{y}  + \dot{z}_{y}}{z\sqrt{\mathcal{R}}}.
\end{eqnarray}

 From (\ref{En2}) and (\ref{intI1}) we get that the turning point $z_*$ is defined from

   \be
  -\mathcal{I}^2 z^{2-\frac{4}{\nu}} \left(f \mathcal{J}^2
   z^{2/n}-f-\mathcal{I}^2 z^2\right) =0.
   \ee
For $\mathcal{I}=0$ this equation is simplified  to give
\begin{eqnarray}\label{constx}
\mathcal{J}^2=\frac{1}{z_*^{^{2/n}}}
\end{eqnarray}
and  we also get
\be\label{z-dot}
\dot z_y=\pm\frac{z^{1-\frac{2}{\nu }} \sqrt{f \left(1-\mathcal{J}^2 z^{2/\nu}\right)}}{\mathcal{J}}.
\ee

 From (\ref{z-dot}) one gets the relation between the ends of geodesic and the thermalization time

\bea\label{ellth2}
\ell&=&
2\,z_{*}^{1/\nu} \int_0^{1} \frac{ w^{-1 + 2/\nu}dw}
{\sqrt{f(wz_*)\left(1-w^{2/\nu}\right) }},\label{ell}\\\nn\\ \label{tau2}
t_{\rm therm} &= &z_*\int _{0}^{1}\frac{dw}{f(z_*w)}.
\eea
Here we remove the regularization since $\nu>0$.
The dependence (\ref{tau2}) on  (\ref{ellth2}) is given in Fig.~\ref{fig:1}.{B}. We see that the thermalization time in the transversal direction depends on the anizotropic parameter $\nu$. In particular, for $\nu=2$ the thermalization process is more then twice faster as compared to the longitudinal direction. By increasing  $\nu$ we make the thermalization in the transversal direction faster.  We also see that for  larger values of  $\nu$  the dependence on the mass $m$ becomes more essential.

\section{Entanglement entropy}\label{Sect:4}

In this section we explore the evolution of  entanglement entropy in the context of the holographic prescription.
 We perform calculations using  both black brane
(\ref{2.2})-(\ref{2.2a}) and Vaidya-Lifshitz  time dependent backgrounds (\ref{V3.2a})-(\ref{V3.2b}).

The  entanglement entropy can be useful to probe correlations in the background  measuring an entanglement of a quantum system. If the system is divided into two spatially disjoint parts $A$ and $B$, the entanglement entropy $S(A)$ gives an estimation of the amount of information loss corresponding to the restriction of an $A$.
It seems not to be simple to calculate the entanglement entropy from the  strongly coupled system side.
However, one can compute its holographic dual using the suggestion from works \cite{RT1}-\cite{HRT}.
The holographic formula for the entanglement entropy of a subsystem $A$

\begin{eqnarray}\label{4.1}
S =  \frac{\mathcal{A}}{4G_{5}},
\end{eqnarray}
where $\mathcal{A}$ is the area of the minimal three-dimensional surface whose boundary coincides with the boundary of the region $A$. The area of the surface is defined by the relation

\begin{equation}\label{4.1a}
\mathcal{A} = \int d^{3}\sigma \sqrt{|\det g_{MN}\partial_{\alpha}X^{M}\partial_{\beta}X^{N}|},
\end{equation}
where
\begin{equation}
\sigma^{1} = x, \quad \sigma^{2} = y_{1}, \quad \sigma^{3} = y_{2}.
\end{equation}
(\ref{4.1}) is known as the holographic entanglement entropy. It is useful to represent   (\ref{V3.2})-(\ref{V3.2c}) and (\ref{V3.2a})-(\ref{V3.2b})
 in the generic form
\begin{eqnarray}\label{GFM}
g_{MN} = g_{00}dt^{2} + g_{11}dx^{2} + g_{22}dy^{2}_{1} + g_{33}dy^{2}_{2} + g_{44}dz^{2}.
\end{eqnarray}
From  (\ref{GFM}) we see that the entanglement entropy depends on the direction along which the subsystem is delineated.
There are two possible cases for the subsystems both for the black brane and the thin shell we have we study and compare each other.

\subsection{Entanglement entropy in a time-independent background}

To begin with, we compute the entanglement entropy for the black hole  (\ref{V3.2})-(\ref{V3.2c}).
Here we present the results for the two subsystems cut out both along longitudinal and transversal directions.

\subsubsection{Subsystem delineated along the longitudinal direction}

First, consider the subsystem $A$ cut out along $x$-direction, say, the belt is located as
\begin{eqnarray}\label{4.1b}
x\in [0, l_{x} < L_{x}], \quad y_{1} \in [0, L_{y_{1}}], \quad  y_{2} \in [0, L_{y_{2}}].
\end{eqnarray}
We assume  that the minimal area surface is invariant under the $y_1$ and $y_2$ planar directions and the embedding function
is the function of only one coordinate, $z = z(x)$.
Thus, the three-dimensional minimal surface is defined by
\begin{eqnarray}\label{MS1}
\mathcal{A} =2 \int^{l_{x}}_{0}dx\int^{L_{y_{1}}}_{0}dy_{1}\int^{L_{y_{2}}}_{0}dy_{2}\sqrt{g_{22}g_{33}\left(g_{11} + g_{44}(z^{'})^{2}\right)}.
\end{eqnarray}
Taking into account (\ref{V3.2}), one has
\begin{eqnarray}\label{MS1.a}
\mathcal{A}
= 2 L_{y_{1}}L_{y_{2}}\int ^{l_{x}}_{0}dx\mathcal{L},\,\,\,{\mbox {where}}\,\,\,\,\,\mathcal{L}=\frac{1}{z^{1+2/\nu}}\sqrt{1+ \frac{z^{\prime 2}}{f(z)}},
\end{eqnarray}
where it is supposed that $\prime = \displaystyle{\frac{d}{dx}}$.

The integral of motion corresponding to the system with the Lagrangian $\mathcal{L}$ reads
\begin{equation}
-\frac{z^{-1-2/\nu}}{\sqrt{1+\frac{z'^{2}}{f(z)}}}=\mathcal{C}.
\end{equation}

The function $z= z(x)$ that minimizes the surface area is then given by the equation of motion
\begin{eqnarray}\label{dz1}
z^{'} = \pm \sqrt{f(z)\left(\left(\frac{z_{*}}{z}\right)^{2(1+ 2/\nu)} - 1\right)},
\end{eqnarray}
where the turning  point $z_{*}$ is related with $\mathcal{C}$ as $z^{1+2/\nu}_{*} = \mathcal{C}^{-1}$. The length scale $l_{x}$ can be found from
\begin{eqnarray}\label{lx1}
\frac{l_{x}}{2} = \displaystyle{\int^{z_{*} - \epsilon}_{z_{0}} \frac{dz}{z^{'}}} = \int^{z_{*} - \epsilon}_{z_{0}}\left(\frac{z}{z_{*}}\right)^{1+\frac{2}{\nu}} \frac{dz}{\sqrt{f(z)\left[1 - \left(\frac{z}{z_{*}}\right)^{2(1+2/\nu)}\right]}}.
\end{eqnarray}

 We can remove $\epsilon$ from the upper limit in (\ref{lx1}) under the assumption that the turning point $z_*$ is above the horizon. Indeed, if the function $f$ is given by (\ref{V3.2c}) with the horizon defined by
\be
\label{m-zh}
z_h=\frac{1}{m^{\nu/(2 +2\nu)}},
\ee
the integrand in (\ref{lx1}) near $z=z_*$ can be represented as
\be\label{sing.h1}
\frac{1}{\sqrt{f\left(1 - \frac{z^{2 +4/\nu}}{z^{2 + 4/\nu}_{*}}\right)} }= \sqrt{\frac{\nu  z_*}{2(\nu
   +2)}} \,\frac{1}{ \sqrt{\left( z_*-z  \right)\left(1-m z_*^{2+\frac{2}{\nu}}\right)}}+{\cal {O}}\left(\sqrt{ z-z_*}  \right),\ee
thus, we have the integrable singularity for  $z_*<z_h$. However, for $z_*=z_h$
one obtains
\be
\frac{1}{\sqrt{\left(1-\left(\frac{z}{z_*}\right)^{2+ \frac{2}{\nu }}\right)
   \left(1-\left(\frac{z}{z_*}\right)^{2+ \frac{4}
   {\nu }}\right)}}
   =\frac{1}{2 \sqrt{\frac{\nu ^2+3 \nu +2}{\nu ^2
   z_*^2}} (z-z_*)}+\frac{-\nu -3}{4 \nu
   z_* \sqrt{\frac{\nu ^2+3 \nu +2}{\nu ^2
   z_*^2}}}+O\left((z-z_*)^1\right),\ee
  which leads to the logarithmic singularity.  For calculations of the entropy in the black hole background we assume that the turning point is below the horizon, while for the case of the shell we present the results for the case when the horizon is crossed.

Substituting  (\ref{dz1}) into (\ref{MS1.a}) and coming to the integration with respect to $z$-variable, one has
\begin{eqnarray}\label{HE.c1.bh}
\mathcal{A} = 2 L_{y_{1}}L_{y_{2}}\int ^{z_{*} }_{z_0}dz \, \mathfrak{a}(z),\,\,\,\,\,\, \mathfrak{a}(z)=\frac{1}{z^{1 + 2/\nu} \sqrt{f(z)\left(1- (z/z_{*})^{2(1 +2/\nu)}\right)}}.
\end{eqnarray}

In (\ref{HE.c1.bh})  we remove  $\epsilon$ assuming $m\neq 1$, but keep UV regularization  $z_0$.

The latter expression can be represented in terms of the dimensionless variable $w = z/z_{*}$ as
\begin{eqnarray}
\frac{\mathcal{A}}{2L_{y_{1}}L_{y_{2}}}  =  \frac{1}{z^{2/\nu}_{*}}\int^{1}_{z_0} \frac{ dw}{w^{1 + 2/\nu}\sqrt{f( z_{*}w)\left(1-w^{2(1 +2/\nu)}\right)}}.
\end{eqnarray}
The renormalized functional for the minimal surface reads
\begin{eqnarray}\label{A-ren}
\frac{\mathcal{A} _{ren}}{2L_{y_{1}}L_{y_{2}}}= \frac{1}{z^{2/\nu}_{*}}\int^{1}_0{\frac{dw}{w^{1 +2/\nu}}\left[\frac{1}{\sqrt{f\left(1 - w^{2 + 4/\nu}\right)}} - 1\right]} - \frac{  \nu}{2z^{2/\nu}_{*}}.
\end{eqnarray}
Taking into account (\ref{sing.h1}) one can also rewrite (\ref{lx1}) in the $w$-variable
\begin{eqnarray}\label{LSC1}
l_{x} = 2 \int^{1}_{0}z_{*}w^{1 + 2/\nu}\frac{dw}{\sqrt{f(z_{*}w)\left(1 - w^{2(1 +2/\nu)}\right)}},
\end{eqnarray}
for $z_* \neq z_{h}$.

One can see that the relation for the entanglement entropy is proportional to the area of the boundary
 $\partial A = L_{y_{1}}L_{y_{2}}$ which is in agreement to the area law.

\begin{figure}[tbp]
\centering
 \includegraphics[scale=0.44]{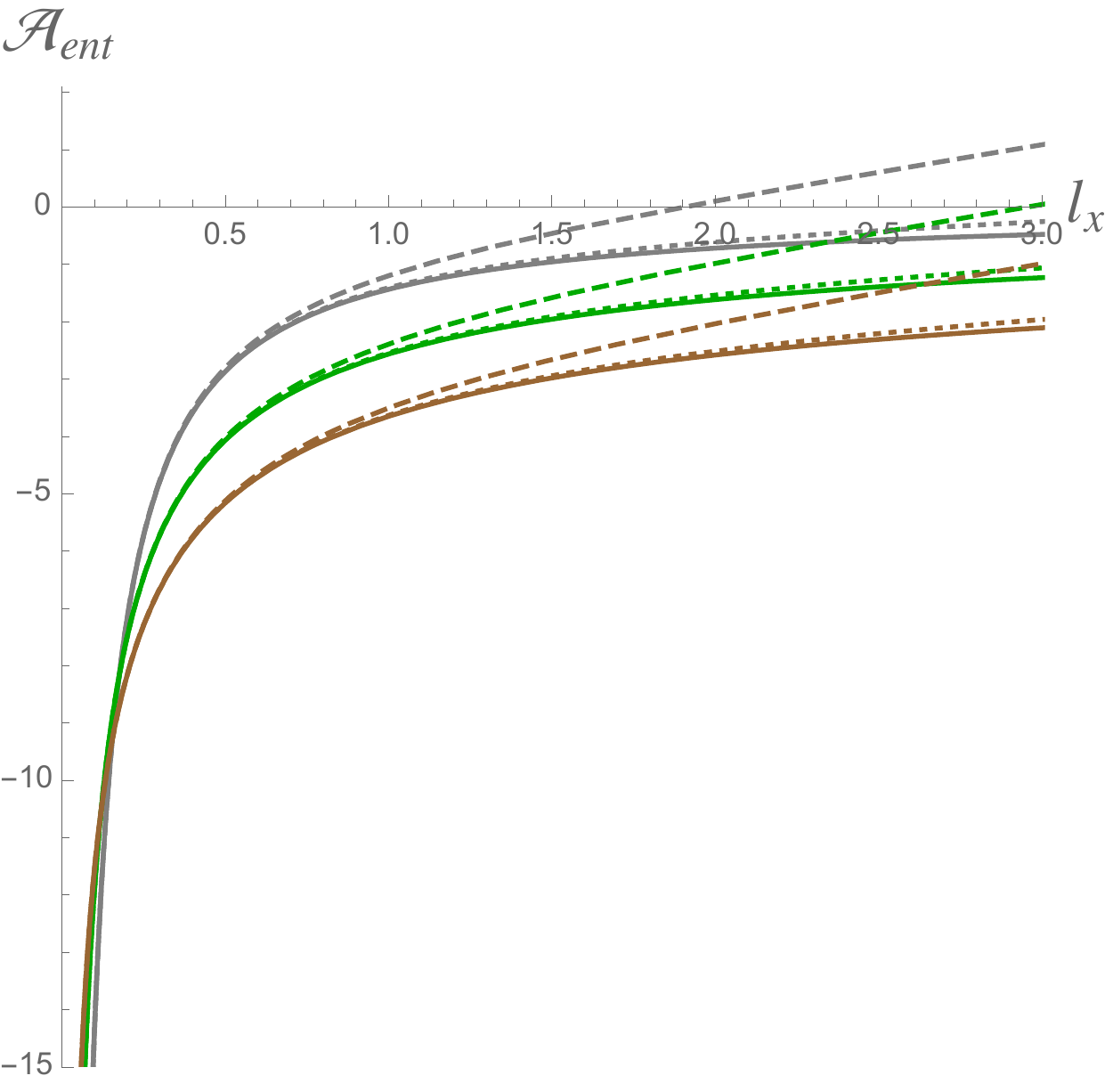}A$\,\,\,\,\,\,\,$
\includegraphics[scale=0.45]{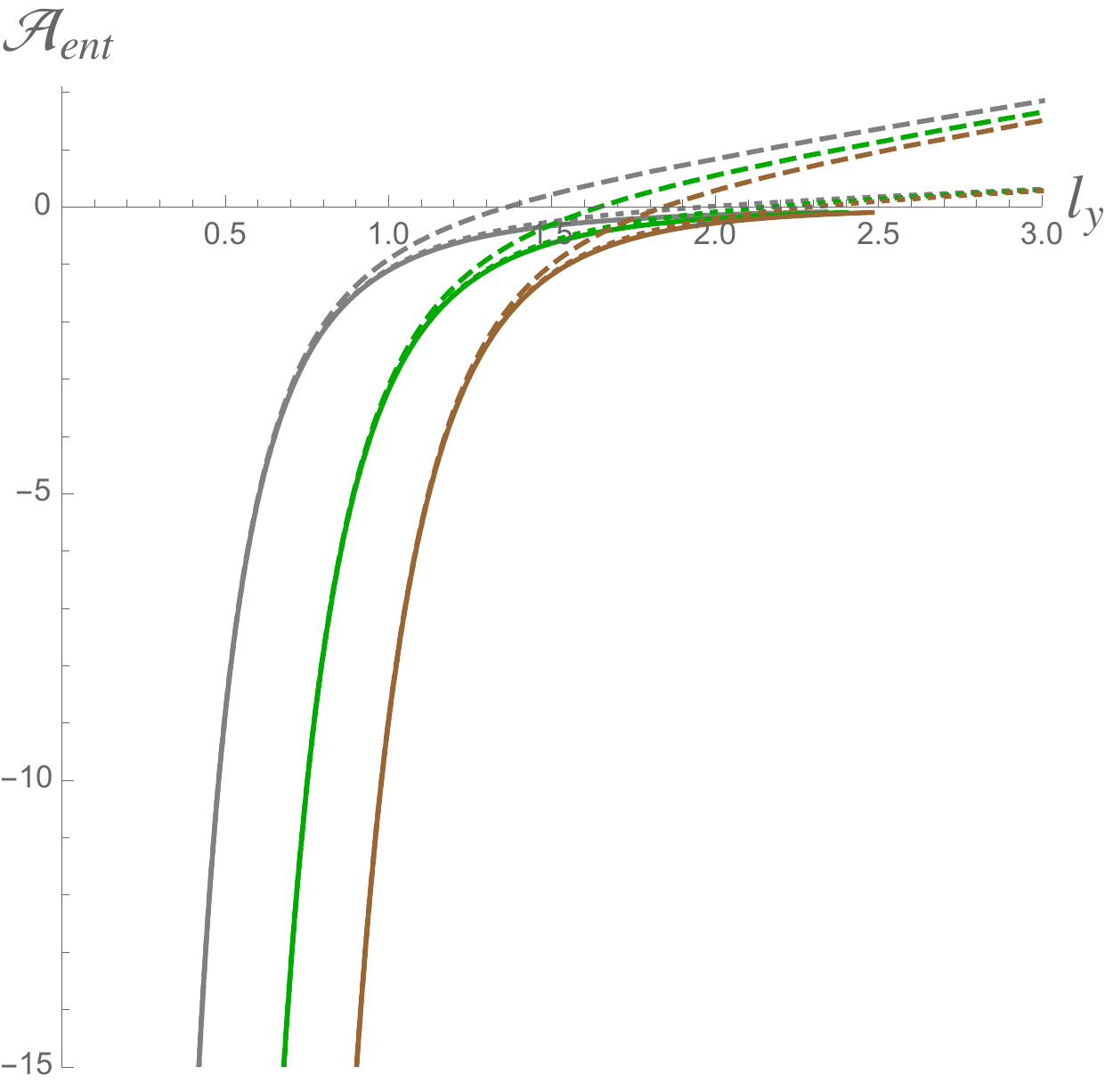}B
 \caption{{\bf A:} The  $\displaystyle{\frac{\mathcal{A}_{ren}}{2L_{y_{1}}L_{y_{2}}}}$  as a function of $\ell$  (\ref{LSC1}) in the 5d Lifshitz black brane background (\ref{V3.2})-(\ref{V3.2c}) for $\nu =2,3,4$ (the upper gray, middle green and lower brown curves, respectively). {\bf B:} Dependencies of $\displaystyle{\frac{\mathcal{A}_{ren}}{2L_{x}L_{y_{2}}}}$ (\ref{EntEnt_2}) on $\ell$ (\ref{LSC2}) in the black brane (\ref{V3.2})-(\ref{V3.2c}) for $\nu =2,3,4$ (from left to right, respectively). For both cases we plot for $m=0$ (solid lines), $m=0.1$(dotted lines) and $m=0.9$(dashed lines).}
 \label{fig:3}
\end{figure}

The behavior of the area (\ref{A-ren})  is presented in Fig.\ref{fig:3} A. 
To get  the dependence of the entanglement entropy of  the length for small values of $\ell$ one can consider the massless case. We see that  for $m=0$
the integrals  \eqref{A-ren} and \eqref{LSC1} can be calculated explicitly.  By analogy with \cite{ALT}  one gets 
\begin{eqnarray}\label{small-ell}
\mathcal{A}_{ren} \propto -\frac{1}{\ell^{2/\nu}}.
\end{eqnarray}
From numerical calculations we see that  for large $\ell$
\begin{eqnarray}\label{large-ell}
 \frac{\mathcal{A}_{ren}}{L_{y_{1}}L_{y_{2}}}  \approx\gamma_L(m) \ell +...
\end{eqnarray}
To keep the correct dimension we have to assume
\be
  \gamma_L(m)\propto \,m^{\frac{2+\nu}{2(1+\nu)}} \ee
  It should also be noted, that from Fig.\ref{fig:3}.A the dependence on the mass of the black brane for the intermediate $\ell$  is rather small.
The physical meaning of  estimation \eqref{large-ell} is that our surface for large $\ell$ becomes like a smothered parallelepiped almost touching to the horizon. 


\subsubsection{Subsystem delineated along the transversal direction}
Another possible subsystem $A$ can be divided along the $y_{1}$-direction (which is equivalent to
dividing it along $y_{2}$). It is also assumed that $z = z(y_{1})$ and
\begin{equation}\label{DS2}
x\in [0, L_{x}], \quad y_{1} \in [0, l_{y_{1}} < L_{y_{1}}], \quad  y_{2} \in [0, L_{y_{2}}].
\end{equation}

The three-dimensional minimal surface bordering on $\partial A$ has the form
\begin{eqnarray}\label{MS2}
\mathcal{A} = 2L_{x}L_{y_{2}} \int^{l_{y_{1}}}_{0} dy_{1}\mathcal{L},\,\,\,{\mbox {where}}\,\,\,\,\, \mathcal{L}=\frac{1}{z^{1+1/\nu}}\sqrt{\frac{1}{z^{2/\nu}} + \frac{(z^{'})^{2}}{f(z)z^{2}}}.
\end{eqnarray}

This dynamical system  has the following integral of motion
\begin{eqnarray}
 - \frac{z^{-(1+3/\nu})}{\sqrt{\frac{1}{z^{2/\nu}}+ \frac{1}{z^{2}f(z)}(z^{'})^{2}}} = \mathcal{C},
\end{eqnarray}
where $\prime = \displaystyle{\frac{d}{dy_{1}}}$.

The corresponding equation of motion reads
\begin{eqnarray}\label{TPEE2}
z^{'} = \pm z^{1-1/\nu} \sqrt{f(z)\left(\left(\frac{z}{z_{*}}\right)^{-2(1 + 2/\nu)} -1\right)},
\end{eqnarray}
where $z_{*}$ is related with ${\cal C}$ as $z_{*}^{1+2/\nu}={\cal C}^{-1}$.
The length scale $l_{y_{1}}$ can be defined in the following way
\begin{eqnarray}\label{LSC2}
l_{y_{1}} = 2 z^{1/\nu}_{*}\int^{1}_{0}\frac{w^{3/\nu} dw}{\sqrt{f(w,z_{*})\left(1-w^{2(1+2/\nu)}\right)}}.
\end{eqnarray}
We note that for the lower limit in (\ref{LSC2}) one can take $z_{0} = 0$. At the same time,  we can remove $\epsilon$ for the upper limit of (\ref{LSC2}) for $z_{*}< z_{h}$ , by the same reason as above in (\ref {lx1}).

Owing to (\ref{TPEE2}) the relation (\ref{MS2}) in terms of the dimensionless $w$-variable takes the form
\begin{eqnarray}\label{MS3}
\frac{\mathcal{A}}{2L_{x}L_{y_{2}}}= \frac{1}{z^{(1+ 1/\nu)}_{*}}\int^{1}_{z_{0}/z_{*}} \frac{1}{w^{(2+1/\nu)}}
\frac{1}{\sqrt{f(w z_{*})\left(1 - w^{2(1+2/\nu)}\right)}}  dw.
\end{eqnarray}

The renormalized functional for the minimal surface (\ref{MS3}) reads
\begin{eqnarray}\label{EntEnt_2}
\frac{\mathcal{A}_{ren}}{2L_{x}L_{y_{2}}} = \frac{1}{z^{1 + 1/\nu}_{*}}\,\left[ \int^{1}_{0}\frac{dw}{w^{2 +1/\nu}}\left(
\frac{1}{\sqrt{f\left(1 - w^{2 + 4/\nu}\right)}} - 1\right) - \frac{\nu}{1+ \nu}\right].
\end{eqnarray}

Numerical results for the entanglement entropy density (\ref{EntEnt_2}) for different values of $\nu$  are shown in Fig.~\ref{fig:3}. In a similar way with \cite{ALT} one can estimate for small $\ell$
 \begin{eqnarray}\label{small-ell2}
 \mathcal{A}_{ren} \propto -\frac{1}{\ell_{y}^{1+\nu}}
 \end{eqnarray}
 and numerical calculations  approximately give 
 \begin{eqnarray}\label{large-ell2}
 \mathcal{A}_{ren}\approx\gamma_T(m) \ell +...
 \end{eqnarray}
 when $\ell$ is large.
 To keep the correct dimension we have to write 
\be
 \gamma_{T}(m) \propto 
\,m^{\frac{2\nu + 1}{2(1+\nu)}}  
  \ee
for large $\ell$.  We see the dependence on the mass in Fig.\ref{fig:3}.B  for large $\ell$.  Note that the functions $\gamma_{L}(m)$ (\ref{large-ell})  and  $\gamma_{T}(m)$ (\ref{large-ell2}) are different.

\subsection{Entanglement entropy in a time-dependent background}

Now we come to studies of the evolution of entanglement entropy in the Lifshitz-Vaidya background (\ref{V3.2a})-(\ref{V3.2b}), describing the infalling shell. As before we will consider subsystems delineated along both transversal and longitudinal directions.

\subsubsection{Subsystem delineated along the longitudinal direction}
We once again start from the consideration of a subsystem $A$ extending along  $x$-direction, assuming that the minimal surface area is parameterized by
\begin{equation}\label{5.1}
v =v(x), \quad z = z(x).
\end{equation}

Taking into account (\ref{4.1b}), the volume functional corresponding to  the minimal three-dimensional surface is given by
\begin{eqnarray}\label{5.2}
\mathcal{A}
&=&2L_{y_{1}}L_{y_{2}}\int^{l_{x}}_{0} dx \mathcal{L},\\
\mathcal{L}&=& \frac{1}{z^{2/\nu +1}}{ \sqrt{1- f(z,v)(v')^{2} -2 v' z'}} .\label{lagr}
\end{eqnarray}
Here we suppose that $\prime \equiv \frac{d}{dx}$.
Substituting (\ref{5.2}) in (\ref{4.1})  we get the expression for the holographic entanglement entropy.

The Lagrangian $\mathcal{L}$ in (\ref{5.2}) has the integral of motion given by
\begin{eqnarray}\label{5.2a}
\mathcal{J} = - \frac{1}{z^{1 + \frac{2}{\nu}}\sqrt{\mathcal{R}}},
\end{eqnarray}
where we denote
\begin{eqnarray}\label{5.2c}
\mathcal{R} =  1 - f(z,v)(v')^{2} - 2 v'z'.
\end{eqnarray}
From  (\ref{5.2a}) we immediately obtain
\begin{eqnarray}\label{5.3}
z_*^{2 + 4/\nu} =z^{2 + 4/\nu}\,\mathcal{R},
\end{eqnarray}
 where $z_*$ is the turning point defined from the requirements $z'=v'=0$ and related with $\mathcal{J}$ as $z^{1 + 2/\nu}_{*} = \mathcal{J}^{-1}$.

The equations of motion  following from the Lagrangian $\mathcal{L}$  (\ref{lagr}) are
\bea\label{5.3a}
2\nu zv''&=& \nu z\,\frac{\partial f}{\partial z}\,v'^{2} + 2(2 +\nu)\left(1 - f(v')^{2} - 2v'z'\right),
\\\nn \label{5.3b}
2\nu zz''&=& - 2(2 + \nu)f + 4f^{2}v'^{2} + 2\nu f^{2} v'^{2} + 4(2 + \nu)fv'z' - \nu z v'^{2} \frac{\partial f}{\partial v} \\ \nn &-& \nu z f v'^{2}\frac{\partial f}{\partial z} 
-  2\nu z v'z'\frac{\partial f}{\partial z},
   \eea
which coincide with those from  \cite{CVJ} for $\nu=1$.

Taking into account (\ref{5.3}) the equations of motion (\ref{5.3a})-(\ref{5.3b}) can be rewritten as
\bea
2\nu z v'' &=& \nu z\frac{\partial f}{\partial z}v'^{2} + 2(2 + \nu)\frac{z_{*}^{2(1 + 2/\nu)}}{z^{2(1 +2/\nu)}},\\
2\nu z z''& = & - \left[2(2 + \nu)f \frac{z_{*}^{2(1 + 2/\nu)}}{z^{2(1 +2/\nu)}} + \nu zv'^{2}\frac{\partial f}{\partial z} + \nu z f v'^{2}\frac{\partial f}{\partial v} + 2\nu zv'z' \frac{\partial f}{\partial z} \right].
\eea

\begin{figure}[h!]
\centering
\begin{picture}(250,110)
\put(-80,0){\includegraphics[scale=0.33]{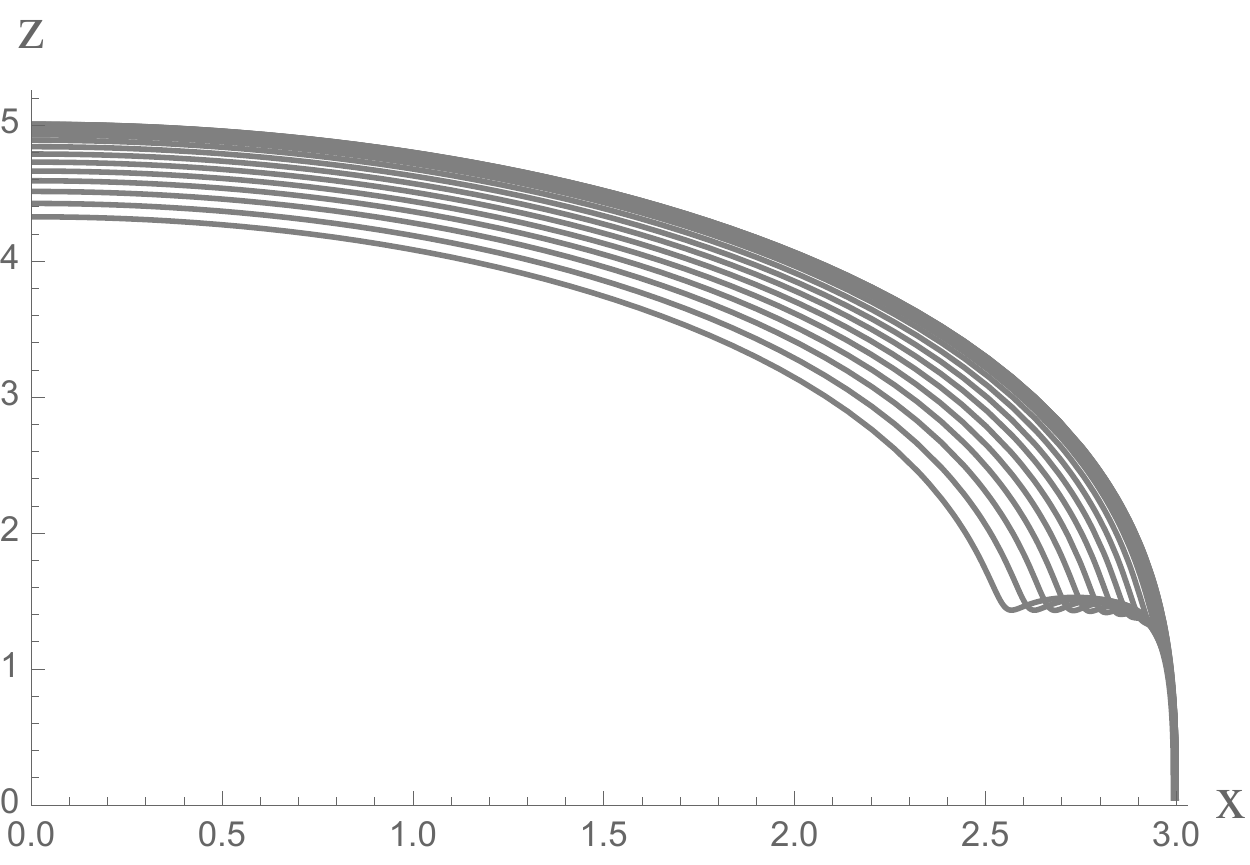}A}
\put(70,0){\includegraphics[scale=0.33]{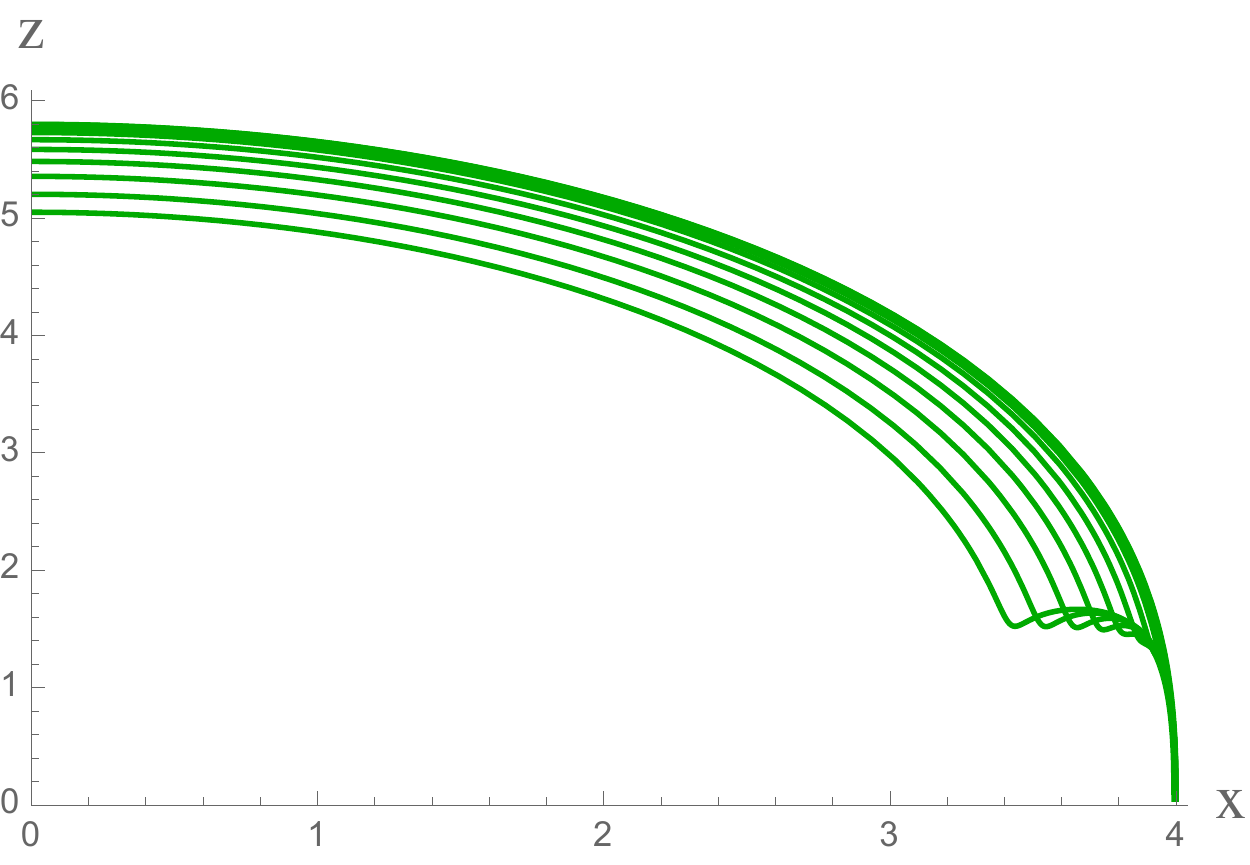}B}
\put(220,0){\includegraphics[scale=0.33]{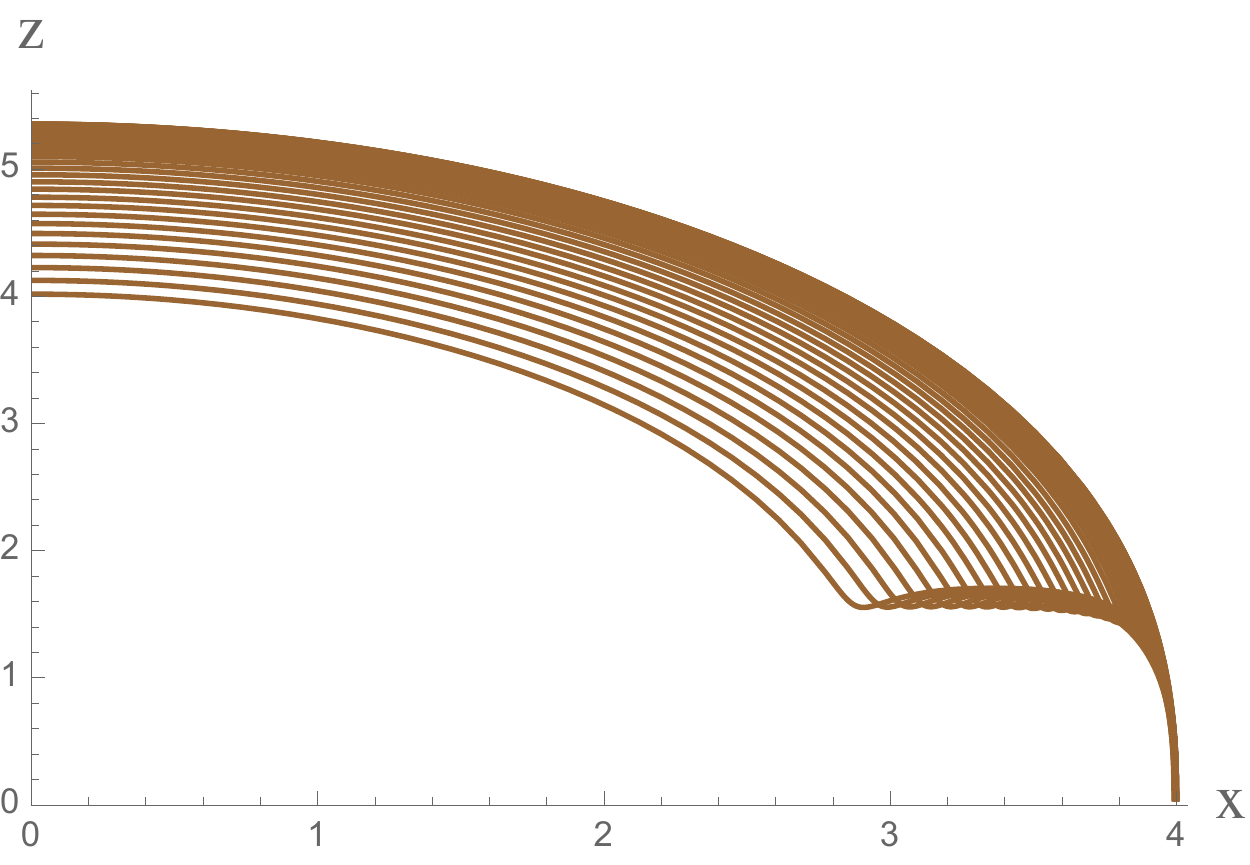}C}
\end{picture} \caption{The behavior of the solution for $z(x)$ to eqs. (\ref{5.3a})-(\ref{5.3b}). {\bf A:} $\nu=2$, $z(3) = 0$. {\bf B:} $\nu=3$, $z(4)=0$. {\bf C:} $\nu=4$, $z(4) = 0$.}
        \label{fig:zh}
  \end{figure}

Here we assume that the function $f$  has the form (\ref{f}). We can solve these equation numerically using the following initial conditions
\bea\label{5.3abc}
z(0)&=&z_*,\,\,\,\,\,z'(0)=0,\\ \label{5.3abc2}
v(0)&=&v_*,\,\,\,\,\,v'(0)=0.
\eea
We are interested in solutions which reach $z=0$ at some point $x_s$, similar boundary conditions have been proposed  for example in \cite{1110.1607,1512.05666} . The point $x_s$ is, in fact, a singular point of the solutions.
Solutions to eqs. (\ref{5.3a})-(\ref{5.3b}) obeying (\ref{5.3abc})-(\ref{5.3abc2}) are presented in Fig.\ref{fig:zh}.
We can observe that there are two types of such solutions that have a $z_*$  below and above the horizon.
It is useful to study a domain of the initial data, where these solutions can exist, see Appendix~\ref{App:B}.

It is also instructive to see the behavior of the quantity $fv'+z'$. An assumption, that the function
$f$ does not depend on $v$, yields to the fact that $fv'+z'$ is some conserved quantity.
At the same time for $f$ defined by (\ref{f}) it changes that we observe on  Fig.\ref{fig:z-mN}.
However, it can also be seen that at the end of the curve $z=z(x)$ at $x=x_{UV}$,
this quantity does not vary significantly and admits the approximation $\partial _v f=0$.

Owing to (\ref{5.3}) the minimal three-dimensional surface (\ref{5.2}) can be  represented  as
\begin{eqnarray}\label{5.4N}
\mathcal{A} &=&2 L_{y_{1}}L_{y_{2}}\int ^{\ell_x}_{0}\frac{dx}{z^{1+2/\nu}}
 \left(\frac{z_*}{z}\right)^{1 + 2/\nu}. \end{eqnarray}

Coming to the $z$-variable one can rewrite  (\ref{5.4N}) in the following form
\begin{eqnarray}\label{5.4NN}
\frac{\mathcal{A}}{2 L_{y_{1}}L_{y_{2}}} &=&-\int ^{z_*}_{z_0}dz\, \mathfrak{a}(z),
\eea
with $\mathfrak{a} (z)$ defined by
\bea\label{GotAC1}
\mathfrak{a} (z)&=&\frac{1}{z'z^{1+2/\nu}}  \,\left(\frac{z_*}{z}\right)^{1+2/\nu}.
\end{eqnarray}
In (\ref{GotAC1}) the RHS is taken with the negative sign since  $z'<0$ for the solutions of our interest.

To calculate the entanglement entropy we have to study the behavior of the integrand $\mathfrak{a}$ in (\ref{5.4N}).
For $z\sim 0$ we expect the following behaviour
\be
\mathfrak{a} (z)\sim\frac{1}{z^{1+2/\nu}} .
\ee
It is also convenient to introduce the quantity $\mathfrak{b} (z)$, defined by
\be\label{b-fr}
\mathfrak{b} (z)=\frac{1}{z'}  \,\left(\frac{z_*}{z}\right)^{1+2/\nu}.
\ee
We study the behaviour of function $\mathfrak{b}(z)$ on the solution to eqs.(\ref{5.3a})-(\ref{5.3b}) for $\nu =2$ and different masses is shown in Fig.\ref{fig:b-fr-n} (Appendix~\ref{App:B}).
We see that $\mathfrak{b} (z)\to C \neq 0$ for any value of mass, and therefore, we have $\mathfrak{a} (z)\sim C/z^{1+2/\nu}$.
From Fig.\ref{fig:b-fr-n} one can see that for $z_*=1$ we have $C=1$. Hence, the UV divergence is similar to the shell free case and
one can perform the similar renormalization
\begin{eqnarray}\label{A-renN}
\frac{\mathcal{A} _{ren}^{Shell}}{2L_{y_{1}}L_{y_{2}}}&=&-\left(\int ^{z_*}_{z_0}\frac{\left[\mathfrak{b}(z)-\mathfrak{b}(z_0)\right]}{z^{1+2/\nu}} \, dz
\,-\frac{\nu}{2}\,\frac{\mathfrak{b}(z_0)}{z_*^{2/\nu}}\right). \label{A-renNC}
\end{eqnarray}

Returning to the variable $x$  we obtain  the finite contribution to the entanglement entropy of the shell
 \bea\label{A-ren-ff}
\frac{\mathcal{A} _{ren}^{Shell}}{2 L_{y_{1}}L_{y_{2}}}&=&\int ^{\ell_x}_{\varepsilon}\frac{dx}{z^{1+2/\nu}}
 \left(\frac{z_*}{z}\right)^{1 + 2/\nu}
-\frac{\nu}{2} \frac{{\mathfrak{b}}(z_0)}{z_0^{2/\nu}}.
\eea

Now we can define the quantity $\Delta \mathcal {A}^{Shell-LV}$
\begin{eqnarray}\label{r}
\frac{\Delta \mathcal{A}^{Shell-LV}}{2 L_{y_{1}}L_{y_{2}}} &=&\frac{ \mathcal{A} ^{Shell} - \mathcal{A}^{LV}}{2 L_{y_{1}}L_{y_{2}}}.
  \eea
It should be noted that the holographic entanglement entropy for the Lifshitz-Vaidya background depends on two parameters,
 $z_*$ and $v_*$, whereas for the pure Lifshitz case it depends only on $z_*$.
One takes $z_*$ in the second term in such a way that it gives the same distance $\ell$ as in the first term.
From (\ref{LSC1}) and  for $f=1$ one gets
\begin{eqnarray}\label{LSC1m}
\nu=2&:\,&l_{sing}=l_{x}/2 \approx  0.59907  z_*, \\
\nu=3&:\,&l_{sing}=l_{x}/2\approx 0.68978 z_*,\\
\nu=4&:\,&l_{sing}=l_{x}/2\approx 0.74687z_{*}.
\end{eqnarray}

Taking into account that
\be
 \frac{\mathcal{A}^{LV}}{2 L_{y_{1}}L_{y_{2}}}=\int _{z_0}^{z_*}\frac{dz}{z^{1+2/\nu}}
\frac{1}{\sqrt{1-\left(\frac{z}{z_*}\right)^{2+ 4/\nu} }}=
\frac{a_{\nu, ren}}{z_*^{2/\nu}}+ \frac{\nu}{2}\frac{1}{z_0^{2/\nu}},
\label{Af1}
\ee
where $a_{2, ren}=-0.5991$, $a_{3, ren}=-1.03468$, and $a_{4, ren}=-1.49367$, we can explicitly write down $\Delta \mathcal{A}^{Shell-LV}$.
Thus, we have
\bea
\label{DeltaVEnu2}
\nu=2:\,\,\,\,\frac{\Delta \mathcal{A}^{Shell-LV}}{2 L_{y_{1}}L_{y_{2}}} &=&
\int ^{\ell_x-\epsilon}_{\varepsilon}\frac{z_*^2dx}{\left ( z_{_{f=f(z,v)}}(x)\right)^{4}}
-\frac{1}{z_0}+\frac{0.5991\cdot 0.59907}{\ell_{sing}},\\
\nn
\\
\nn
\\
\nn
\nu=3:\,\,\,\,\frac{\Delta \mathcal{A}^{Shell-LV}}{2 L_{y_{1}}L_{y_{2}}} &=&
\int ^{\ell_x-\epsilon}_{\varepsilon}\frac{z^{5/3}_{*}dx}{\left ( z_{_{f=f(z,v)}}(x)\right)^{10/3}}
-\frac{3}{2z^{2/3}_{0}}+\frac{1.03468\cdot (0.68977)^{2/3}}{\ell^{2/3}_{sing}},\\
\label{DeltaVEnu3}
\\
\nn
\nu=4:\,\,\,\,\frac{\Delta \mathcal{A}^{Shell-LV}}{2 L_{y_{1}}L_{y_{2}}} &=&
\int ^{\ell_x-\epsilon}_{\varepsilon}\frac{z_*^{3/2}\,dx}{\left ( z_{_{f=f(z,v)}}(x)\right)^{3}}
- \frac{2}{z^{1/2}_{0}}+\frac{1.49367\cdot (0.74687)^{1/2}}{\ell^{1/2}_{sing}}.
\\\label{DeltaVEnu4}
\eea

 Fig.\ref{EE-shelm1n2}  shows the behavior of the entropy density as a function the length $\ell$ for different values of the anisotropic exponent $\nu$. We see that the entanglement entropy increases in a linear regime at small distances like it was observed for the black brane case.  The dependence on the critical exponent grows with the reaching  the saturation value of the entropy.   In Figs. \ref{EE-shellm1n234}, \ref{EE-shellm1n234t25} we demonstrate the evolution of the entanglement entropy in time.
 Note that in Figs. \ref{EE-shellm1n234} we show the difference between the entropy in the current time and the initial value of the entropy at $t=0$, i.e. the value of the entropy in the Lifshitz vacuum.  Figs.\ref{EE-shellm1n234t25}  show the difference between the entropy in the current time and the  value of the entropy at very large time (time when the thermalization has already taken place), i.e. 
 the difference between the entropy in the current time and thermal  entropy. We observe the kink in the evolution which was considered for Lifshitz  ($\nu=2$) and  $AdS$  ($\nu=1$) backgrounds in \cite{KKVT} and \cite{CVJ}, respectively. From Figs. \ref{EE-shellm1n234}, \ref{EE-shellm1n234t25} we see that the entanglement entropy increases almost linearly with time.  We note that after the saturation point had been reached the entropy flattens out. It  should also be mentioned that the saturation is faster for small values of $\ell$ and is almost independent on the anisotropic parameter $\nu$.

 \begin{figure}[h]
\centering
  \includegraphics[scale=0.4]{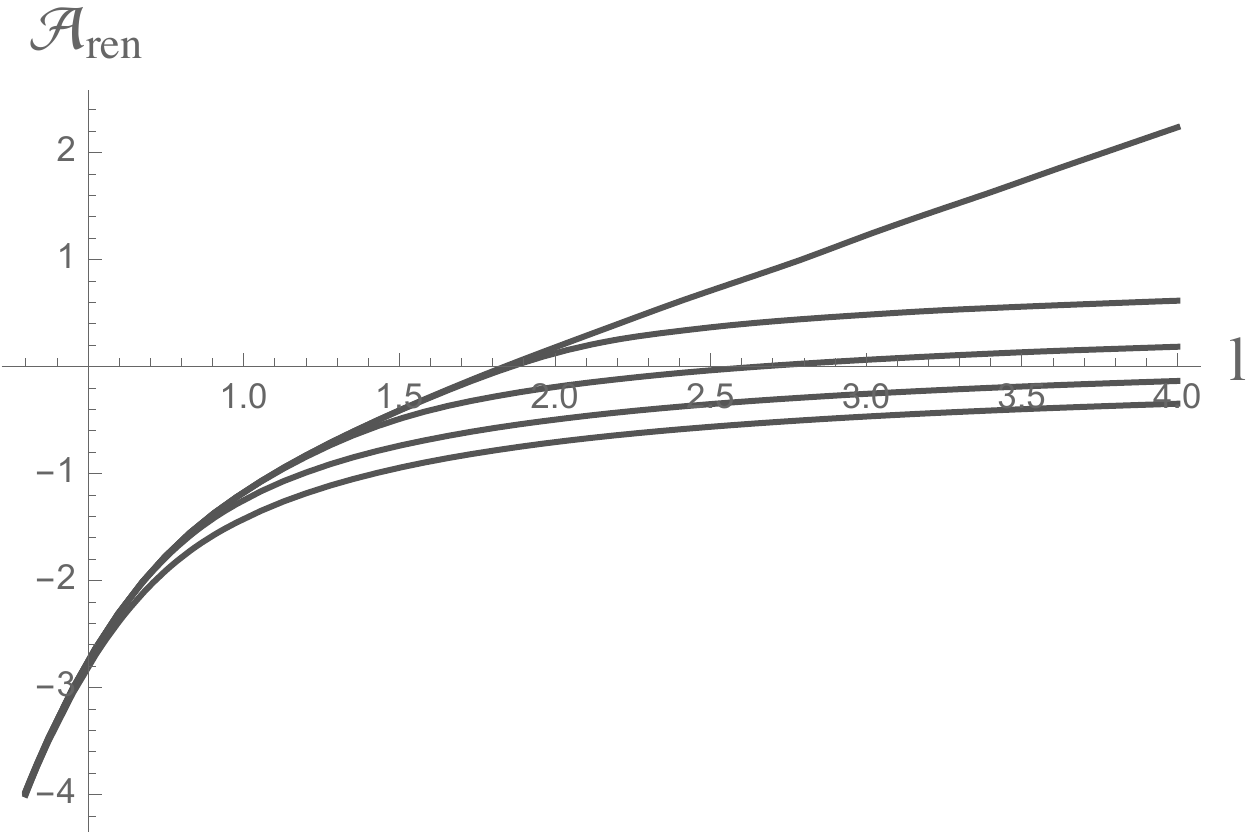}A$\,\,\,\,\,\,\,\,\,\,\,\,$
        \includegraphics[scale=0.4]{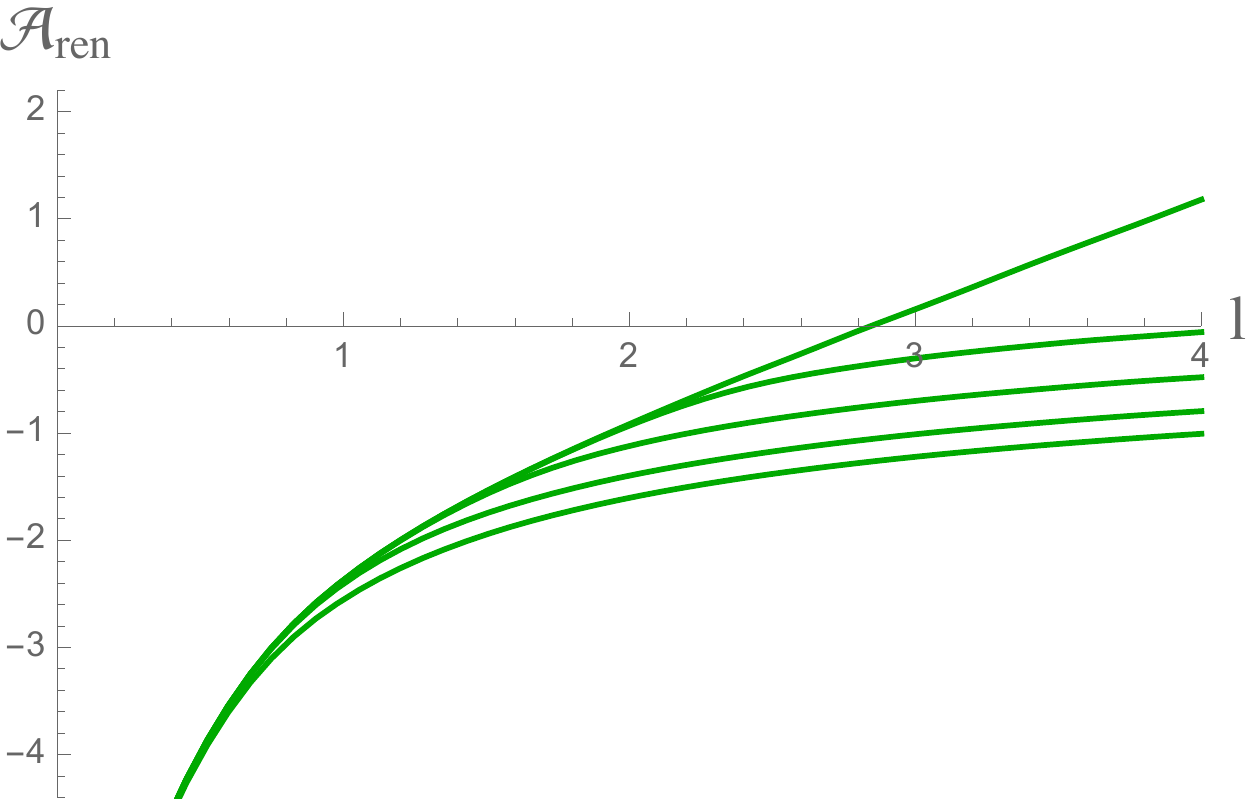}B\\
        \vspace{5mm}
      \includegraphics[scale=0.4]{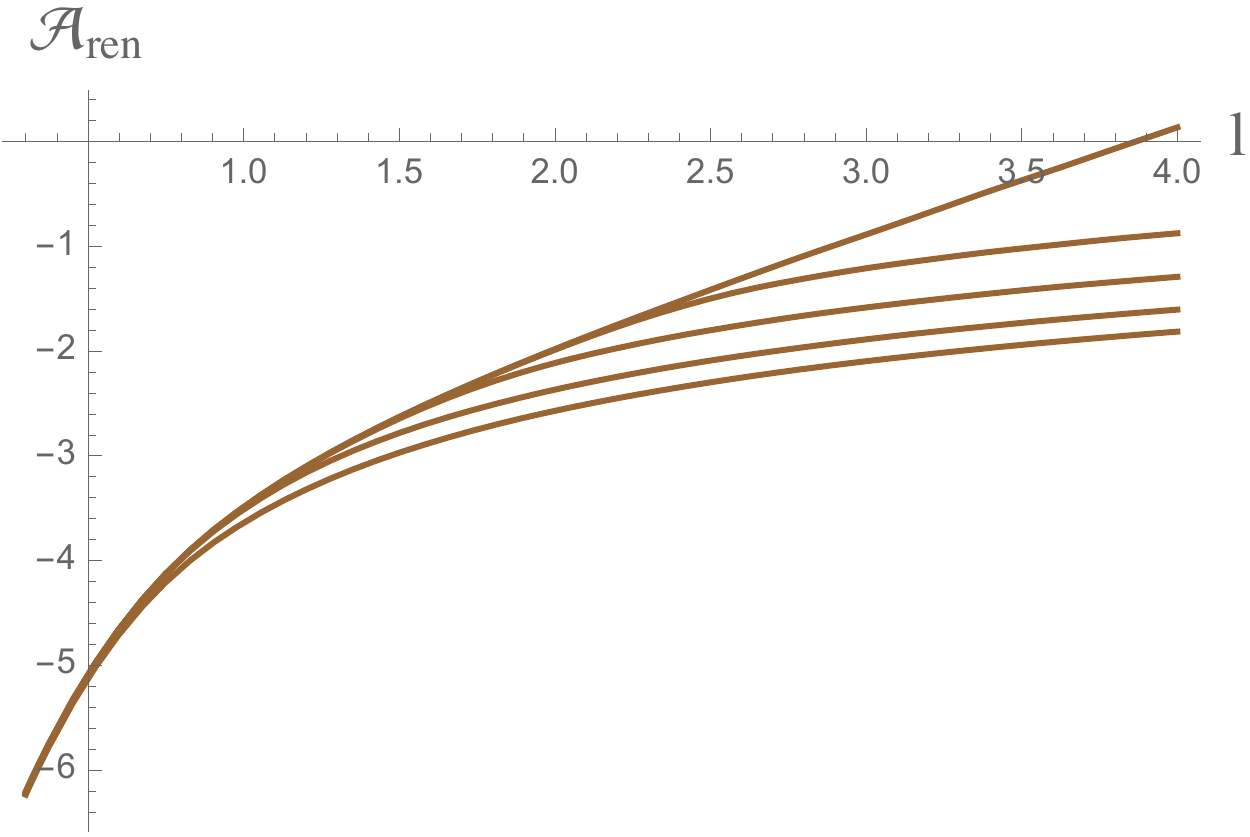}C$\,\,\,\,\,\,\,\,\,\,\,\,$
          \includegraphics[scale=0.4]{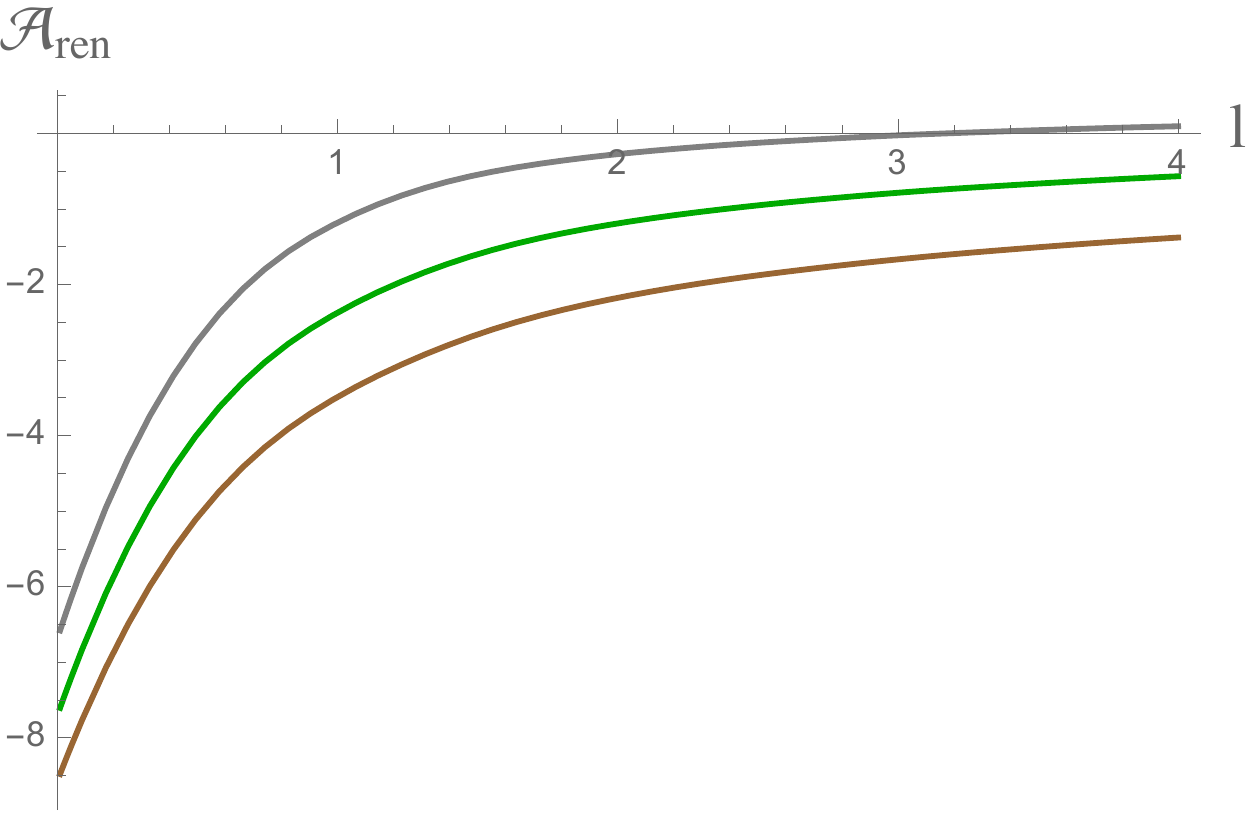}D
        \caption{The renormalized entanglement entropy at fixed $t = 0,0.6,1,1.4,3$, as a function of $\ell$ for a subsystem delineated along the longitudinal direction, $\nu =2,3,4$ ({\bf A, B, C}, respectively). In {\bf D} we plot the renormalized entanglement entropy as a function of $\ell$ at $t = 0.9$.The different curves correspond to the values $\nu =2,3,4$ from top to bottom.}
        \label{EE-shelm1n2}
  \end{figure}

 \begin{figure}[h]
\centering
  \includegraphics[scale=0.45]{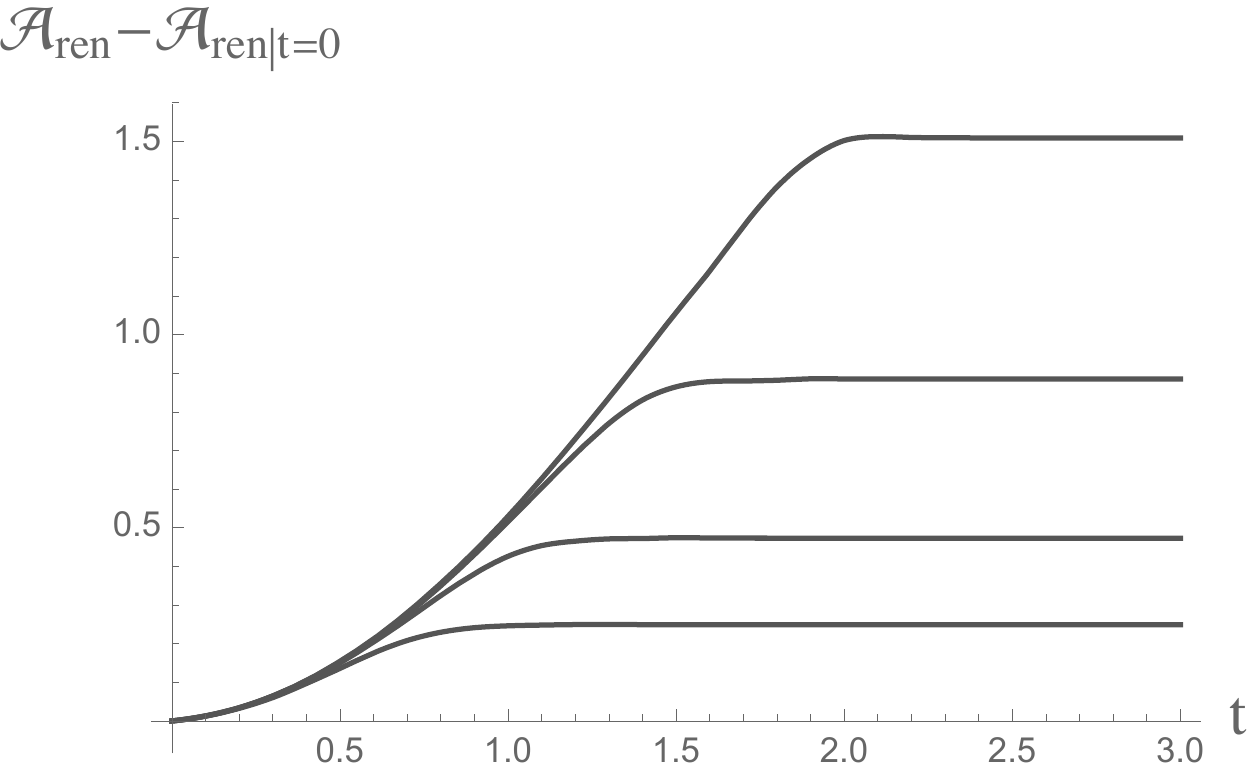}A$\,\,\,\,\,\,\,\,\,\,\,\,$
        \includegraphics[scale=0.45]{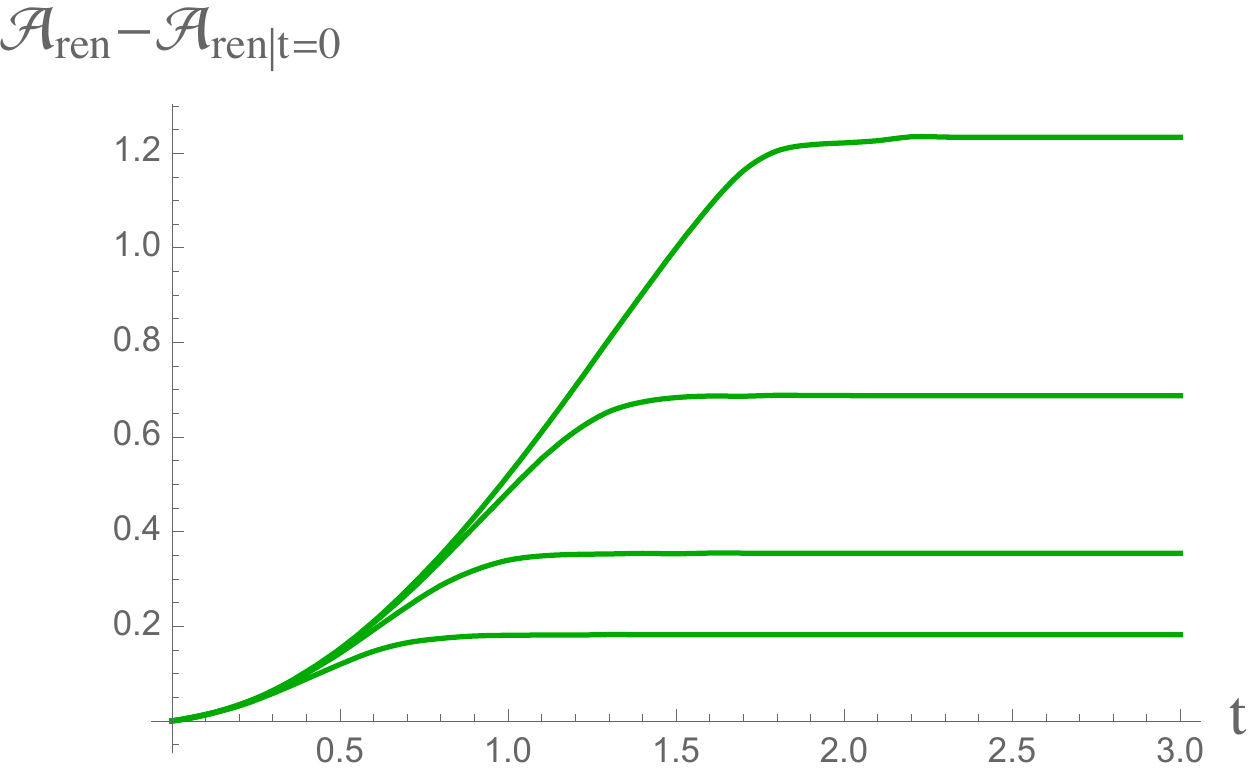}B\\
        \vspace{5mm}
      \includegraphics[scale=0.45]{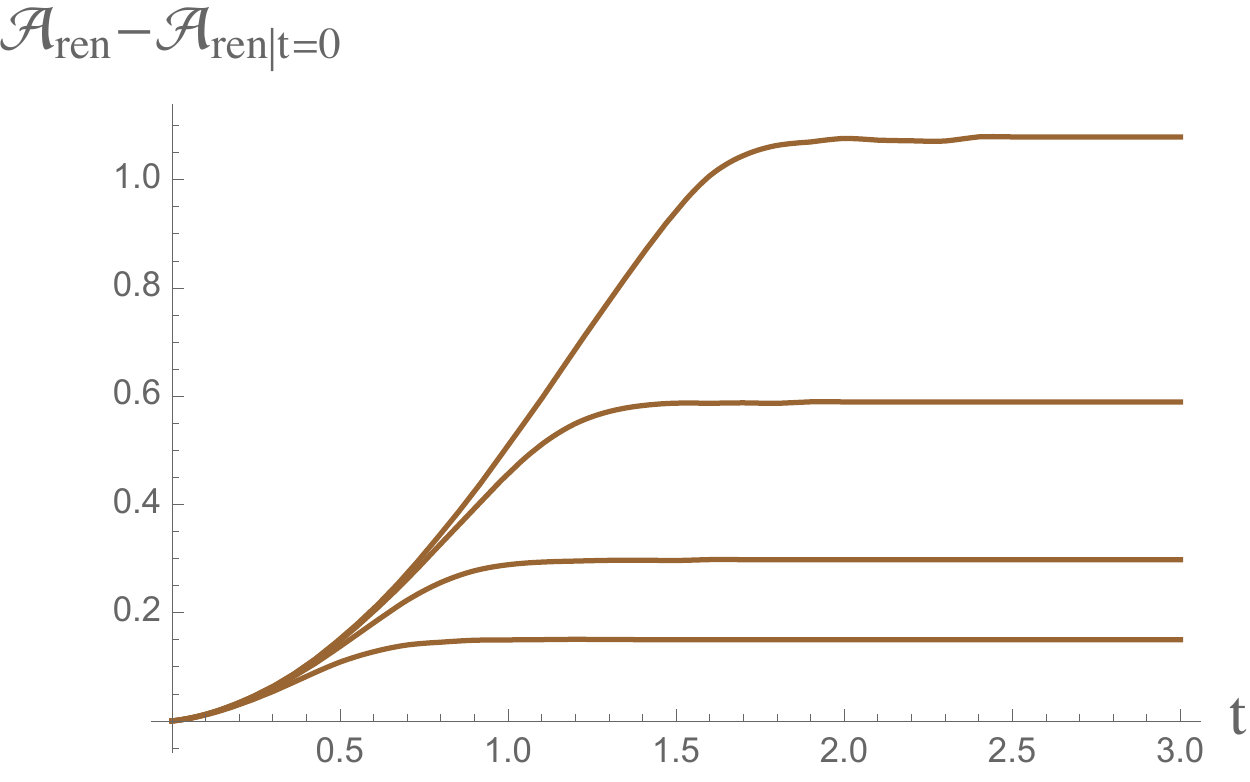}C$\,\,\,\,\,\,\,\,\,\,\,\,$
          \includegraphics[scale=0.45]{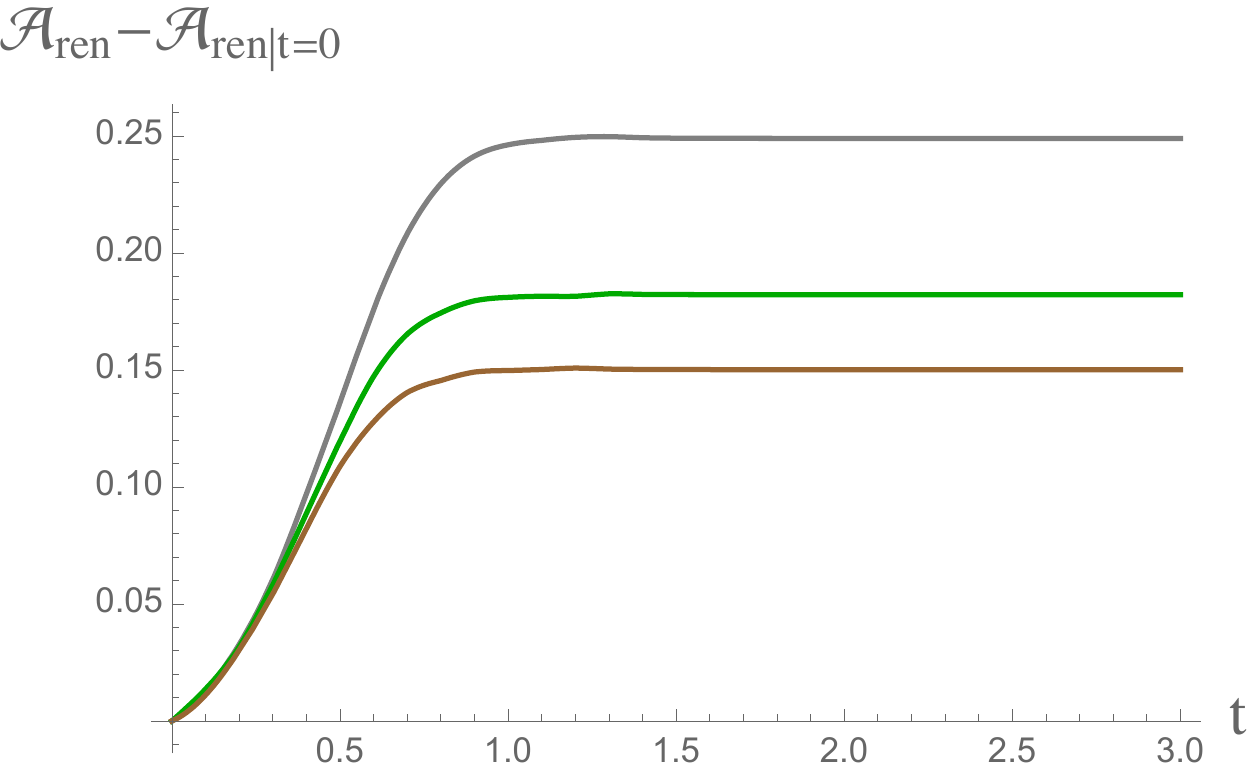}D\\     
          \caption{The time dependence of the holographic entanglement entropy  $\mathcal{A}_{ren}$ after the corresponding initial state subtraction ($t=0$)  at fixed $l = 1,1.4,2, 2.8$ for a subsystem delineated along the longitudinal direction,({\bf A, B, C}, respectively). In {\bf D} we plot the  time dependence of  $\mathcal{A}_{ren} - \mathcal{A}_{ren}|_{t=0}$ at $\ell = 1$. The different curves correspond to the values $\nu =2,3,4$ from top to bottom.}
        \label{EE-shellm1n234}
  \end{figure}
 \begin{figure}[h]
\centering
\includegraphics[scale=0.45]{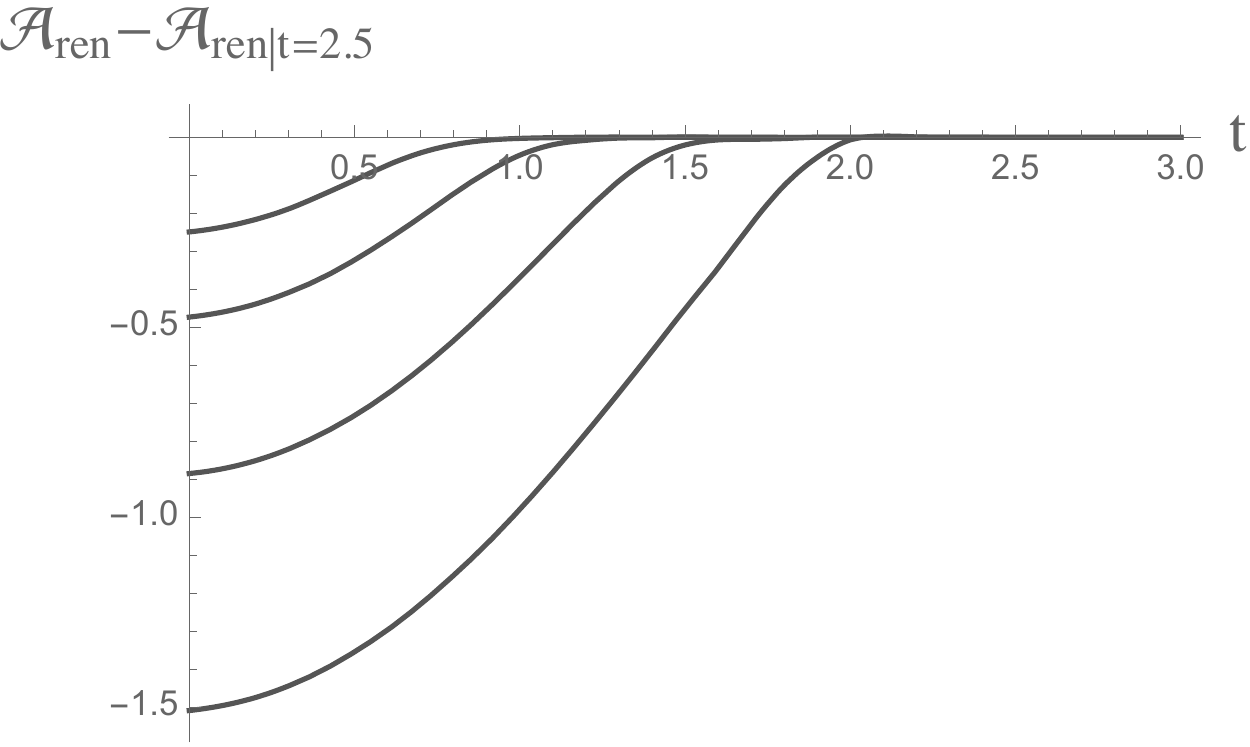}A$\,\,\,\,\,\,\,\,\,\,\,\,$
\includegraphics[scale=0.45]{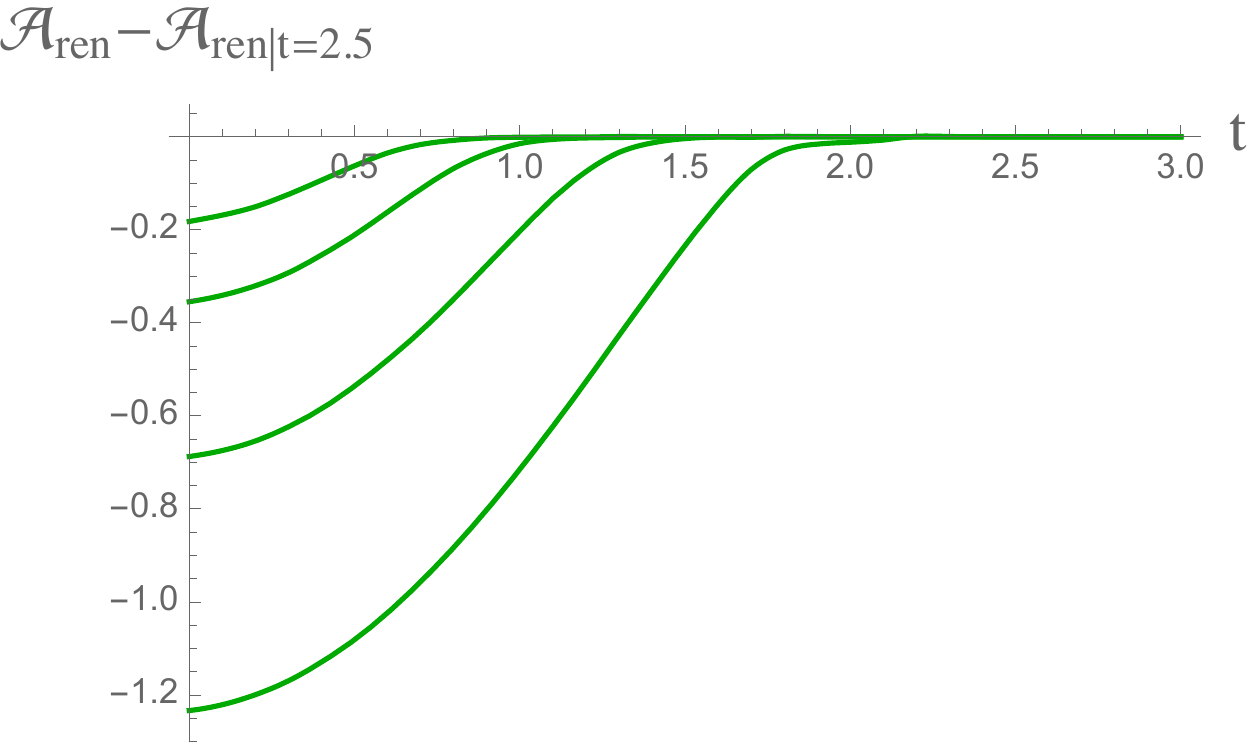}B\\
      \vspace{5mm}
\includegraphics[scale=0.45]{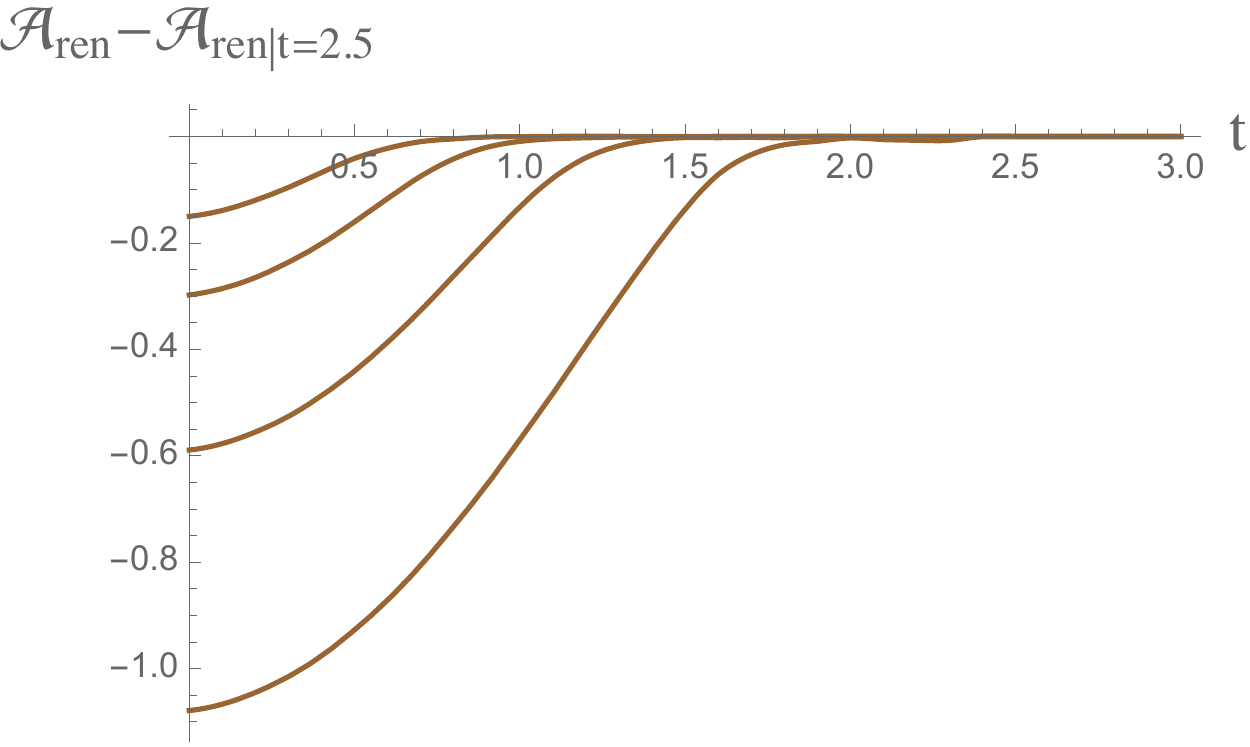}C$\,\,\,\,\,\,\,\,\,\,\,\,$
\includegraphics[scale=0.45]{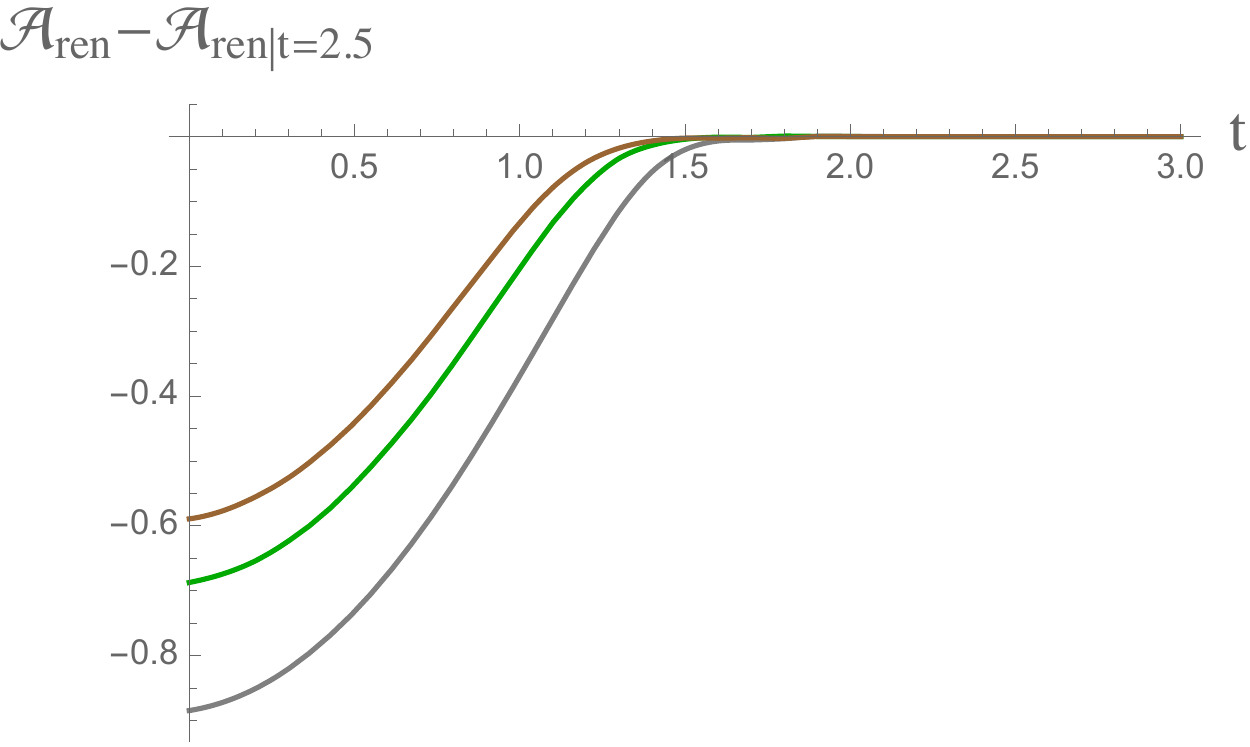}D        
          \caption{The time dependence of the holographic entanglement entropy  $\mathcal{A}_{ren}$  for the Lifshitz-Vaidya metric  after the corresponding subtraction of the state when the black brane has already been formed ($t = 2.5$) at fixed $l = 1,1.4,2, 2.8$ for a subsystem delineated along the longitudinal direction, ({\bf A, B, C}, respectively). In {\bf D} we plot the time dependence of $\mathcal{A}_{ren} - \mathcal{A}_{ren}|_{t=2.5}$ at $\ell = 2$  for the values of  $\nu =2,3,4$ from bottom to top.}
        \label{EE-shellm1n234t25}
  \end{figure}

\subsubsection{Subsystem delineated along the transversal direction} \label{Sect:4.2.2}

Now we turn to the case when  a subsystem $A$ is delineated along $y_{1}$ ($y_2$)-direction.
Parameterizing the minimal surface are by $v = v(y_{1})$, $z = z(y_{1})$ with
(\ref{DS2}),
we have
\begin{eqnarray}\label{6.1}
\mathcal{A} = 2L_{x}L_{y_{2}}\int^{l_{y_{1}}}_{0} dy_{1}\mathcal{L},
\eea
 with
 \bea\mathcal{L}=\frac{1}{z^{1+1/\nu}}\sqrt{\frac{1}{z^{2/\nu}} - \frac{1}{z^{2}}f(z,v)(v')^{2} - \frac{2}{z^{2}} v' z'},
\end{eqnarray}
where it is supposed that $\prime \equiv \frac{d}{dy_{1}}$.

The integral of motion corresponding to the system with Lagrangian $\mathcal{L}$  (\ref{6.1}) is
\begin{eqnarray}\label{6.1a}
\mathcal{J} = - \frac{1}{z^{1 + 3/\nu}\sqrt{\mathcal{R}}},
\end{eqnarray}
with
\begin{eqnarray}
\mathcal{R} =  \frac{1}{z^{2/\nu}} - \frac{1}{z^{2}}f(z,v)(v')^{2} - \frac{2}{z^{2}} v' z'.
\end{eqnarray}
Denoting
\be
\mathcal{J} = - \frac{1}{z_*^{1 + 2/\nu}},
\ee
we get  the conserved quantity
\be\label{6.3}
 z^{6/\nu}(z^{2 - 2/\nu} -v^{'2}f-2z'v')=z_*^{2 + 4/\nu}.
 \ee

The EOM corresponding to (\ref{6.1}) can be presented in the form
\bea\label{EOM-1-2}
2\nu z^{1+ \frac{2}{\nu}}v''&=& \nu z^{1+\frac{2}{\nu}} \frac{\partial f}{\partial z} v'^{2} - 2 f z^{2/\nu} v'^{2}\\ \nn
   &-&4 f \nu z^{2/\nu}v'^{2} - 4 \nu z^{2/\nu} v' z' - 8 z^{2/\nu} v' z'+2 \nu z^2 + 4 z^2,\\ \label{EOM-2-2}
 -2 \nu z^{1+\frac{2}{\nu}} z''&=& - 2 f^{2} z^{2/\nu}v'^{2} - 4 \nu f^{2} z^{2/\nu}v'^{2} +
  \nu z^{1 + 2/\nu}\frac{\partial f}{\partial v} v'^{2} + \nu z^{1 + 2/\nu}f \frac{\partial f}{\partial z}  v'^{2}  \\ \nn
  &-& 4fz^{2/\nu}v'z' - 8\nu fz^{2/\nu}v'z' + 2\nu z^{1+2/\nu}\frac{\partial f}{\partial z}v'z'+4z^{2/\nu}z'^{2}  - 4\nu z^{2/\nu}z'^{2} + 4fz^{2} + 2\nu fz^{2}.
  \label{EOM-2-2}
\eea

Taking into account (\ref{6.3}) the equations of motion (\ref{EOM-1-2})-(\ref{EOM-2-2}) can be re written as
\bea
2\nu z^{1+ \frac{2}{\nu} } v'' &=& \nu z^{\frac{2}{\nu} +1}\frac{\partial f}{\partial z}v'^{2} + 2(2 + \nu)z^{2}\frac{z_{*}^{2(1 + 2/\nu)}}{z^{2(1 +2/\nu)}} + 2(1- \nu)fz^{2/\nu}v'^{2},\\
 -2\nu z^{1+\frac{2}{\nu}} z''& = & 2(2\nu+ 1)f z^{2}\frac{z_{*}^{2(1 + 2/\nu)}}{z^{2(1 +2/\nu)}} + \nu z^{1+ \frac{2}{\nu}} v'^{2}\frac{\partial f}{\partial z} + \nu z^{1+ \frac{2}{\nu} }  f v'^{2}\frac{\partial f}{\partial v} \\
 &+& 2\nu z^{1+ \frac{2}{\nu}} v'z' \frac{\partial f}{\partial z}  + 4(1-\nu)z^{2/\nu}z'^{2}.\nn
\eea

Here we assume that the function $f$  is  given by (\ref{f})  as in the previous section
and we solve (\ref{EOM-1-2})-(\ref{EOM-2-2}) with the same boundary conditions
\bea\label{EOM-1-2bc}
z(0)&=&z_*,\,\,\,\,\,z'(0)=0,\\ \label{EOM-1-2bc2}
v(0)&=&v_*,\,\,\,\,\,v'(0)=0.
\eea

We consider once again only solutions that reach $z=0$ at some point $x_s$, that
 is in fact a singular point of the solutions. In Fig.\ref{fig:zx2m} we plot  solutions to (\ref{EOM-1-2})-(\ref{EOM-2-2}) with (\ref{EOM-1-2bc})-(\ref{EOM-1-2bc2}). We also present domains of the initial data plane, where these solutions can
exist  in Fig.\ref{fig:sing-area}-\ref{fig:sing-areah}, see Appendix~\ref{App:B}. We show  the position of the singular point corresponding to the solution with given $z_*$ and varying $v_*$. We present more details about solutions to eqs. (\ref{5.3a})-(\ref{5.3b})  in Appendix~\ref{App:B}.

 \begin{figure}[h!]
\centering
\begin{picture}(250,110)
\put(-80,0){\includegraphics[scale=0.33]{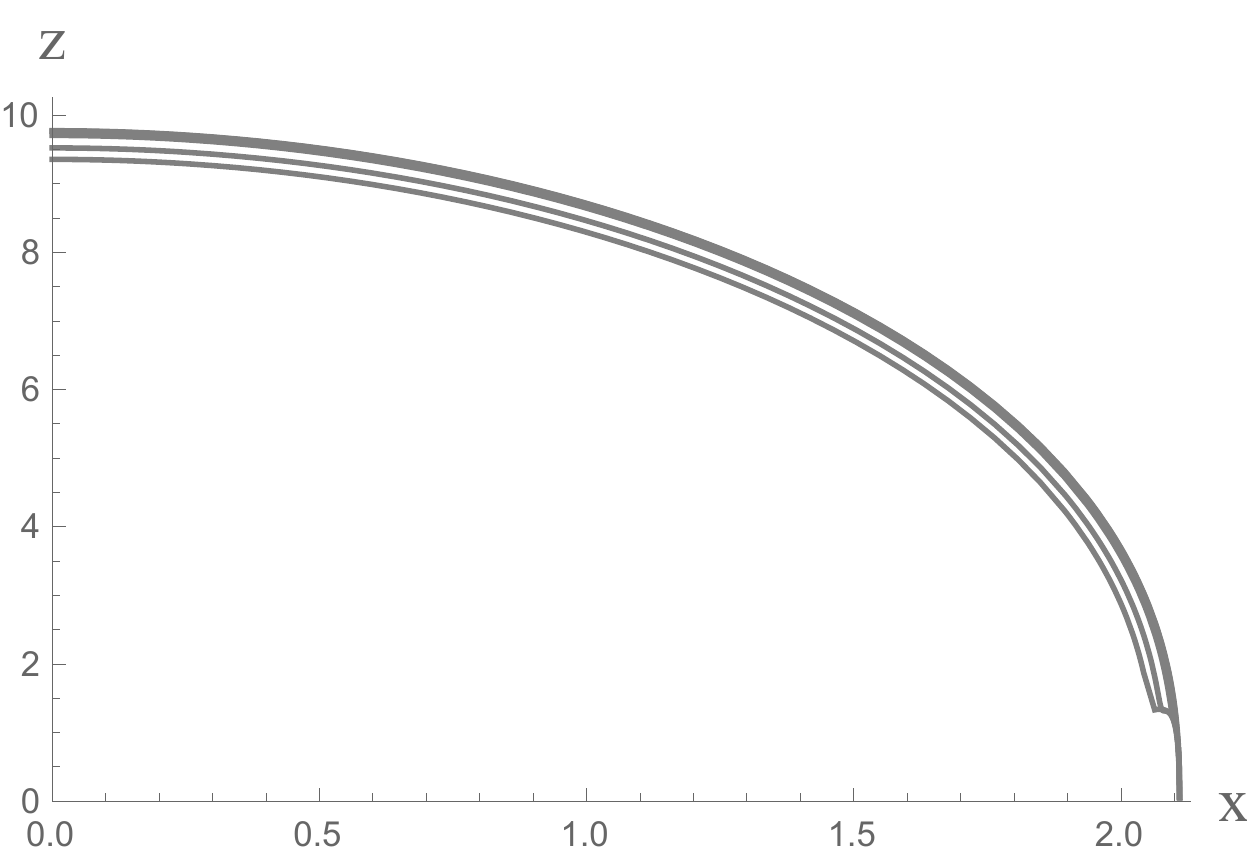}A}
\put(70,0){\includegraphics[scale=0.33]{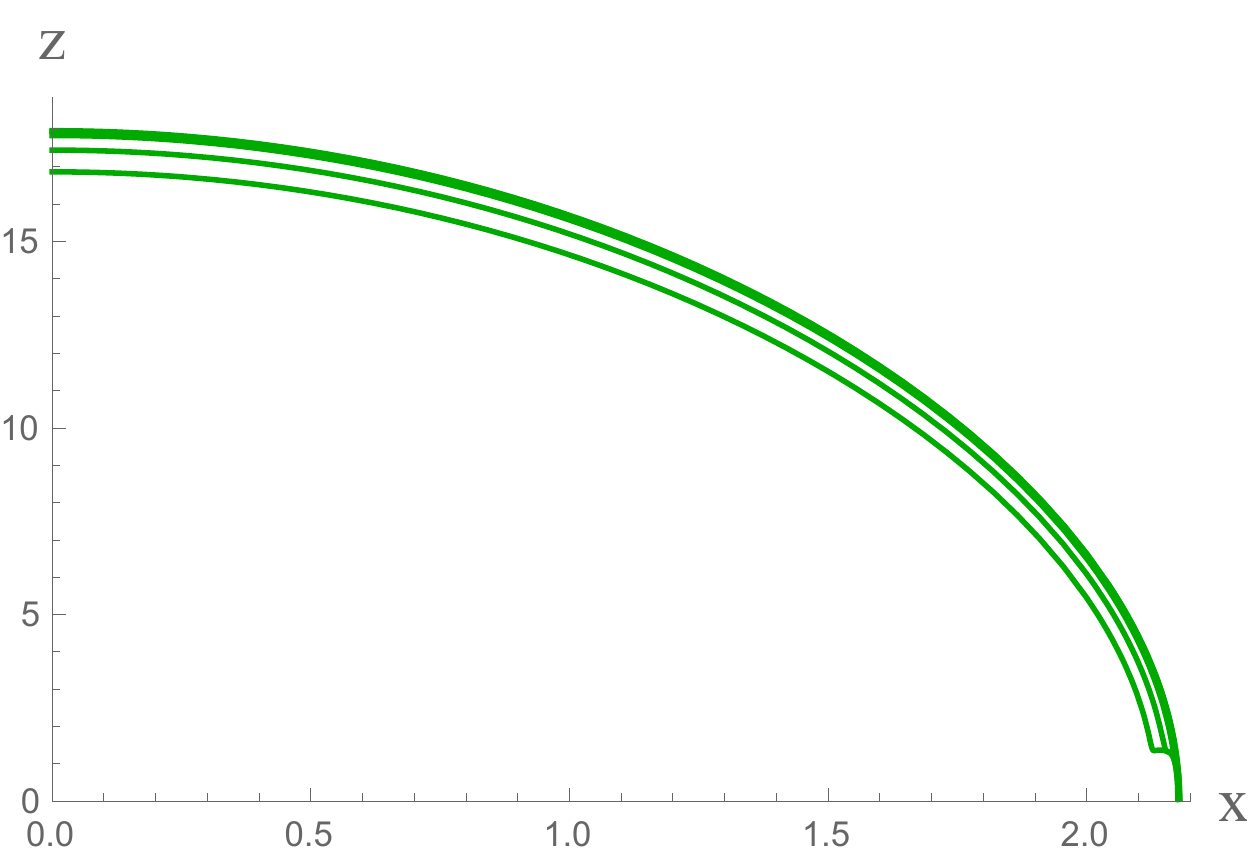}B}
\put(220,0){\includegraphics[scale=0.33]{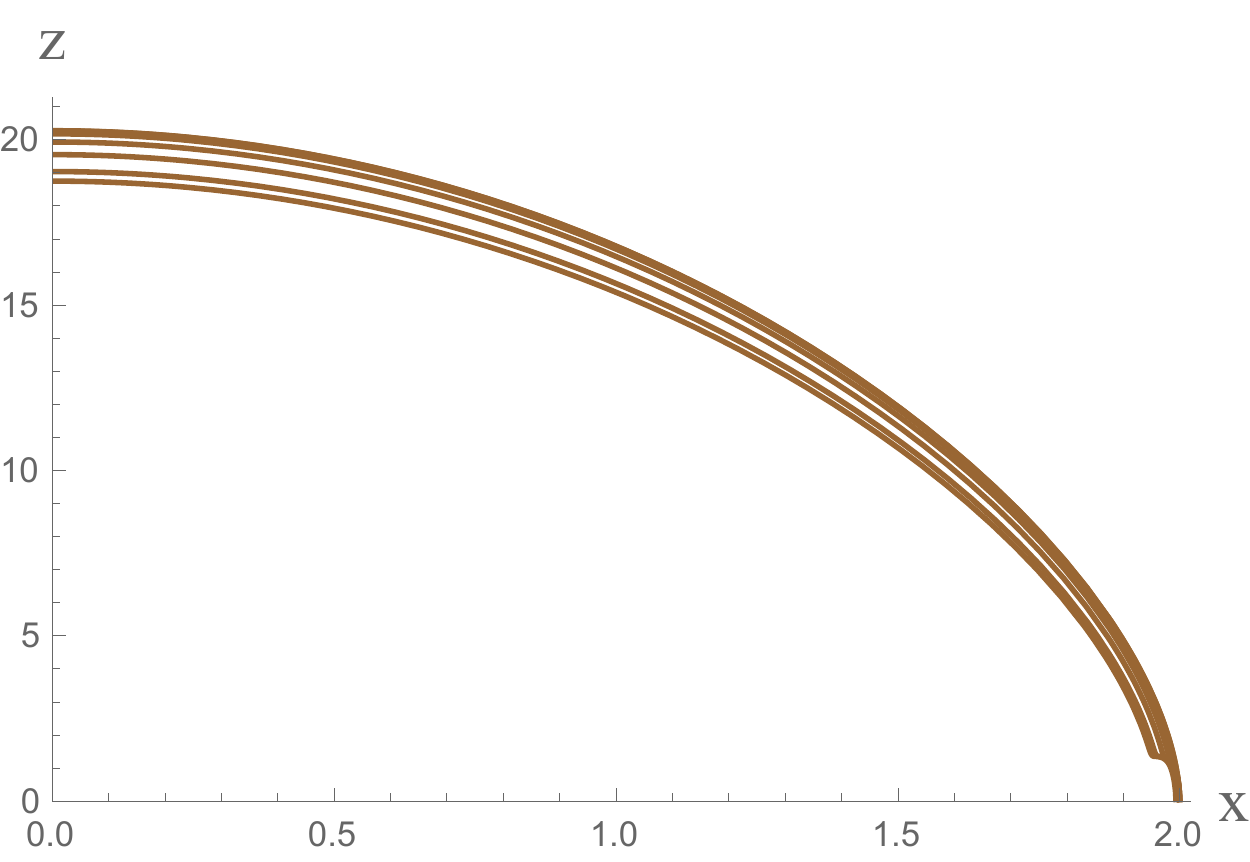}C}
\end{picture}
         \caption{The behavior of the profiles of the solution $z(x)$ to (\ref{5.3a})-(\ref{5.3b}). {\bf A:} $\nu =2$, $z(2.1) = 0$. {\bf B:} $\nu = 3$, $z(2.2) =0$. {\bf C:} $\nu = 4$, $z(2.1) = 0$.}
        \label{fig:zx2m}
 \end{figure}

One  can rewrite  (\ref{6.1}) in the following way
\bea\label{A-2case}
{\cal A} &= &  2L_{x}L_{y_{2}}\int^{l_{y_{1}}}_{0} dy_{1}  \frac{z_*^{1+2/\nu}}{z^{2+4/\nu}}\label{HEE-x}.
\eea

On the solution $z$ the functional (\ref{A-2case}) can be presented as
\bea\label{HEE-xmm}
\frac{{\cal A}}{2L_{x}L_{y_{2}}}
&= & -\int^{z_*}_{z_0}   \,dz\,\mathfrak{a}(z),
\eea
where
\bea
\mathfrak{a} (z)&=&\frac{1}{z^{2+1/\nu}}\mathfrak{b}(z),
\end{eqnarray}
and
\be\label{b-got-2}
\mathfrak{b}(z)= \frac{z_*^{1+2/\nu}}{z'\,z^{3/\nu}}.
\ee
 As in the previous section we derive the factor  $\mathfrak{b}(z)$ thus $\mathfrak{b}(z)\to 1$ (see, Appendix~\ref{App:B}).
 In this case the UV behaviour is the same as in the vacuum case and we can represent the answer  to (\ref{HEE-x})
in the following form
\bea
\frac{{\cal A}_{ren}^{Shell}}{2L_{x}L_{y_{2}}}&=&-\left(
\int ^{z_*}_{z_0}\frac{\left[\mathfrak{b}(z)-\mathfrak{b}(z_0)\right]}{z^{2+1/\nu}} \, dz
\,-\frac{\nu}{1+\nu}\,\frac{\mathfrak{b}(z_0)}{z_*^{1+1/\nu}}\right).
\label{B-renNC1}
\eea
The finite contribution to the holographic entanglement entropy  can be  represented in the following way
\begin{eqnarray}\label{B-renNC}
\frac{\mathcal{A}^{Shell}_{ren}}{2L_{x}L_{y_{2}}} = \int^{l_{x}}_{\epsilon}\frac{dx}{z^{2 + 1/\nu}}\frac{z^{1+2/\nu}_{*}}{z^{3/\nu}} - \frac{\nu}{\nu +1}\frac{\mathfrak{b}(z_{0})}{z^{1 + 1/\nu}_{0}}.
\end{eqnarray}

\begin{figure}[h]
\centering
 \includegraphics[scale=0.4]{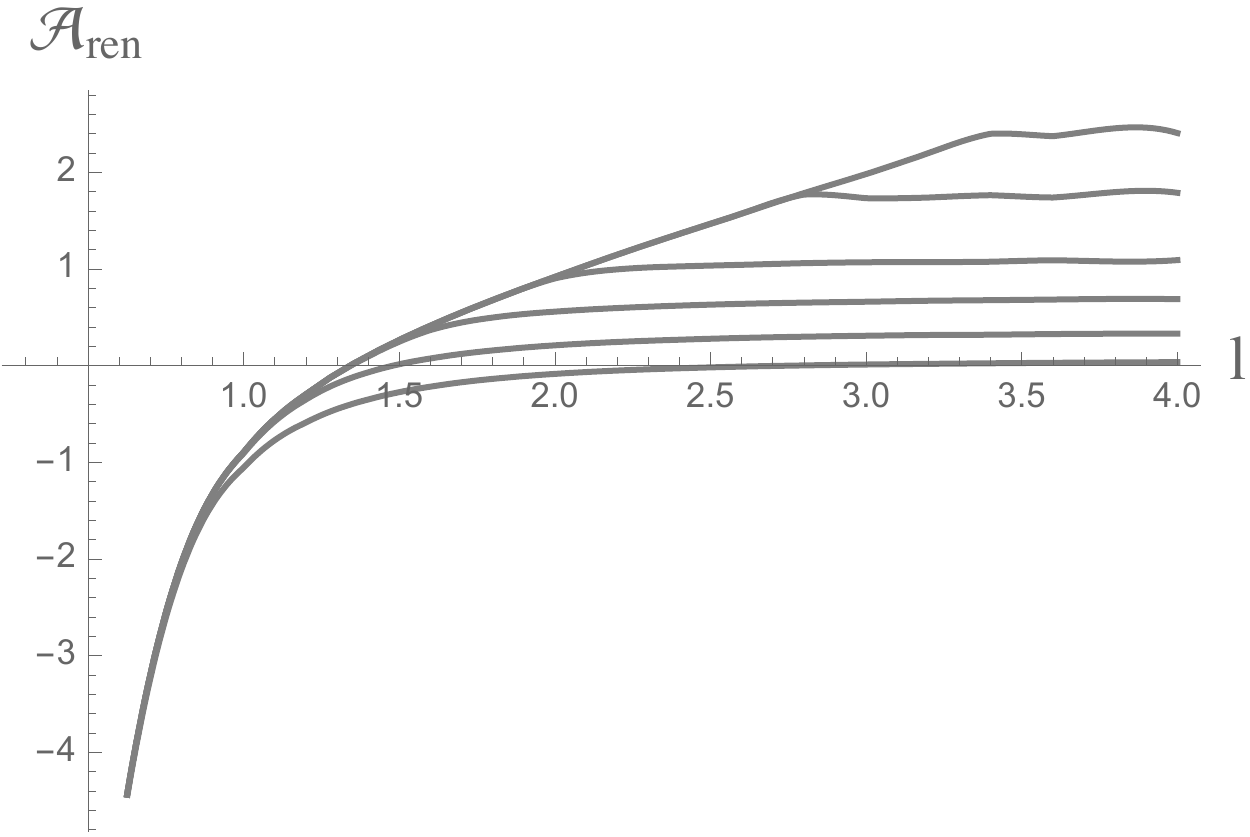}A$\,\,\,\,\,\,\,\,\,\,\,\,$
  \includegraphics[scale=0.4]{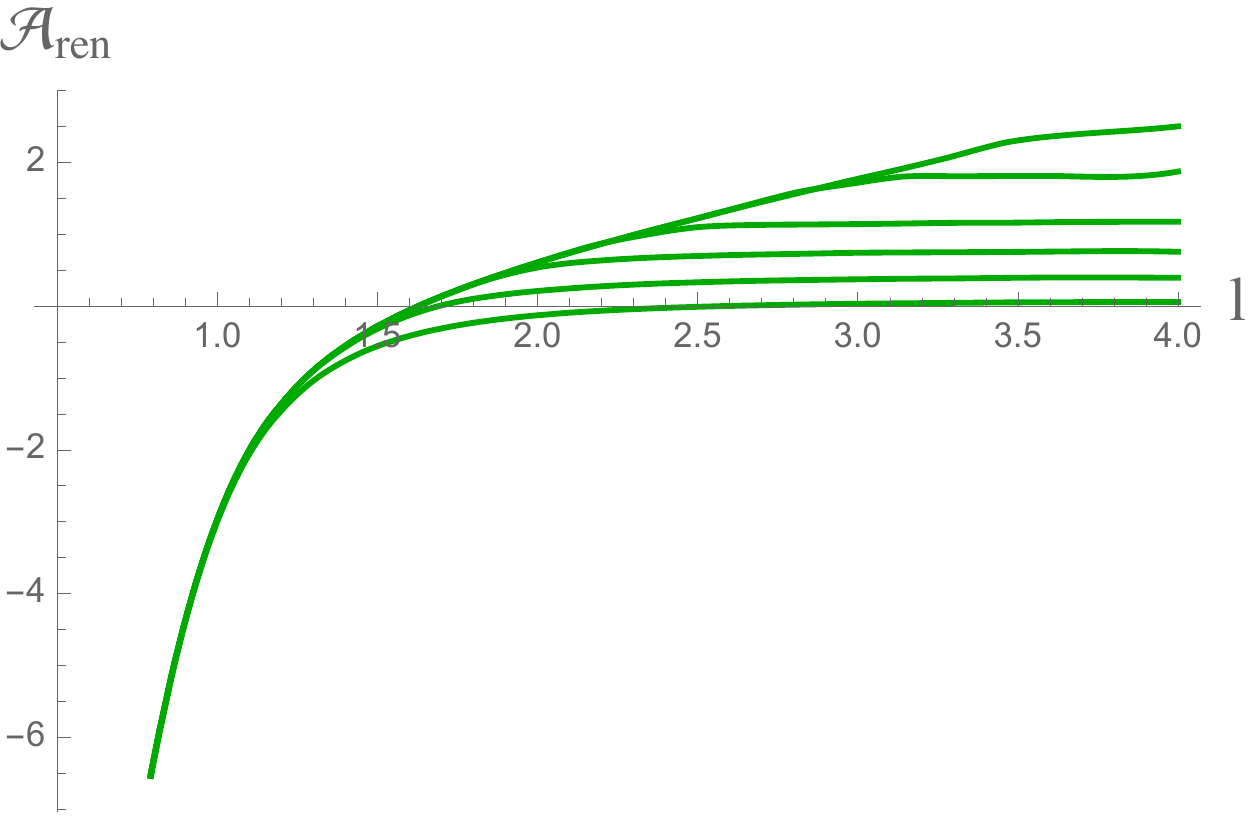}B\\
        \vspace{5mm}
\includegraphics[scale=0.41]{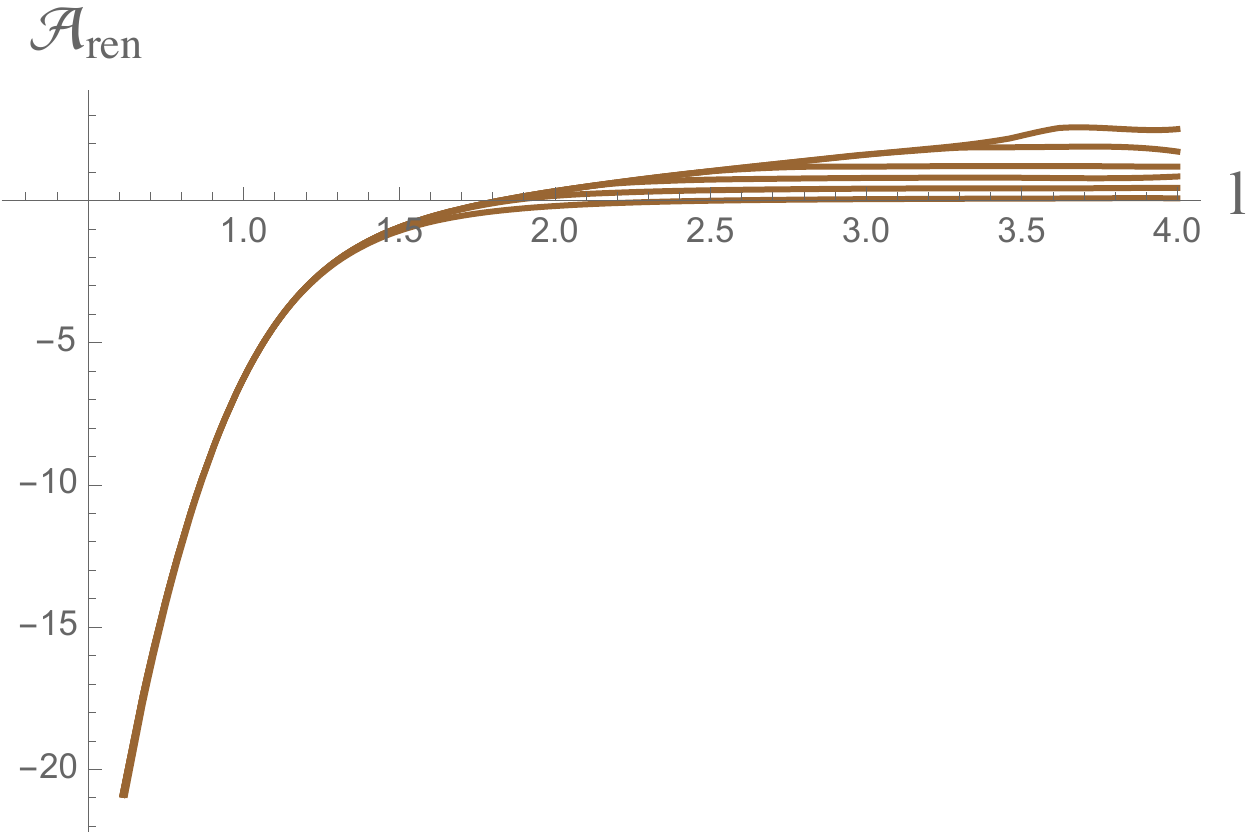}C$\,\,\,\,\,\,\,\,\,\,\,\,$
\includegraphics[scale=0.41]{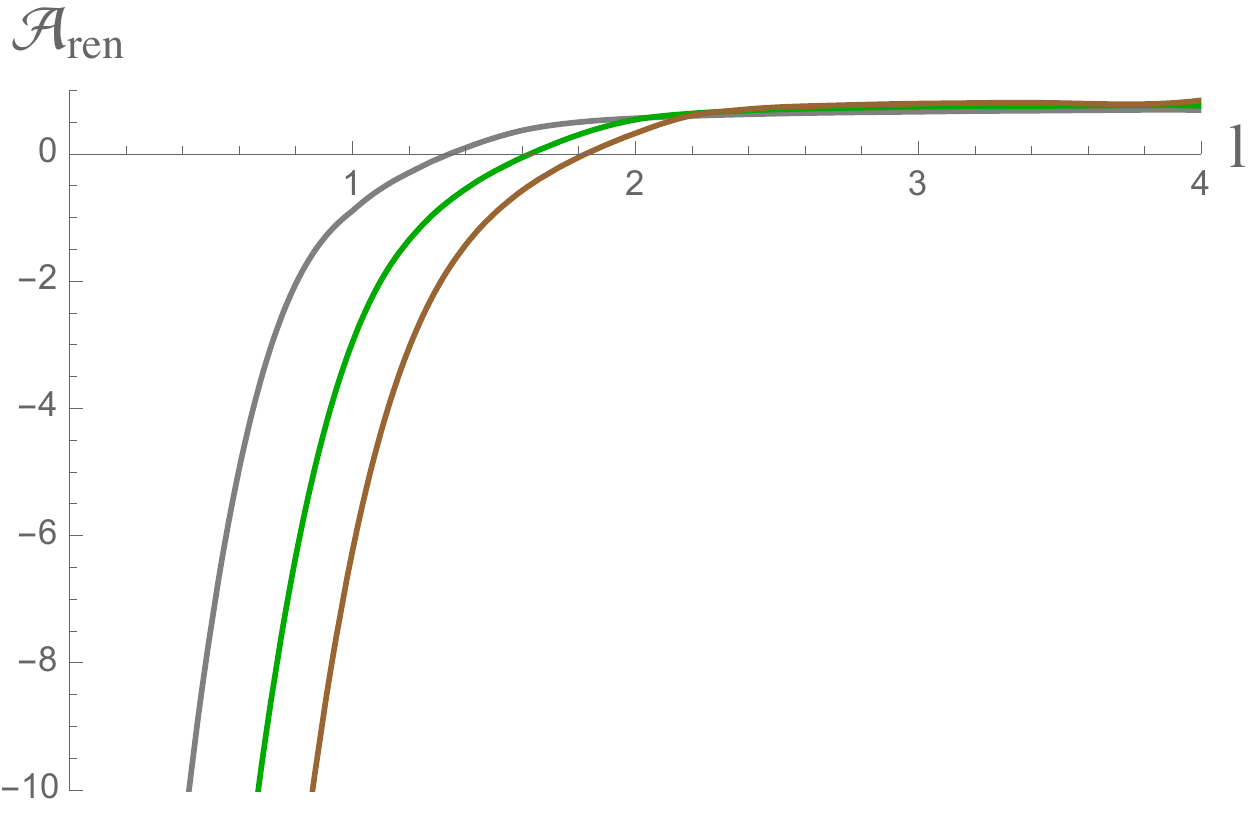}D
        \caption{The renormalized entanglement entropy at fixed $t = 0.1, 0.6, 1, 1.4, 2, 2.6$, as a function of $\ell$ for a subsystem delineated along the transversal direction,  $\nu =2,3,4$ ({\bf A, B, C}, respectively). In {\bf D} we plot the renormalized entanglement entropy as a function of $l$ at $t = 1$  for different values of $\nu$ (from left to right, respectively).}
        \label{fig:DHEEC2zzh}
 \end{figure}

The renormalized entanglement entropy (\ref{B-renNC}) as a funciton of $\ell$ is presented in Fig.\ref{fig:DHEEC2zzh}. From Fig.\ref{fig:3} {\bf B} and Fig.\ref{fig:DHEEC2zzh} {\bf D} one can see the entanglement entropy in time-dependent background has the similar behavior as for the static case. For small $\ell$ we observe the dependence of the entropy on $\nu$, which vanishes for large $\ell$, where  the entropy has linear behavior.

In three left panels of Fig.\ref{fig:A-ren2} we present the renormalized entanglement entropy (\ref{B-renNC}) as a function of $\ell$  and $t$.

Now as above let us define  $\Delta \mathcal{A}^{Shell-LV} _{reg}$ by
\bea \label{DeltaB}
&\,&\frac{\Delta \mathcal{A}^{Shell-LV}}{2 L_{x}L_{y_{2}}}= \frac{\mathcal{A}^{Shell}-
 \mathcal{A}^{LV}}{2L_{x}L_{y_{2}}}.
 \eea
Taking into account that  for ${\cal A}^{LV}_{reg}$ we have
 \bea
\frac{{\cal A}^{LV}}{2L_xL_{y_2}}&=&\frac{\nu}{1+\nu }
\frac{1}{z_0^{1+1/\nu }}+\left(\frac{L(\nu,0)}{l_{y_1}}\right)^{\nu+1}\, a_{\nu, ren}\label{ent2-f1-lm},
\eea
 where  $a_{2, ren} =-0.225$, $a_{3, ren} =-0.208$,$a_{4, ren} = -1.885$  and $L(\nu,0)$ is read from (\ref{LSC2}) with $m=0$, so
\begin{eqnarray}\label{LSC2m0}
\nu=2&:\,& L(2,0) \approx  0.67497,\\
\nu=3&:\,&L(3,0) \approx 0.8324,\\
\nu=4&:\,& L(4,0)\approx 0.9425.
\end{eqnarray}
Now we get
 \begin{eqnarray}\label{DeltaB}
\frac{\Delta \mathcal{A}^{Shell-LV}}{2 L_{x}L_{y_{2}}}&=&
\int ^{\ell_y}_{\varepsilon}\frac{z_*^{1 + 2/\nu}\,dx}{\left ( z_{_{f=f(z,v)}}(x)\right)^{2+4/\nu}}
-
 \frac{\nu}{1+\nu }
\frac{1}{(z(\ell_y))^{\frac{1+\nu}{\nu }}}-\left(\frac{L(\nu,0)}{l_{y_1}}\right)^{\nu+1}\, a_{\nu, ren}.
\label{ent2-f1-lf}
\nn\\
\label{case2-f}\eea
For $\nu=2,3,4$ one can write down explicitly
\begin{eqnarray}\label{DeltaB2}
\nu=2: &\,&\frac{\Delta \mathcal{A}^{Shell-LV} }{2 L_{x}L_{y_{2}}}=
\int ^{\ell_y}_{\varepsilon}\frac{z_*^{2}\,dx}{\left ( z_{_{f=f(z,v)}}(x)\right)^{4}}
-  \frac{2}{3}\frac{1}{z_{0}^{\frac{3}{2}}} + \frac{0.225 \cdot 0.6749^{3} }{l^{3}_{y}},
\\\nn
\\
\nu=3: &\,&
\frac{\Delta \mathcal{A}^{Shell-LV} }{2 L_{x}L_{y_{2}}}=
\int ^{\ell_y}_{\varepsilon}\frac{z_*^{5/3}\,dx}{\left ( z_{_{f=f(z,v)}}(x)\right)^{10/3}}
-  \frac{3}{4}\frac{1}{z_{0}^{\frac{4}{3}}} + \frac{0.208 \cdot 0.8324^{4} }{l^{4}_{y}},\\
\nu=4: &\,&
\frac{\Delta \mathcal{A}^{Shell-LV} }{2 L_{x}L_{y_{2}}}= \int ^{\ell_y}_{\varepsilon}\frac{z_*^{3/2}\,dx}{\left ( z_{_{f=f(z,v)}}(x)\right)^{3}}
-  \frac{4}{5}\frac{1}{z_{0}^{\frac{5}{4}}} + \frac{0.1885 \cdot 0.9425^{5} }{l^{5}_{y}}.
\end{eqnarray}

We present the time evolution of  the entanglement entropy (\ref{B-renNC})  for different values of the critical exponent $\nu$ in Fig.\ref{fig:EE-shell2m1n234} and Fig.\ref{fig:EE-shell2m1n234t25} . 
In Figs.\ref{fig:EE-shell2m1n234} we show the difference between the entropy in the shell background and the value of the entropy in the Lifshitz vacuum.  
The evolution in time of the quantity which represents  the difference between the entropy in the current time and thermal  entropy is demonstrated in  Figs.\ref{fig:EE-shell2m1n234t25}.
For each value of $\ell$ we observe that the entropy grows linearly at small times. Then it approaches saturation and we see a kink in the dependence on time, which is much sharper for greater values $\ell$.  We note that the evolution of the entanglement entropy  has more essential dependence on the anisotropic parameter $\nu$ comparing to the case when the subsystem cut out along the longitudinal direction.  

Three right panels in Fig.\ref{fig:A-ren2} also demonstrate the behavior of the entanglement entropy with subtracted vacuum values as a function of $\ell$ and $t$ for different $\nu$.

  \begin{figure}[h]
\centering
 \includegraphics[scale=0.4]{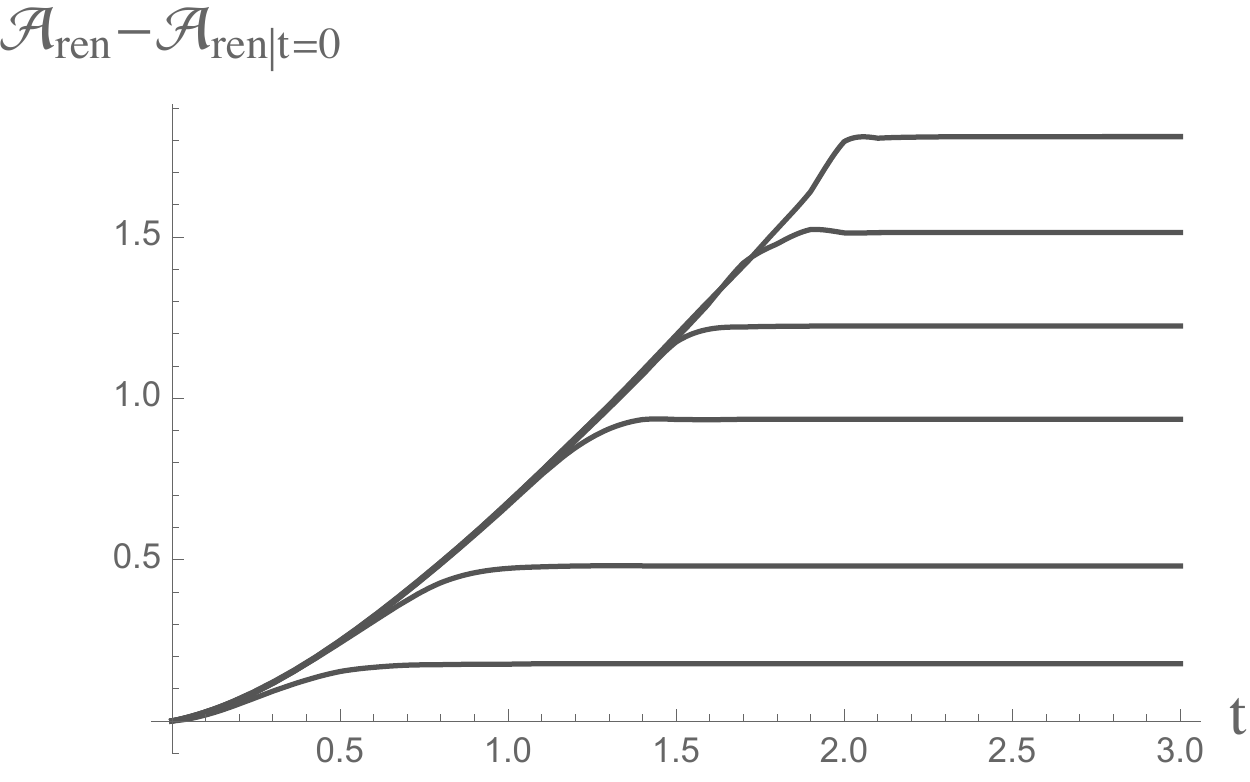}A$\,\,\,\,\,\,\,\,\,\,\,\,$
  \includegraphics[scale=0.4]{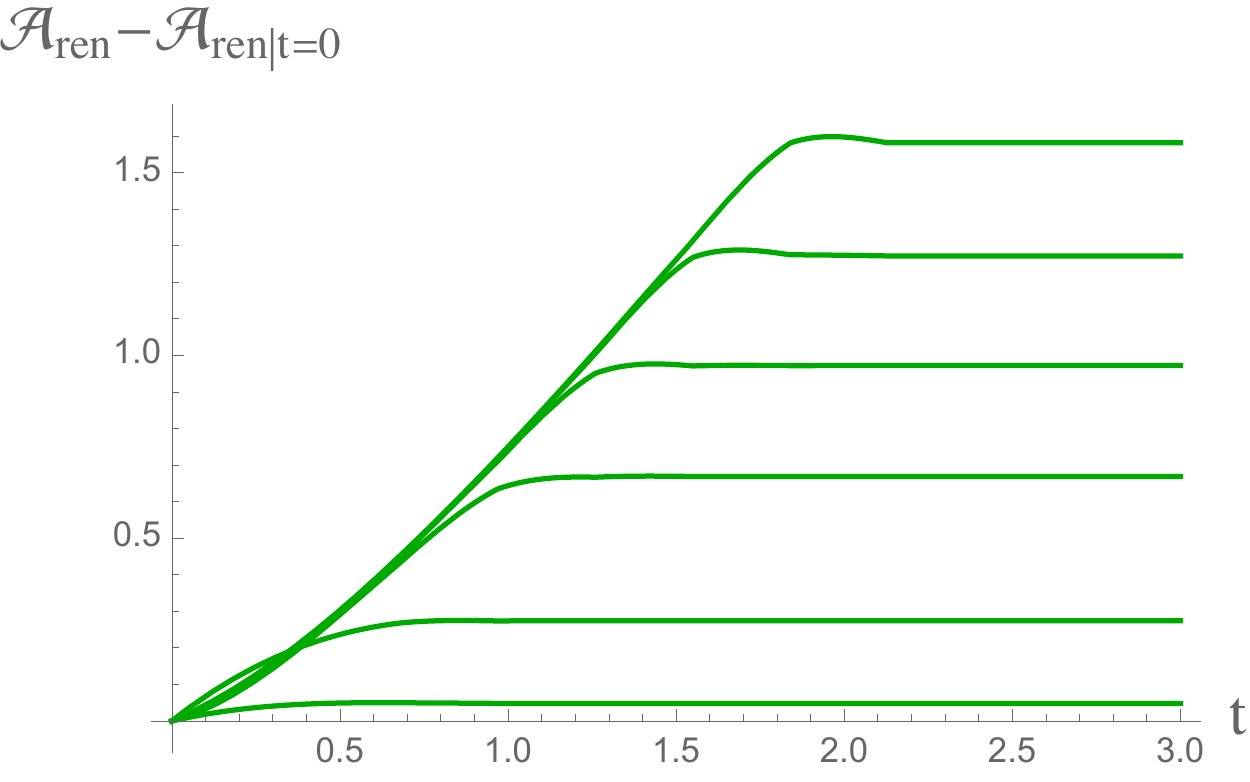}B\\
        \vspace{5mm}
\includegraphics[scale=0.4]{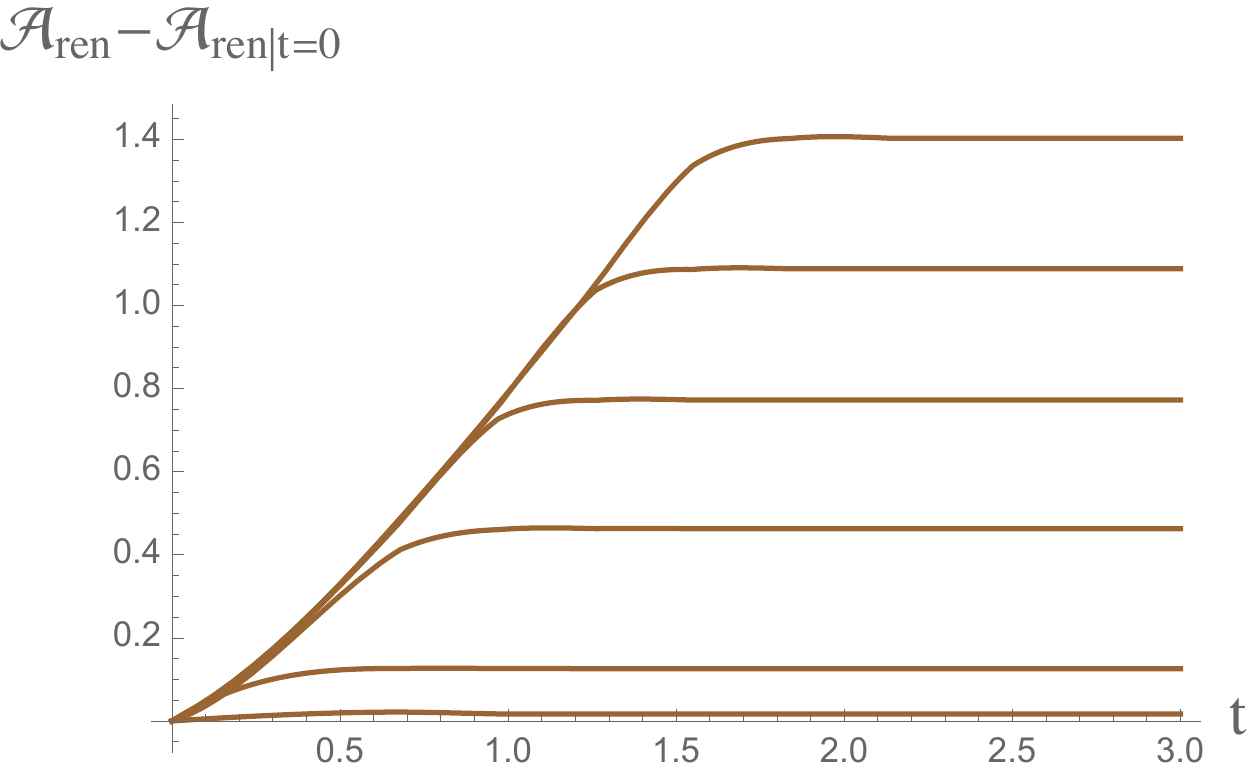}C$\,\,\,\,\,\,\,\,\,\,\,\,$
\includegraphics[scale=0.4]{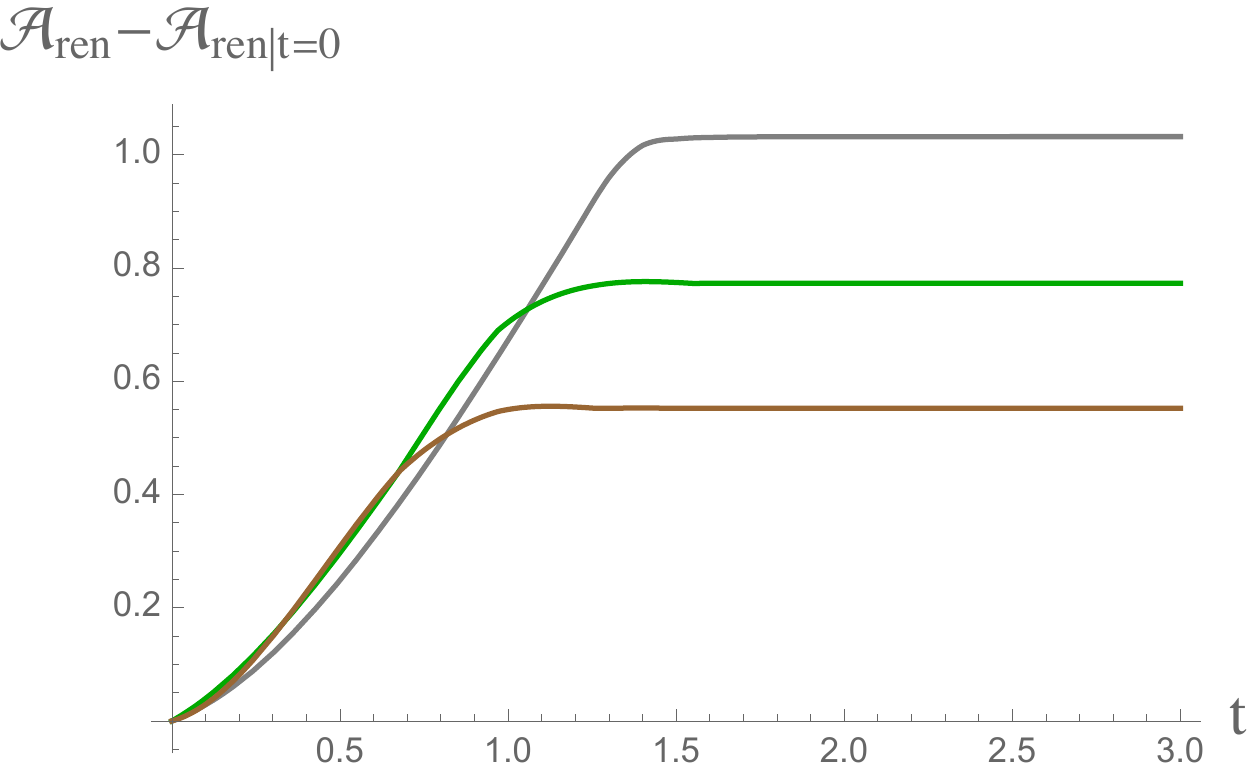}D
        \caption{The time dependence of the holographic entanglement entropy  $\mathcal{A}_{ren}$ after the corresponding initial state subtraction for the Lifshitz-Vaidya metric  at fixed $l =1,1.4, 1.9, 2.2, 2.5, 2.8$ for a subsystem delineated along the transversal direction,  $\nu =2,3,4$ ({\bf A, B, C}, respectively). In {\bf D} we plot the time dependence of $\mathcal{A}_{ren} - \mathcal{A}_{ren}|_{t=0}$ at $l =2$ for $\nu = 2,3,4$ (from top to bottom, respectively)}
        \label{fig:EE-shell2m1n234}
 \end{figure}

  \begin{figure}[h]
\centering
\includegraphics[scale=0.4]{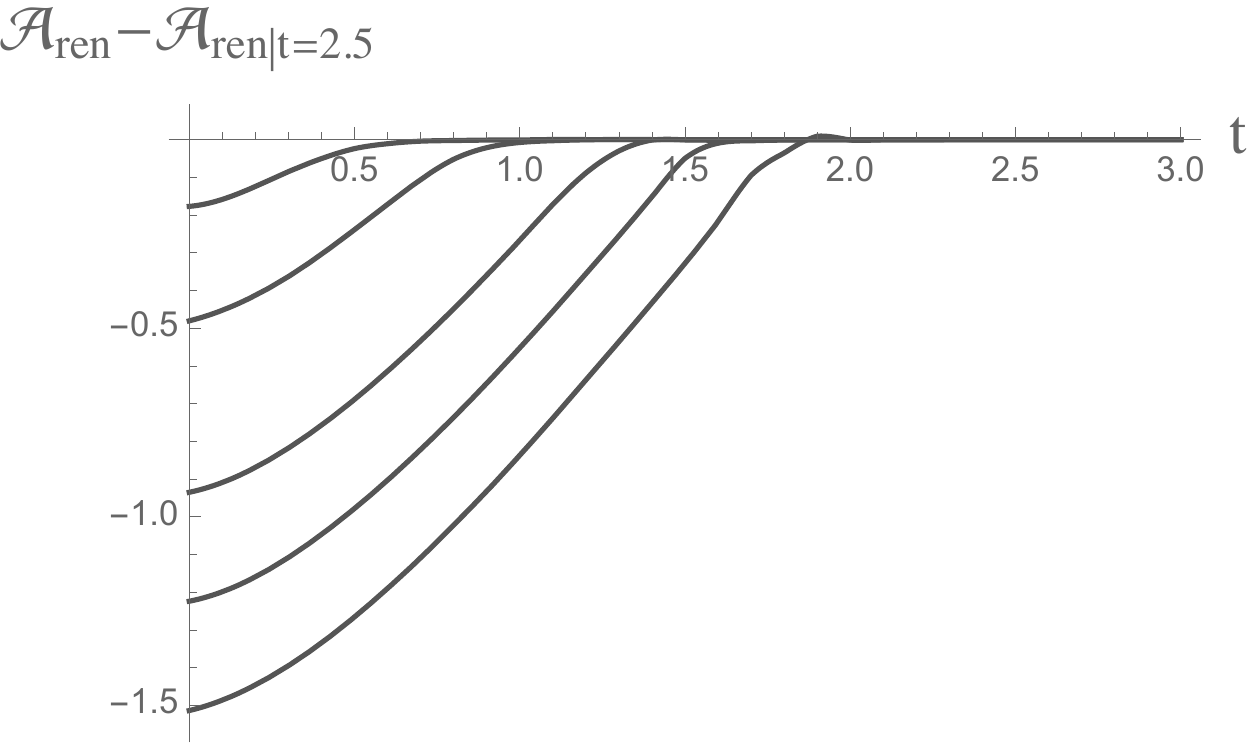}A$\,\,\,\,\,\,\,\,\,\,\,\,$
\includegraphics[scale=0.4]{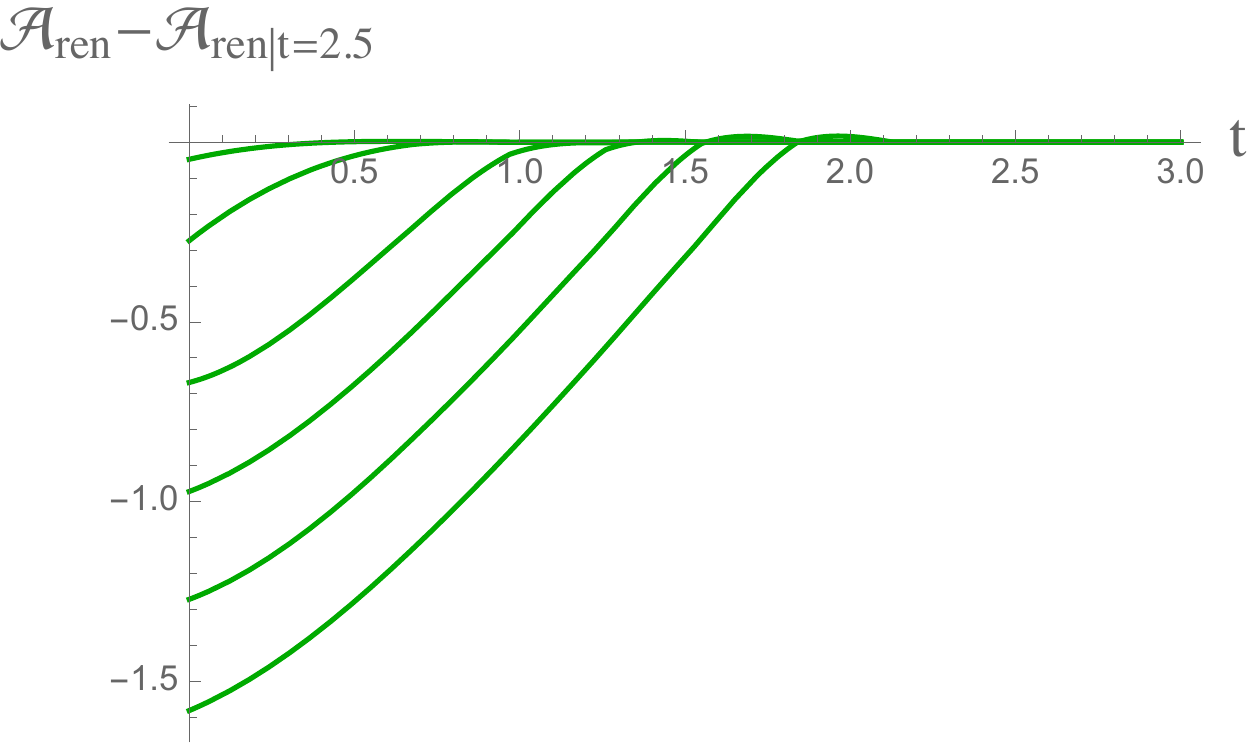}B\\
\includegraphics[scale=0.4]{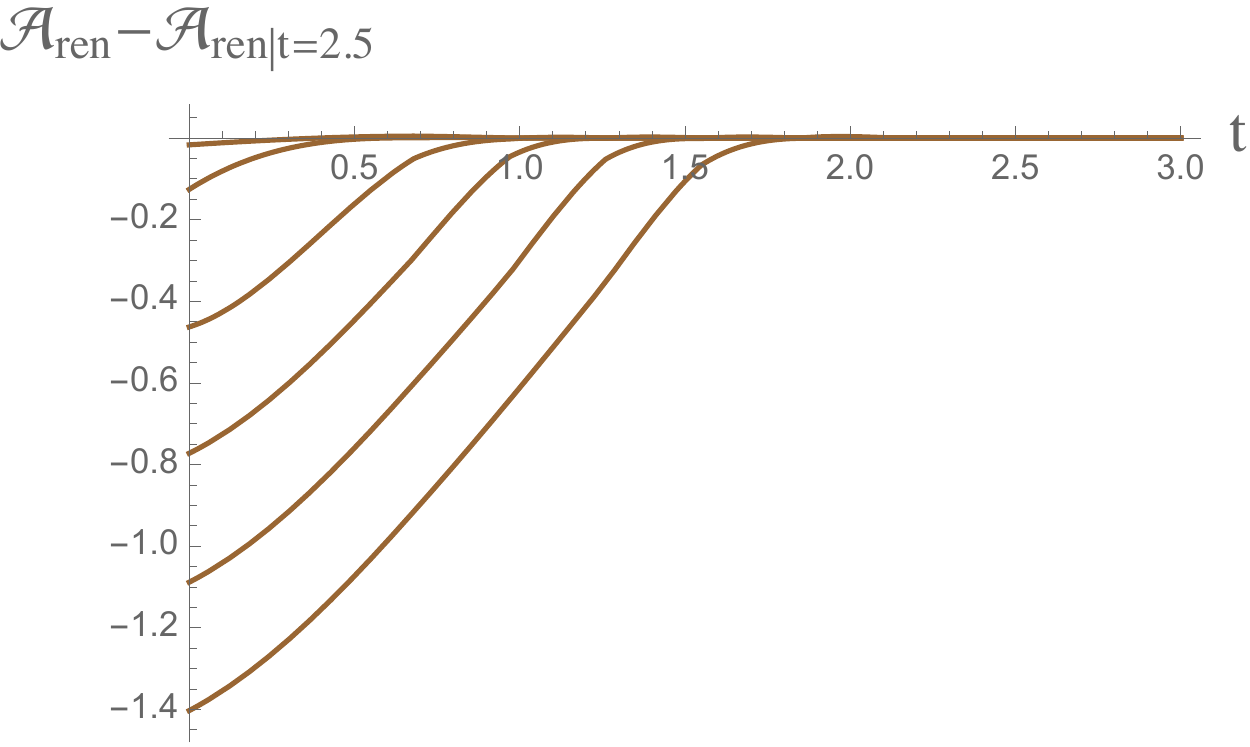}C$\,\,\,\,\,\,\,\,\,\,\,\,$
\includegraphics[scale=0.4]{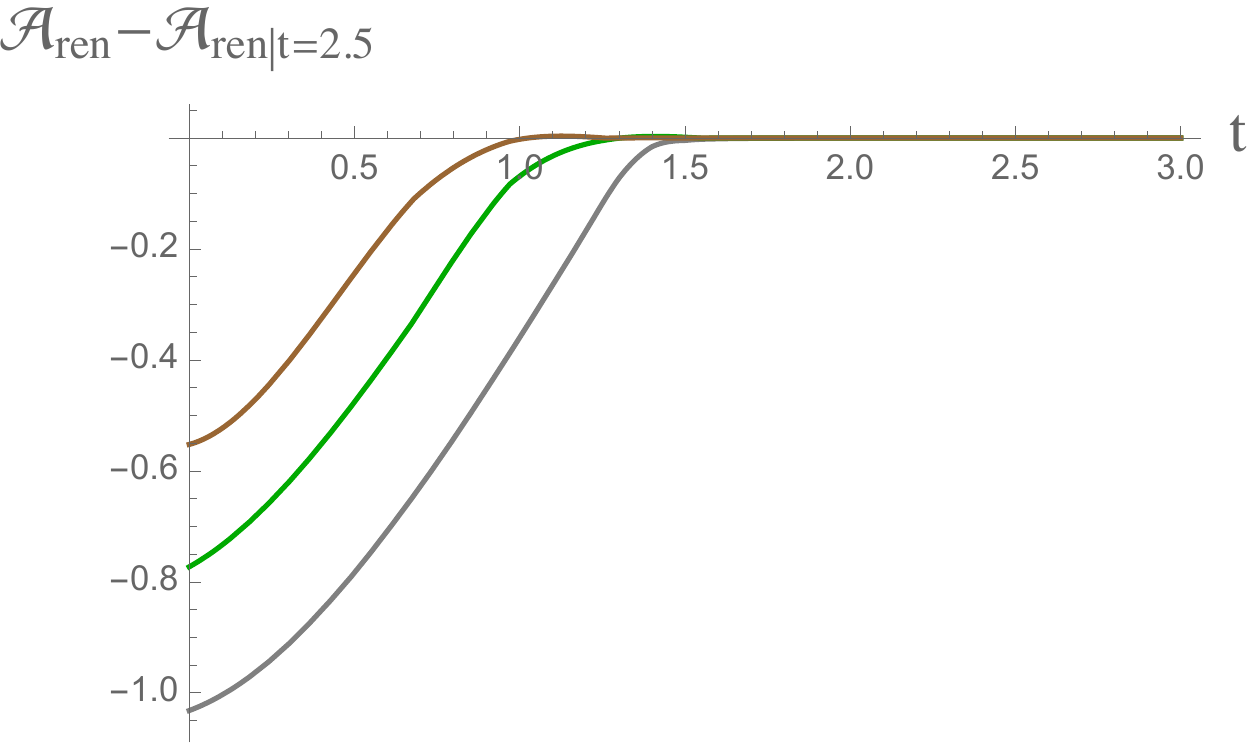}D
        \caption{The time dependence of the holographic entanglement entropy  $\mathcal{A}_{ren}$  for the Lifshitz-Vaidya metric  after the corresponding subtraction of the state when the black brane has already been formed ($t = 2.5$) at fixed $l =1,1.4, 1.9, 2.2, 2.5, 2.8$ for a subsystem delineated along the transversal direction,  $\nu =2,3,4$ ({\bf A,B,C}, respectively). In {\bf D} we plot the time dependence of $\mathcal{A}_{ren} - \mathcal{A}_{ren}|_{t=2.5}$ at $l =2$ for $\nu = 2,3,4$ (from bottom to top, respectively).}
        \label{fig:EE-shell2m1n234t25}
  \end{figure}

 \begin{figure}[h!]
\centering
\begin{picture}(250,100)
\put(-80,-60){ \includegraphics[scale=0.33]{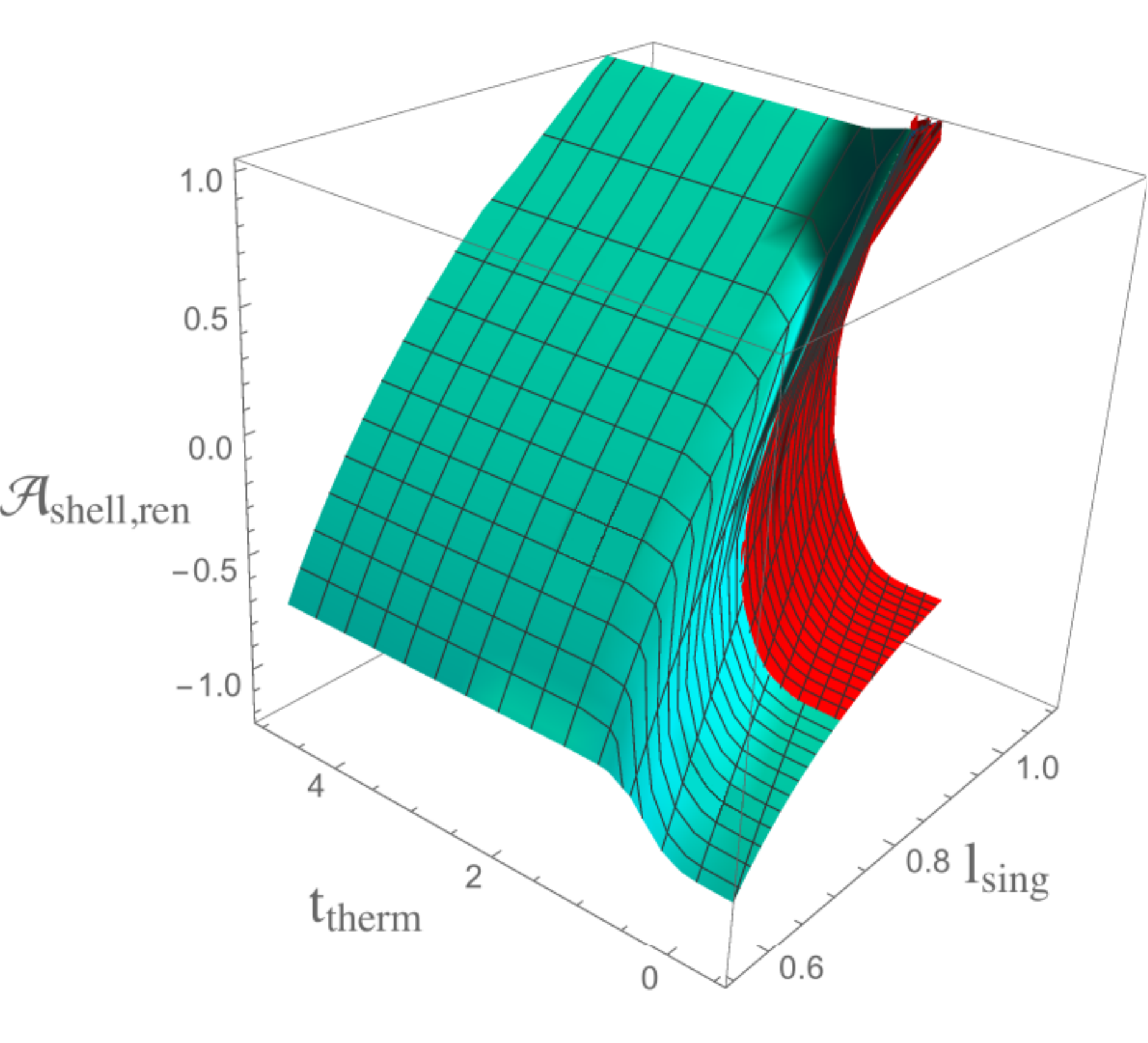}A}
\put(150,-60){\includegraphics[scale=0.25]{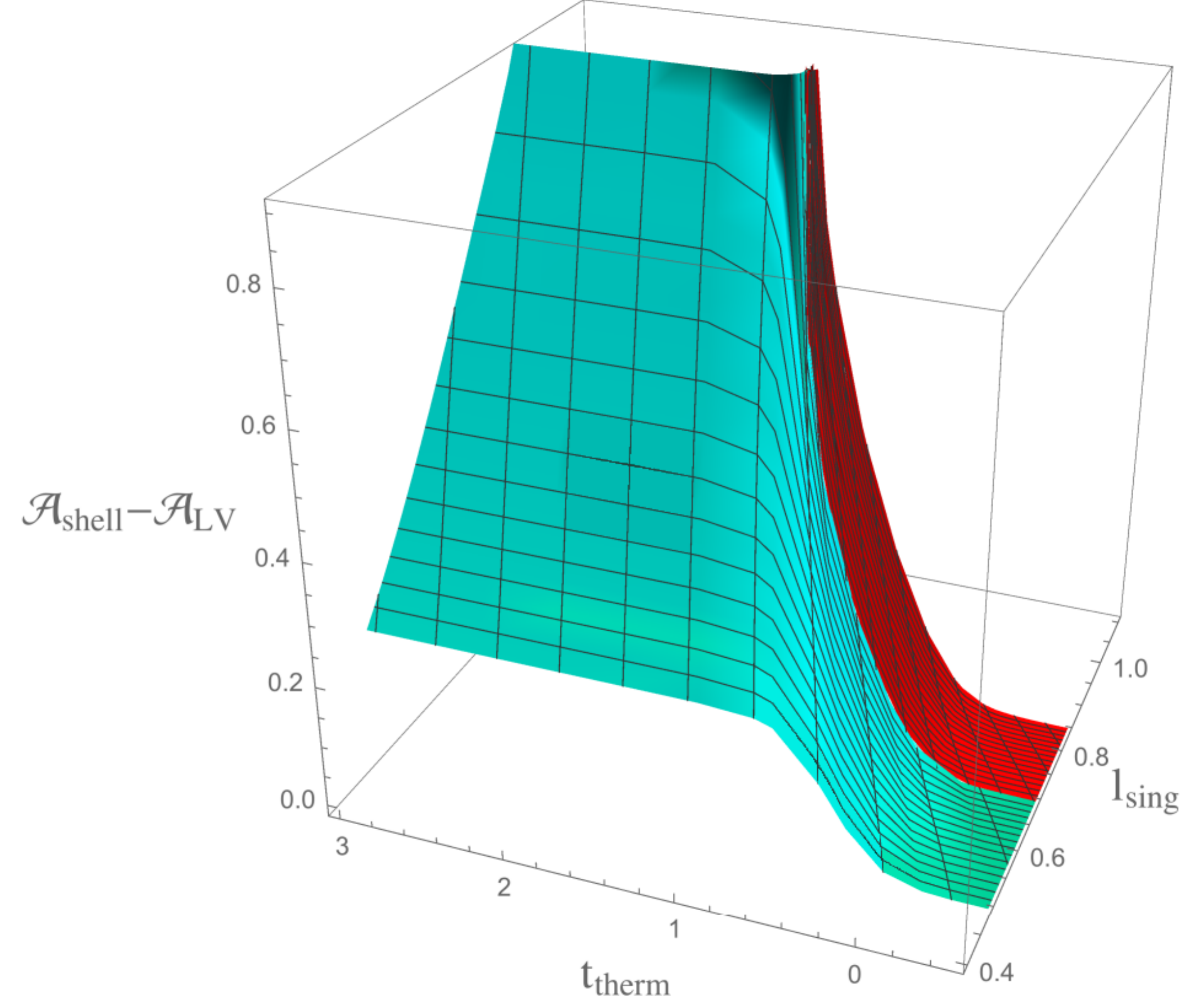}B}
\end{picture}\\$\,$\\
$\,$\\
$\,$\\
$\,$\\
$\,$\\
       \includegraphics[scale=0.3]{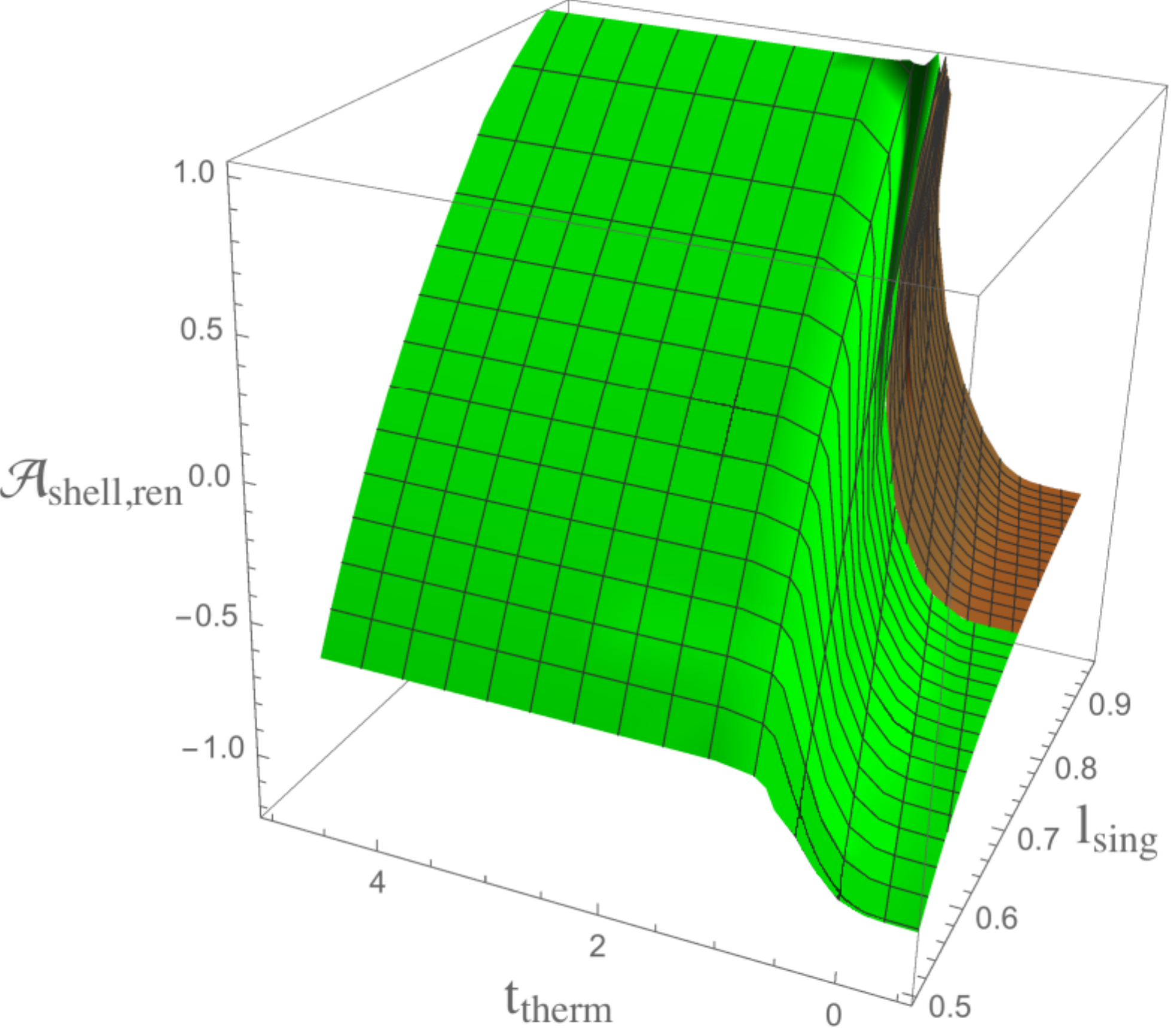}C$\,\,\,\,\,\,\,\,\,\,\,\,$
        \includegraphics[scale=0.3]{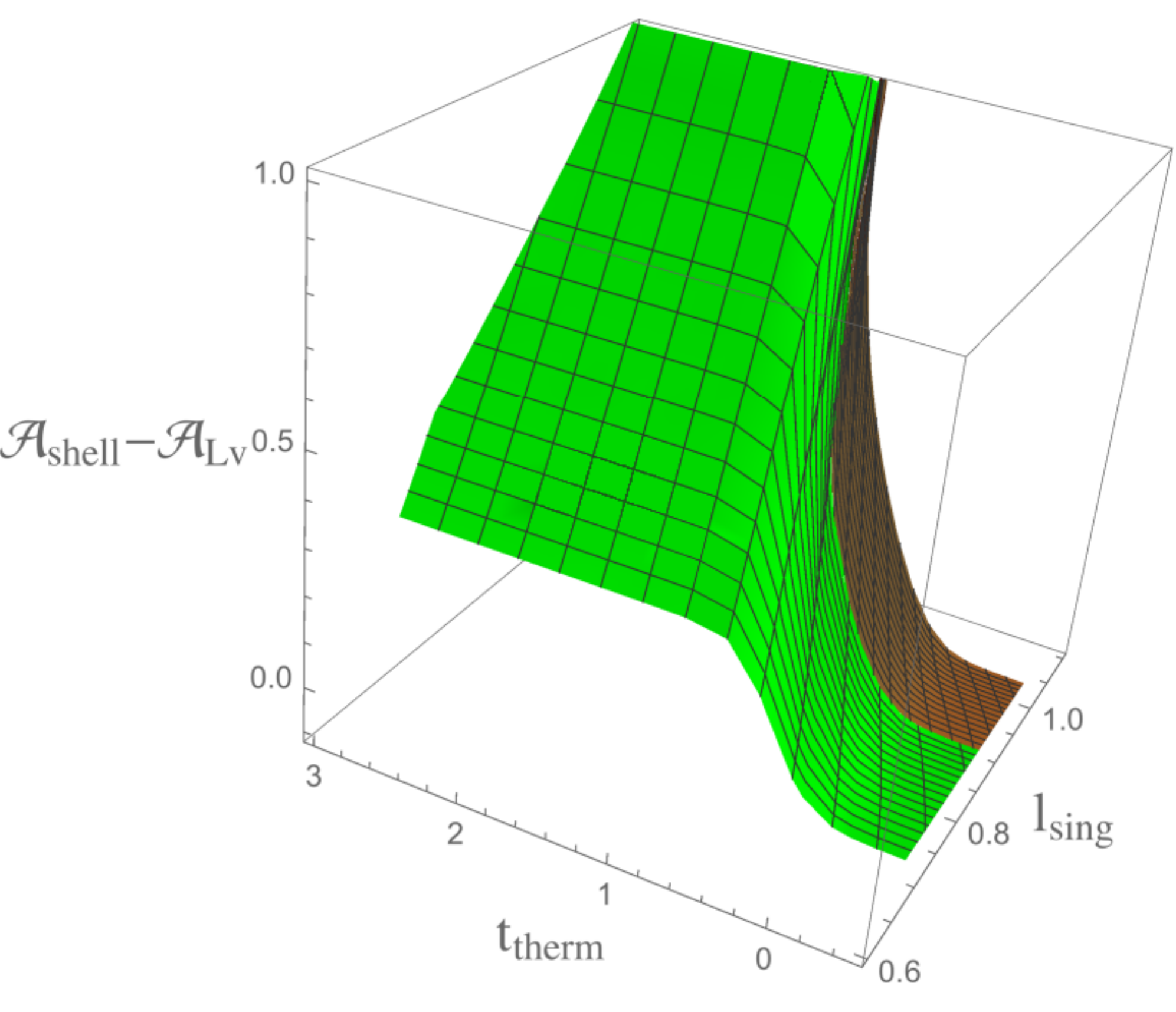}D\\
      \includegraphics[scale=0.33]{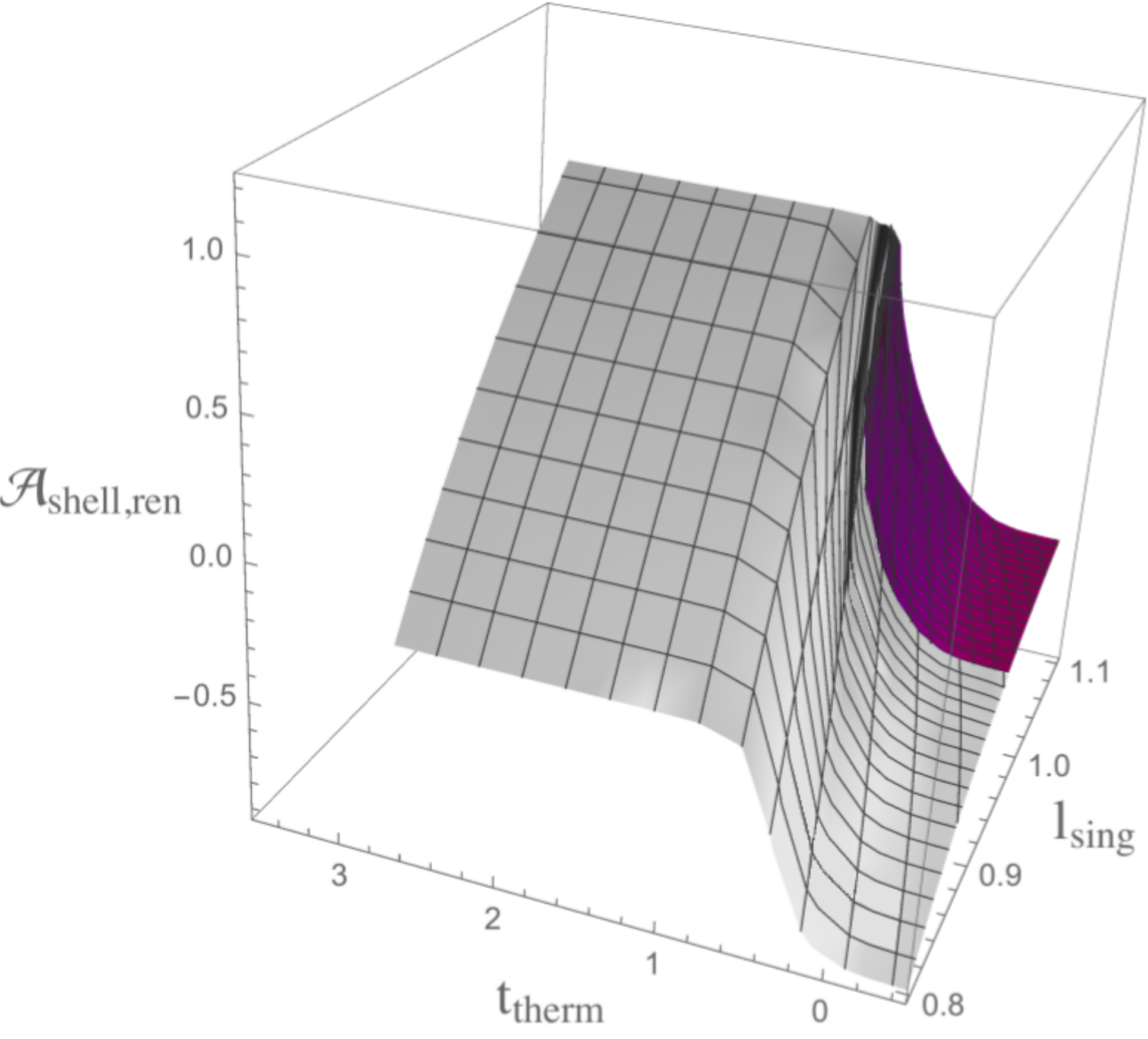}E$\,\,\,\,\,\,\,\,\,\,\,\,$
          \includegraphics[scale=0.34]{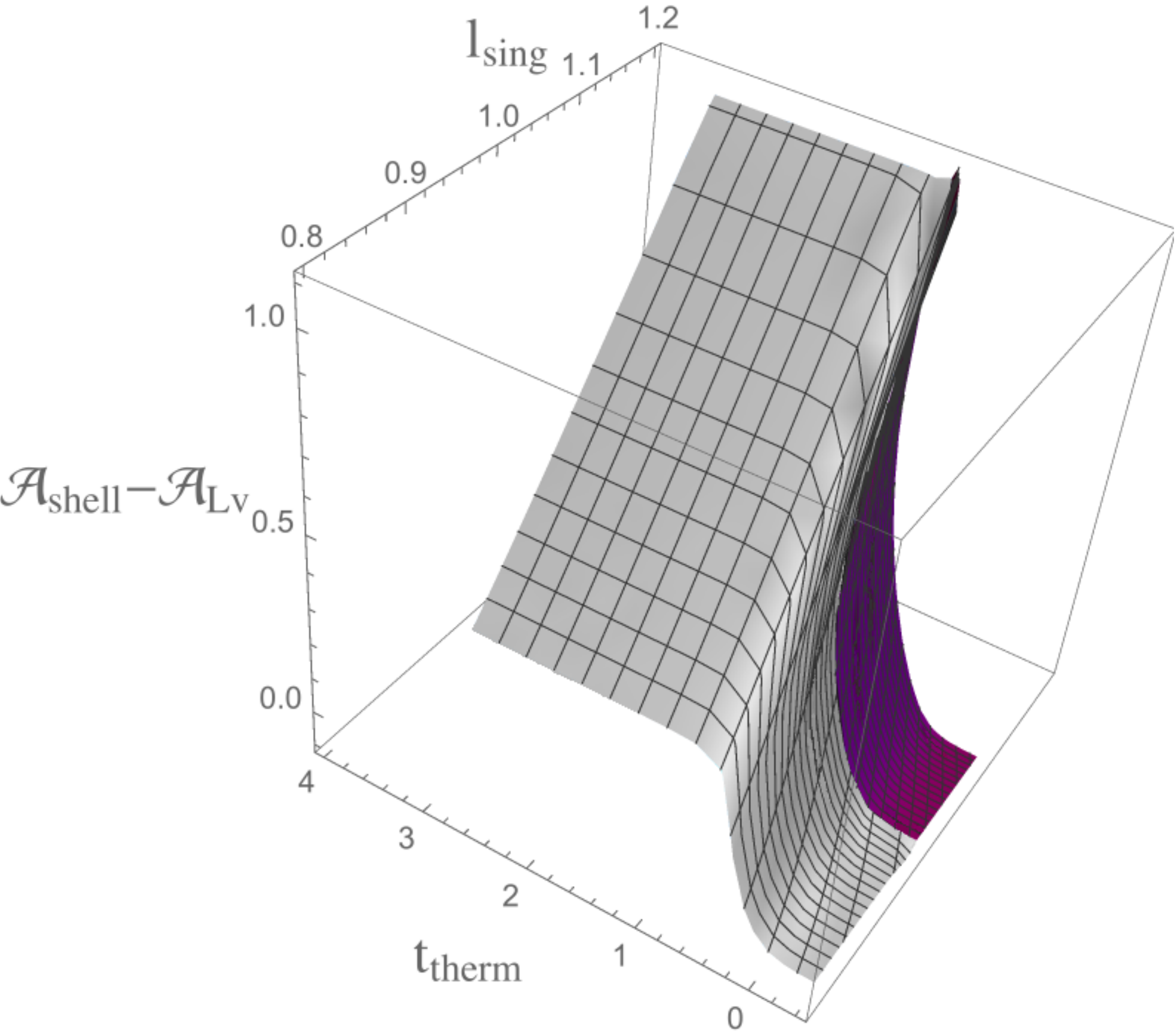}F
         \caption{Left panel: the evolution of $\mathcal{A}^{Shell}_{ren}$
on $\ell$ and $t$ for $\nu=2,3,4$ (from top to bottom, respectively); Right panel:
the evolution of $\Delta\mathcal{A}^{Shell-LV}$ on $\ell$ and $t$ for $\nu=2,3,4$ (from top to bottom, respectively).}
        \label{fig:A-ren2}
 \end{figure}

\section{Conclusions}\label{Sect:6}

In this paper, we have investigated the holographic thermalization process of the quark-gluon plasma in anisotropic backgrounds. For this purpose, we have used an
analytic black brane solution which asymptotes to the Lifshitz-like spacetime with arbitrary critical exponent.  We also have built the corresponding  Lifshitz-Vaidya solution, which metric interpolates between the vacuum Lifshitz-like and the black brane geometries.  This background  has been used to describe the thermalization process as well to model the ``quench'' process.
Let us note that 4-dimensional  Lifshitz spacetimes with black hole are widely used in AdS/CMT models \cite{0911.3586,1005.4690,1105.1162}.

We have considered thermalization processes both in the transversal and longitudinal directions, which differ by the contribution of the anisotropic exponent. The thermalization along the longitudinal direction turned to have the linear regime similar to that in modified AdS models. At the same time, in the transversal direction the thermalization process is much faster  and behaves linearly only for large distances. It  should also be noted that the thermalization along the longitudinal direction is independent of the value of the dynamical exponent, while results obtained for the transversal direction strongly depend on the anisotropic parameter and are more sensitive to the value of mass.

Holographic entanglement entropy have also been studied for the subsystems delineated along both transversal and longitudinal directions. For a subsystem cutting out along the longitudinal direction in the black brane background, we have found that the dependence of the entropy on the critical exponent for small distances was absent and appeared for larger values of $\ell$.  In the transversal direction we have observed that the entropy  depends on $\nu$ at small distances and approaches a linear behavior which is the same  for all $\nu$. Thus, for both subsystems at large $\ell$, the entanglement entropy comes to a linear regime,  which,
depending on the chosen direction, depends or does not on the critical exponent.

The regime is similar to the one found for the Lifshitz metrics in \cite{KKVT},  which, however, is independent on $\nu$. This is related to that the  anisotropy between the spatial coordinates is absent in the Lifshitz backgrounds considered in  \cite{KKVT} unlike the Lifshitz-like metrics suggested in
\cite{TAYLOR, ALT}.

The most interesting results concern the holographic entanglement entropy in the Vaidya-Lifshitz solution that we constructed. Here we again studied subsystems divided along two possible directions.  The common feature of the time evolution of the entropy for both subsystems is the kink observed already for small distances. The entropy increases linearly in time until it approaches the saturation point. We found that the form of the kink is sharper for large values of $\ell$.  The dependence on the critical exponent looks similar to this one in the black brane background. Since the subsystems differ by the contribution of the critical exponent, the rate of approaching saturation and the saturation value of the entanglement entropy were seen to be different for each case.

It would be interesting to study other non-local operators, like two-point correlation functions and Wilson loops operators, in the backgrounds considered in this paper and compare their velocity bounds as well as estimate with experimental data. We shall address these problems in our future work \cite{AGG}.

\section*{Acknowledgments}
We would like to thank Dima Ageev, Misha Khramtsov and Giuseppe Policastro for useful discussions,
as well as Blaise Goutreaux and Elias Kiritsis for the correspondence at early stage of the work. 
We also thank to the JHEP referee for careful reading of our paper and fruitful discussions.
This work was supported by the RFBR grant 14-01-00707
and by the ANR grant ANR-12-BS01-012-01.
I. A. and A.G. thank the Galileo Galilei Institute for Theoretical Physics for the
hospitality and the INFN for partial support during the preparation of
this work.\\
 \appendix

\section{Einstein Equations} \label{App:A}
Here we provide both sides of the Einstein equations derived from the action
(\ref{1.2a}):
\begin{eqnarray}\label{A.0}
R_{mn}  = - \frac{\Lambda}{3}g_{mn} + \frac{1}{2}\partial_{m}\phi \partial_{n}\phi  + \frac{1}{4} e^{\lambda \phi} \left(2F_{mp}F^{\  p}_{n}\right) - \frac{1}{12}e^{\lambda \phi}F^{2}g_{mn}.
\end{eqnarray}
The computations have been checked with SageManifolds \cite{SageManif}, which
is an extension of the free computer algebra system SageMath \cite{Sage}. The
corresponding worksheets are publicly available at the following links:\\
{\footnotesize
\url{https://cloud.sagemath.com/3edbca82-97d6-41b3-9b6f-d83ea06fc1e9/raw/Lifshitz_black_brane.html}\\
\url{https://cloud.sagemath.com/3edbca82-97d6-41b3-9b6f-d83ea06fc1e9/raw/Vaidya-Lifshitz.html}
}

\subsection{The LHS of the Einstein Equations}\label{App:A1}

Without any black brane, the metric reads
\begin{eqnarray}\label{A.1}
ds^{2} = e^{2\nu r}\left(-dt^{2} + dx^{2}\right) + e^{2r}\left(dy^{2}_{1} + dy^{2}_{2}\right) + dr^{2}.
\end{eqnarray}
The non-zero components of the Ricci tensor are
\begin{eqnarray}\label{A.2}
R_{00} &=& 2(\nu^{2}+ \nu)e^{2\nu r}, \quad
R_{11} = -  2(\nu^{2}+ \nu)e^{2\nu r};\\
R_{22} &=& - 2(\nu + 1)e^{2 r}\quad
R_{33} = - 2(\nu+ 1)e^{2 r}, \quad
R_{44}= - 2(\nu^{2}+ 1).
\end{eqnarray}
The scalar curvature is
\begin{eqnarray}\label{A.3}
R = - 6\nu^{2} - 8\nu  - 6.
\end{eqnarray}
The metric for a black brane solution that asymptotes to the Lifshitz background (\ref{A.1}) is given by
\begin{eqnarray}\label{A.4}
ds^{2} = e^{2\nu r}\left(-f(r)dt^{2} + dx^{2}\right) + e^{2r}\left(dy^{2}_{1} + dy^{2}_{2}\right) + \frac{dr^{2}}{f(r)},
\end{eqnarray}
where
\begin{equation}\label{A.4a}
f(r)=1- me^{-(2\nu+2)r}.
\end{equation}
The geometry (\ref{A.4})-(\ref{A.4a}) is supported by
\begin{eqnarray}\label{A.4b}
e^{\lambda\phi} =\mu e^{4r}, \quad F_{(2)} = \frac{1}{2} q \, dy_1 \wedge d y_2 .
\end{eqnarray}

The non-zero components of the Ricci tensor of the metric (\ref{A.4}) are
\begin{eqnarray} \label{A.5}
R_{00} &= &e^{2\nu r}f(r)\left( 2(\nu^{2} + \nu )f(r) + (2\nu +1) \frac{\partial f(r)}{\partial r} + \frac{1}{2}\frac{\partial^{2} f(r)}{\partial r^{2}}\right),\\
R_{11} &= &- 2(\nu^{2} + \nu) e^{2\nu r} f(r) - \nu e^{2\nu r} \frac{\partial f(r)}{\partial r},\\
R_{22} &=& - 2(\nu + 1)e^{2r}f(r) - e^{2r}\frac{\partial f(r)}{\partial r},\\
R_{33} &=& - 2(\nu + 1)e^{2r}f(r) - e^{2r}\frac{\partial f(r)}{\partial r},\\
R_{44}&=  &- 2(\nu^{2} + 1) - \frac{1}{f(r)}\left(2\nu +1 \right) \frac{\partial f(r)}{\partial r} - \frac{1}{2f(r)} \left(\frac{\partial^{2} f(r)}{\partial r^{2}}\right).
\end{eqnarray}

The generalization of (\ref{A.4})-(\ref{A.4a}) to the Vaidya background reads
\begin{eqnarray}\label{A.6}
ds^{2} = -e^{2\nu r}f(v,r)dv^{2} + 2 e^{\nu r} dvdr + e^{2\nu r}dx^{2} + e^{2r}\left(dy^{2}_{1} + dy^{2}_{2}\right),
\end{eqnarray}
where
\begin{equation}\label{A.6a}
f(v,r) = 1 - m(v)e^{-(2\nu+2)r}.
\end{equation}
The solution (\ref{A.6})-(\ref{A.6a}) is supported by the fields (\ref{A.4b}), plus
the infalling shell of null dust, whose energy-momentum tensor is
\begin{equation}
T^{\rm s} = T_{00}^{\rm s} \, dv \otimes dv .
\end{equation}
The  non-zero components of the Ricci tensor of the metric (\ref{A.6}) are
\begin{eqnarray}\label{A.7}
R_{00} &=&e^{2\nu r}f(v,r)\left( 2(\nu^{2} + \nu )f(v,r) + (2\nu + 1) \frac{\partial f(v,r)}{\partial r}  + \frac{1}{2}\frac{\partial^{2} f(v,r)}{\partial r^{2}}\right) \nonumber \\ &-& \frac{(\nu + 2)}{2}e^{\nu r}\frac{\partial f(v,r)}{\partial v},\\
R_{04}& = &-  e^{\nu r}\left(2(\nu^{2} + \nu)f(v,r) + (2\nu +1) \frac{\partial f(v,r)}{\partial r}  +\frac{1}{2} \frac{\partial^{2}f(v,r)}{\partial r^{2}} \right),\\
R_{11} & = &-e^{2\nu r}\nu\left(2(\nu + 1)f(v,r) + \frac{\partial f(v,r)}{\partial r} \right), \\ \label{A.7a}
R_{22} & = & R_{33} = -e^{2r}\left(2(\nu +1) f(v,r) + \frac{\partial f(v,r)}{\partial r} \right), \quad R_{44} = 2(\nu - 1).
\end{eqnarray}

\subsection{The RHS of the Einstein equations}\label{App:A2}
Here we write down the right-hand sides of Einstein equations corresponding to the Vaidya solution (\ref{A.6}) which asymptotes to the Lifshitz-like spacetime (\ref{A.1}):
\begin{eqnarray}\label{A.8}
&{\bf 00:}& -\frac{\Lambda}{3}g_{00} - \frac{1}{6}e^{\lambda \phi} F_{23}F_{23}g^{22}g^{33} g_{00} +T^{\rm s}_{00}  =  \left(\frac{\Lambda}{3} + \frac{1}{24}\mu q^{2}\right)e^{2\nu r}f(r) +  T^{\rm s}_{00}, \\
&{\bf 11:}& -\frac{\Lambda}{3}g_{11} - \frac{1}{6}e^{\lambda \phi} F_{23}F_{23}g^{22}g^{33} g_{11} =  - \left(\frac{\Lambda}{3} + \frac{1}{24}\mu q^{2}\right) e^{2\nu r}, \\
&{\bf 22:}& -\frac{\Lambda}{3}g_{22} + \frac{1}{2}e^{\lambda \phi} F_{23}F_{23}g^{33}  - \frac{1}{6}e^{\lambda \phi} F_{23}F_{23}g^{22}g^{33} g_{22} = -\left(\frac{\Lambda}{3}- \frac{1}{12}\mu q^2\right)e^{2r},\\
&{\bf 33:}&  -\frac{\Lambda}{3}g_{33} +  \frac{1}{2}e^{\lambda \phi} F_{23}F_{23}g^{22}  - \frac{1}{6}e^{\lambda \phi} F_{23}F_{23}g^{22}g^{33} g_{33} =  -\left(\frac{\Lambda}{3}- \frac{1}{12}\mu q^2\right)e^{2r}, \\
&{\bf 44:}&  -\frac{\Lambda}{3}g_{44} + \frac{1}{2} \left(\partial_{4} \phi \right)^{2} - \frac{1}{6}e^{\lambda \phi} F_{23}F_{23}g^{22}g^{33} g_{44} = \frac{1}{2}\left(\frac{\partial \phi}{\partial r}\right)^{2}, \\ \label{A.8a}
&{\bf 04:}&  -\frac{\Lambda}{3}g_{04}   - \frac{1}{6}e^{\lambda \phi} F_{23}F_{23}g^{22}g^{33} g_{04} = -\left(\frac{\Lambda}{3}  + \frac{1}{24}\mu q^{2} \right)e^{\nu r}.
\end{eqnarray}
Substituting (\ref{A.6a}) into (\ref{A.7})-(\ref{A.8a}) and representing the non-vanishing term in the component $R_{00}$ (\ref{A.7}) as
\begin{eqnarray}
- \frac{(\nu + 2)}{2}e^{\nu r}\frac{\partial f(v,r)}{\partial v} = \frac{(\nu + 2)}{2}e^{-(\nu+2)r}\frac{dm}{d v},
\end{eqnarray}
leads to a solution for the constants $\lambda$, $\mu$, $q$ and $\Lambda$
and the component $T^{\rm s}_{00}$ of the shell energy-momentum tensor.
Thus, one finds that the ansatz (\ref{A.4b}) for the fields is valid.
For $\nu=4$, the solution is formed by the values (\ref{2.2b}) for the constants
$\lambda$, $\mu$, $q$ and $\Lambda$, as well as by the following expression of
the shell energy-momentum:
\be
    T^{\rm s}_{00}(v,r) = 3 e^{-6 r} \frac{dm}{dv} .
\ee

\section{Details on solutions to profiles equations} \label{App:B}

\subsection{Equations (\ref{5.3a}),  (\ref{5.3b})} \label{App:B1}

In Fig.\ref{fig:sing-area1} and Fig.\ref{fig:sing-area-all2}  we show the position of the singular point ($x$-axis) of the solution with given $z_*$ ($y$-axis) and varying $v_*$.
From Fig.\ref{fig:sing-area1}  we see that for the  given value of $z_*\leq 1$ varying $v_*$, say from $v_*=v_1$ to
 $v_*=v_2$,
we get different  positions of $l_s$ lying between $l_s(v_1)$ and $l_s(v_2)$. It is interesting to note that for small $z_*$
the position of the singular point is not considerably depends on value of $v_*$.
For $z_*\to 1$ this dependence is more significant. We also see that  one given value of $l_s$ corresponds to different values of $z_*$ and $v_*$.  
 \begin{figure}[t]
\centering
\begin{picture}(200,70)
\put(0,0){\includegraphics[width=6.5cm]{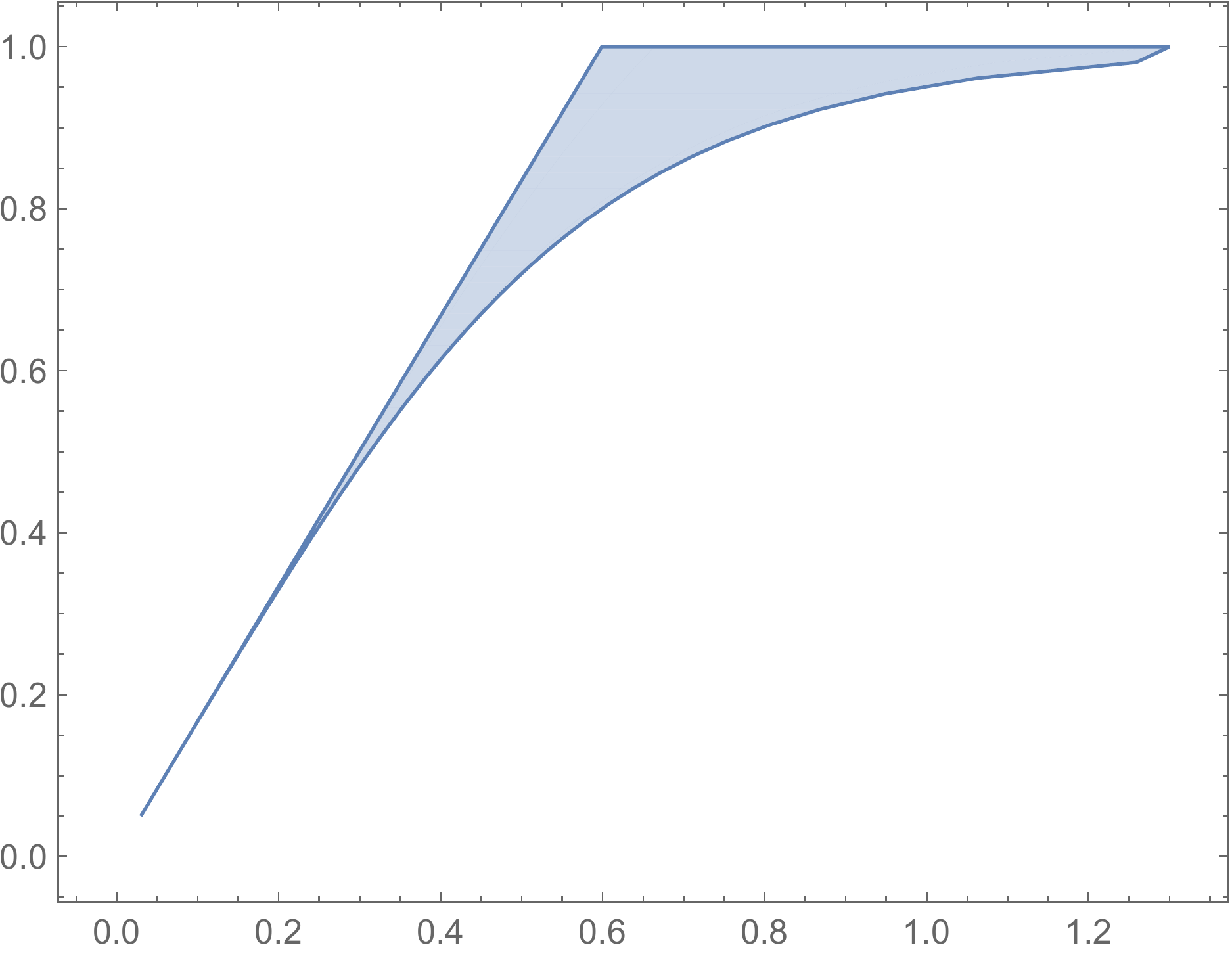}A}

\linethickness{0.1mm}
\put(10,125){\line(1,0){104}}\linethickness{0.1mm}
\put(85,7){\line(0,1){120}}
\put(114,7){\line(0,1){119}}
\put(12,117){$z_*$}
\put(58,15){$l_{s}(v_{1})$}
\put(115,15){$l_{s}(v_{2})$}
\end{picture}
\includegraphics[width=6.5cm]{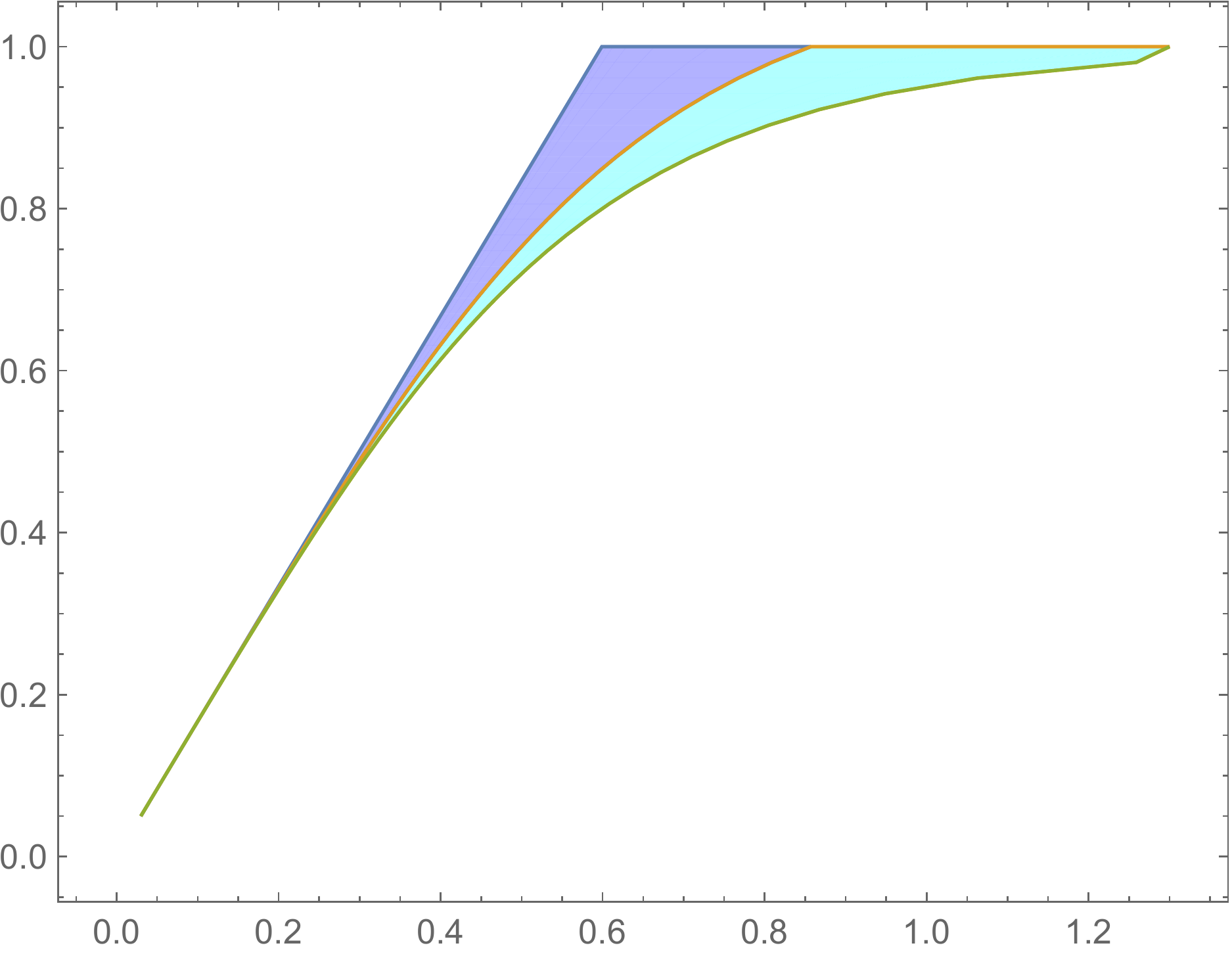}B
 \caption{{\bf A:} Positions of  the singular point  for given $z_* $ and given $v_*$  belonging to the variety $v_1\leq v_*\leq v_2$,
  $l_{s}(v_{1})\leq l_{s}(v_{1})\leq l_{s}(v_{2})$. In this plot $v_1=-5,\,v_2=25$.  $\nu =2$. {\bf B:} Different zones corresponds to different domains for varying $v_*$.}
\label{fig:sing-area1}
 \end{figure}
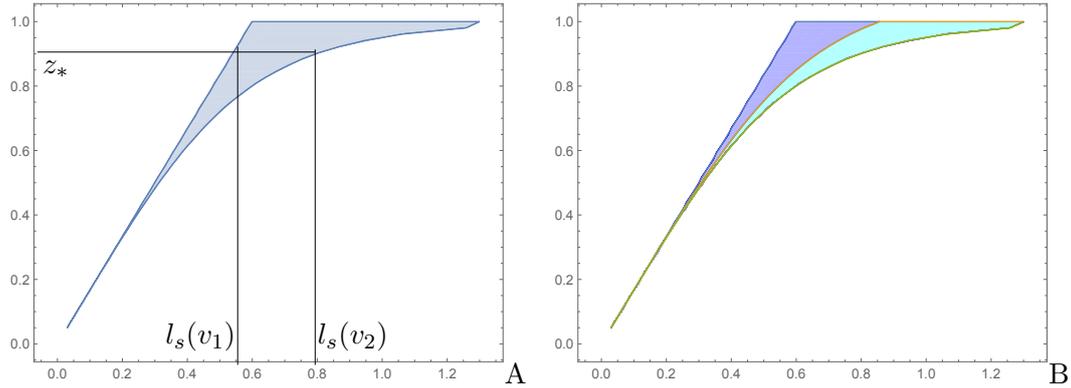
 \begin{figure}[h!]
 \centering
\includegraphics[scale=0.3]{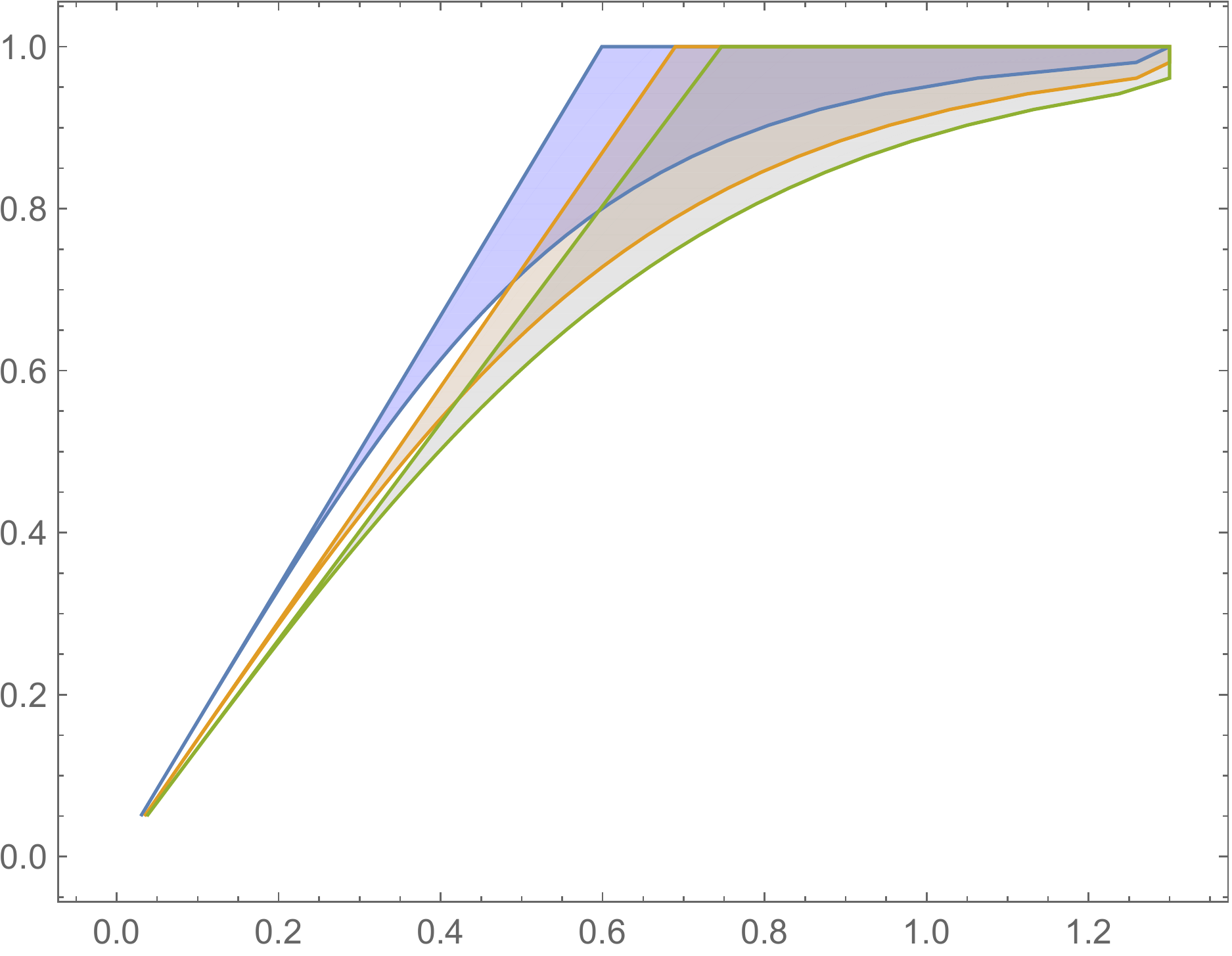}$\,\,$A$\,\,\,\,\,\,\,\,$$\,\,\,\,\,\,\,\,$$\,\,\,\,\,\,\,\,$
\includegraphics[scale=0.36]{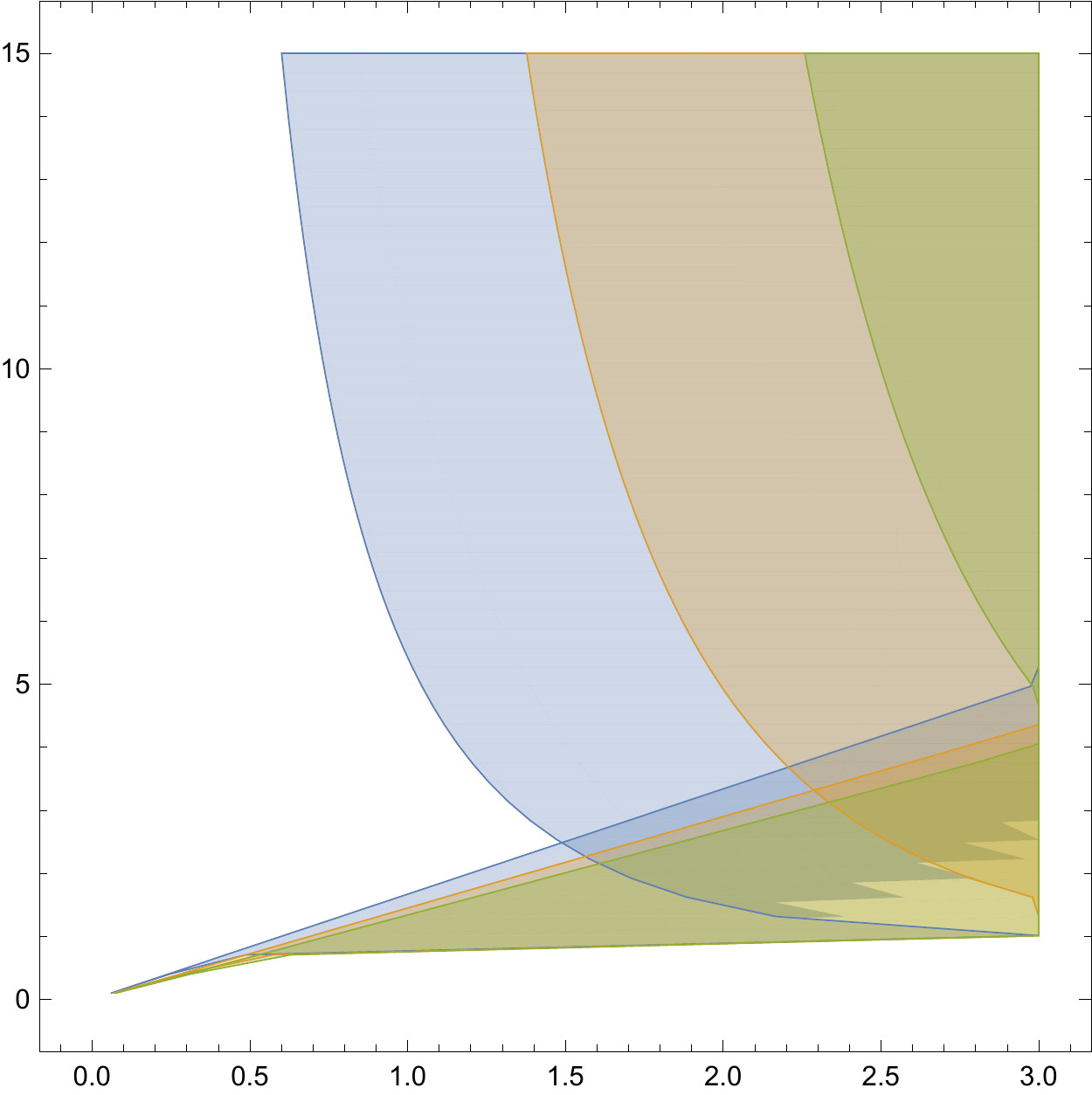}$\,\,$B
  \caption{The same as in Fig.\ref{fig:sing-area1} for $\nu=2,3,4$ (blue,brown,green,respectively),  {\bf A:} $z_*\leq 1$, {\bf B:} $z_*>1$.}
        \label{fig:sing-area-all2}
 \end{figure}
\newpage
The position of the singular point for different values of  critical exponent $\nu$ is presented in Fig.\ref{fig:sing-area-all2}.

 In Fig.\ref{fig:sing1} we present the contour plots for the boundary time and $z(l_{sing})$ as functions of $z_{*}$ and $v_{*}$.
The values of the initial conditions taken from regions of white colour  yields solutions to  eqs. (\ref{5.3a})-(\ref{5.3b}), which do not obey the boundary constraints.
 
  \begin{figure}[h!]
 \centering
\includegraphics[width=5.5cm]{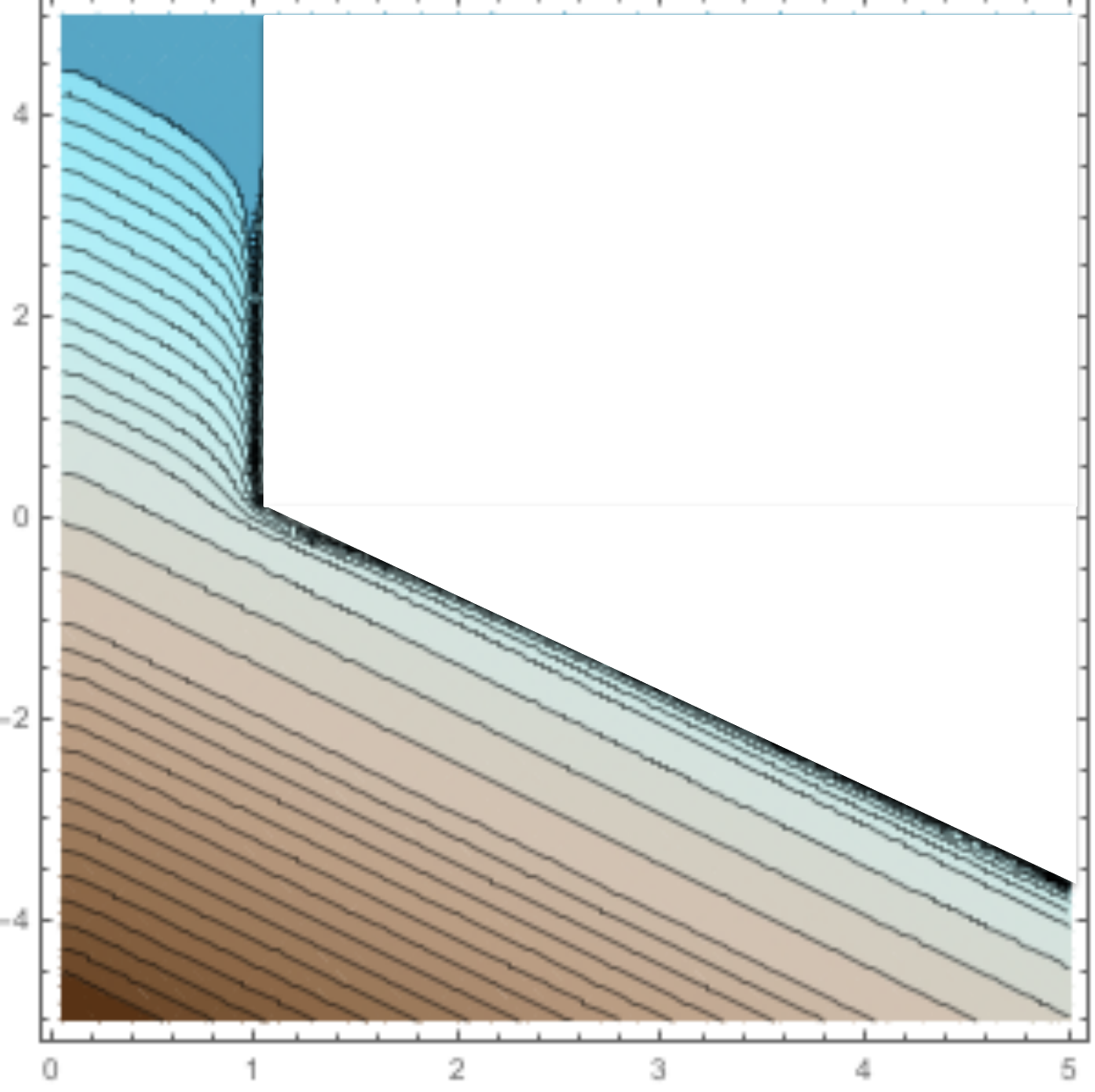}$\,\,$A$\,\,\,\,\,\,\,\,$$\,\,\,\,\,\,\,\,$$\,\,\,\,\,\,\,\,$
\includegraphics[width=5.5cm]{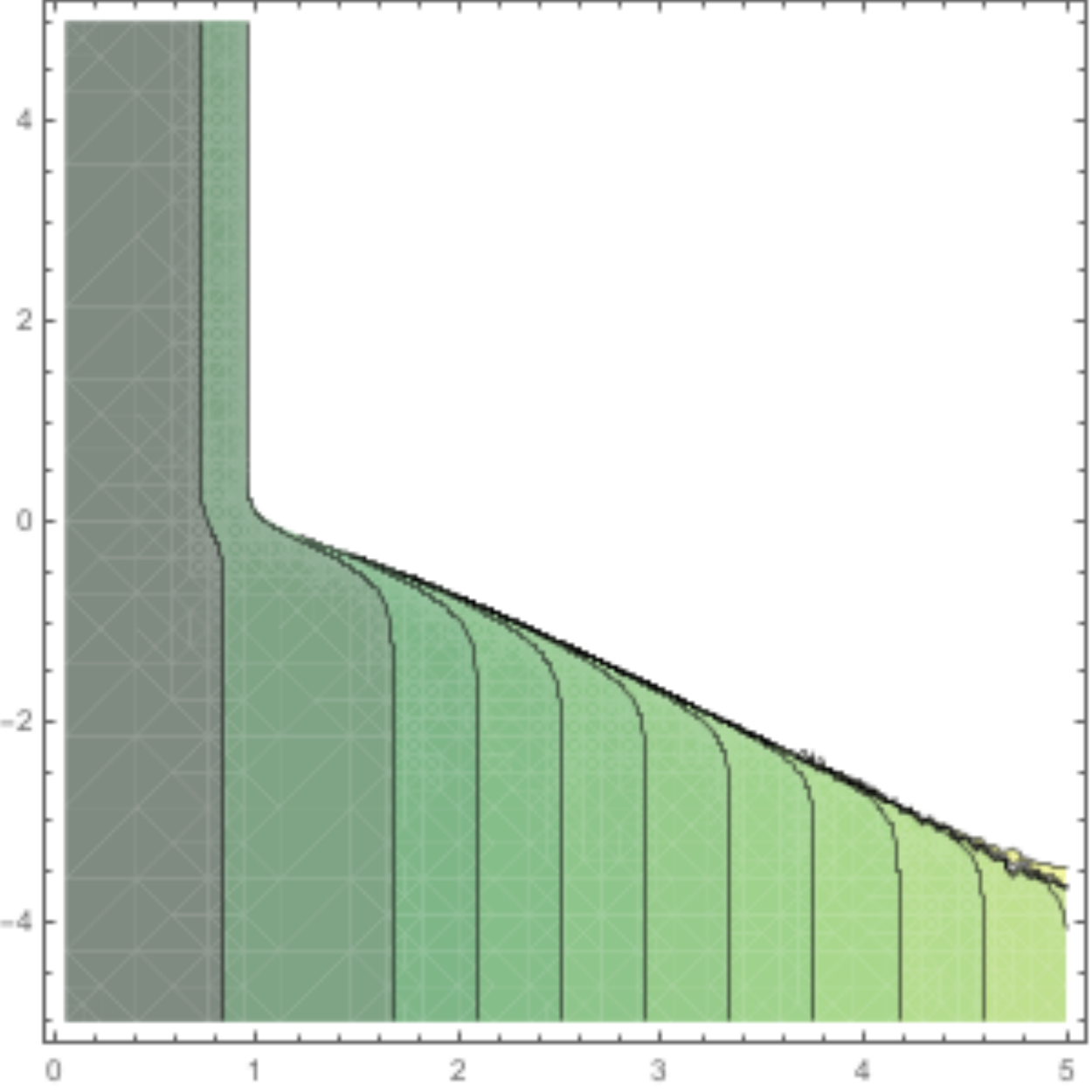}$\,\,$B
 \includegraphics[width=5.5cm]{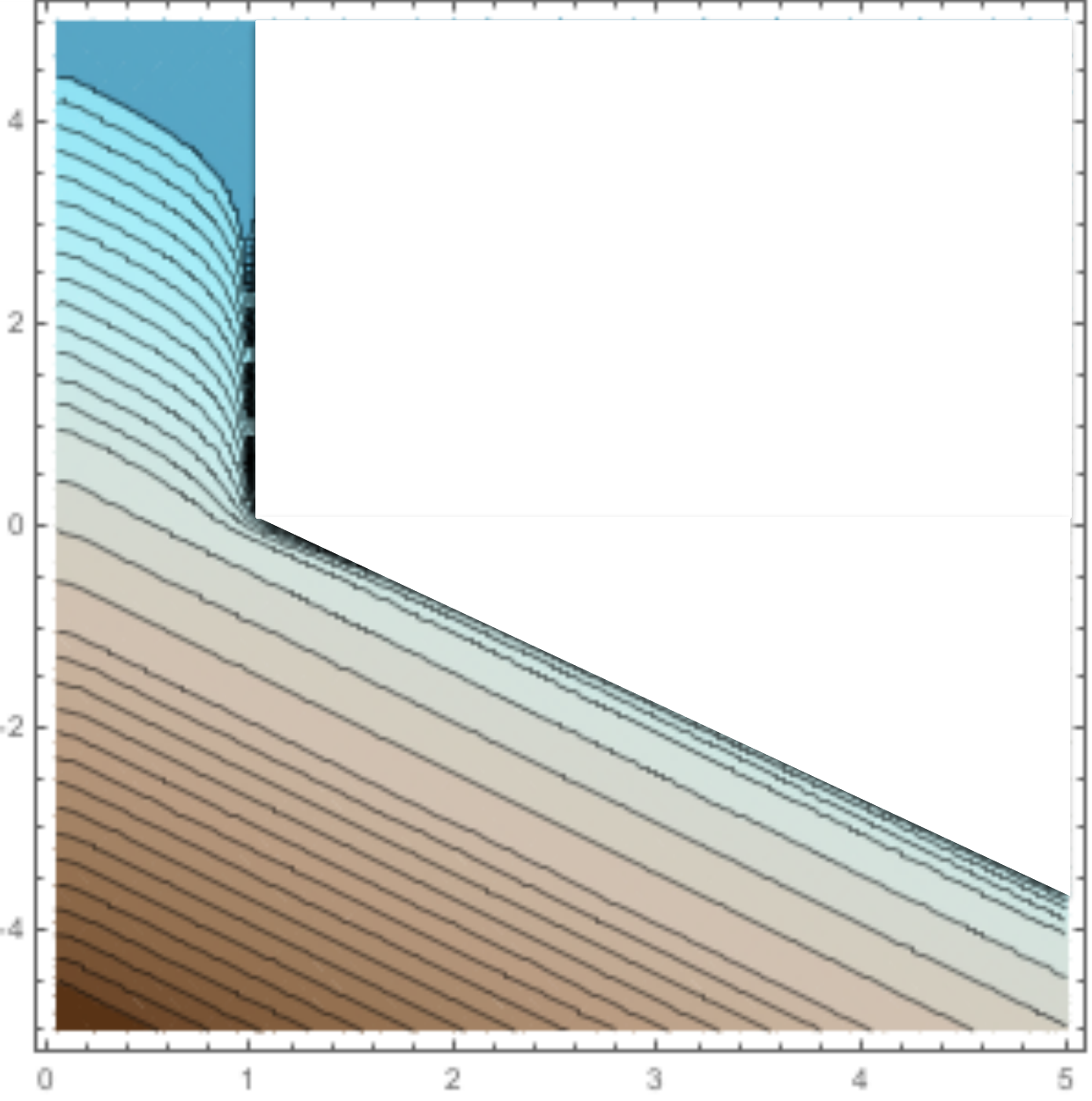}$\,\,$C$\,\,\,\,\,\,\,\,$$\,\,\,\,\,\,\,\,$$\,\,\,\,\,\,\,\,$
\includegraphics[width=5.5cm]{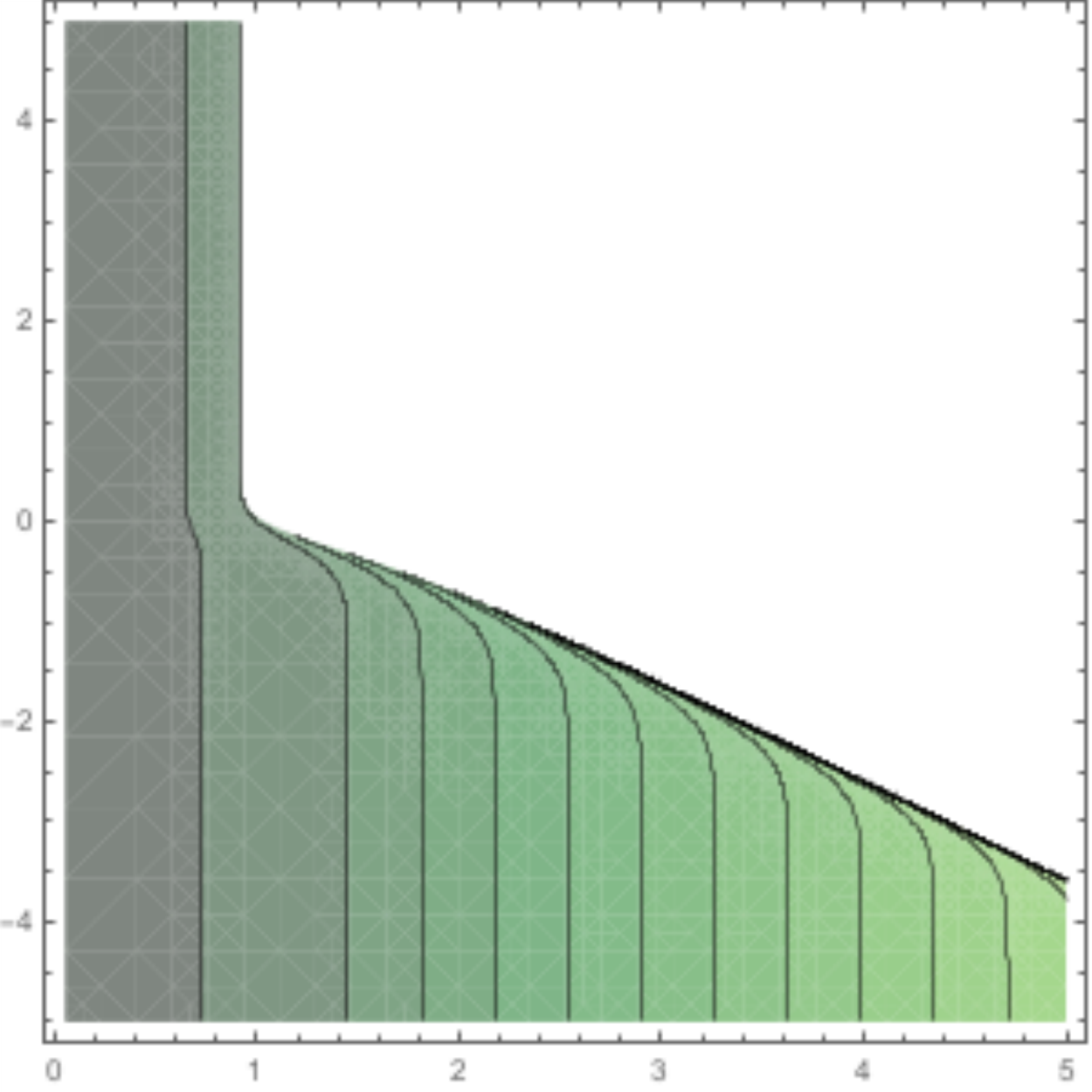}$\,\,$D
\includegraphics[width=5.5cm]{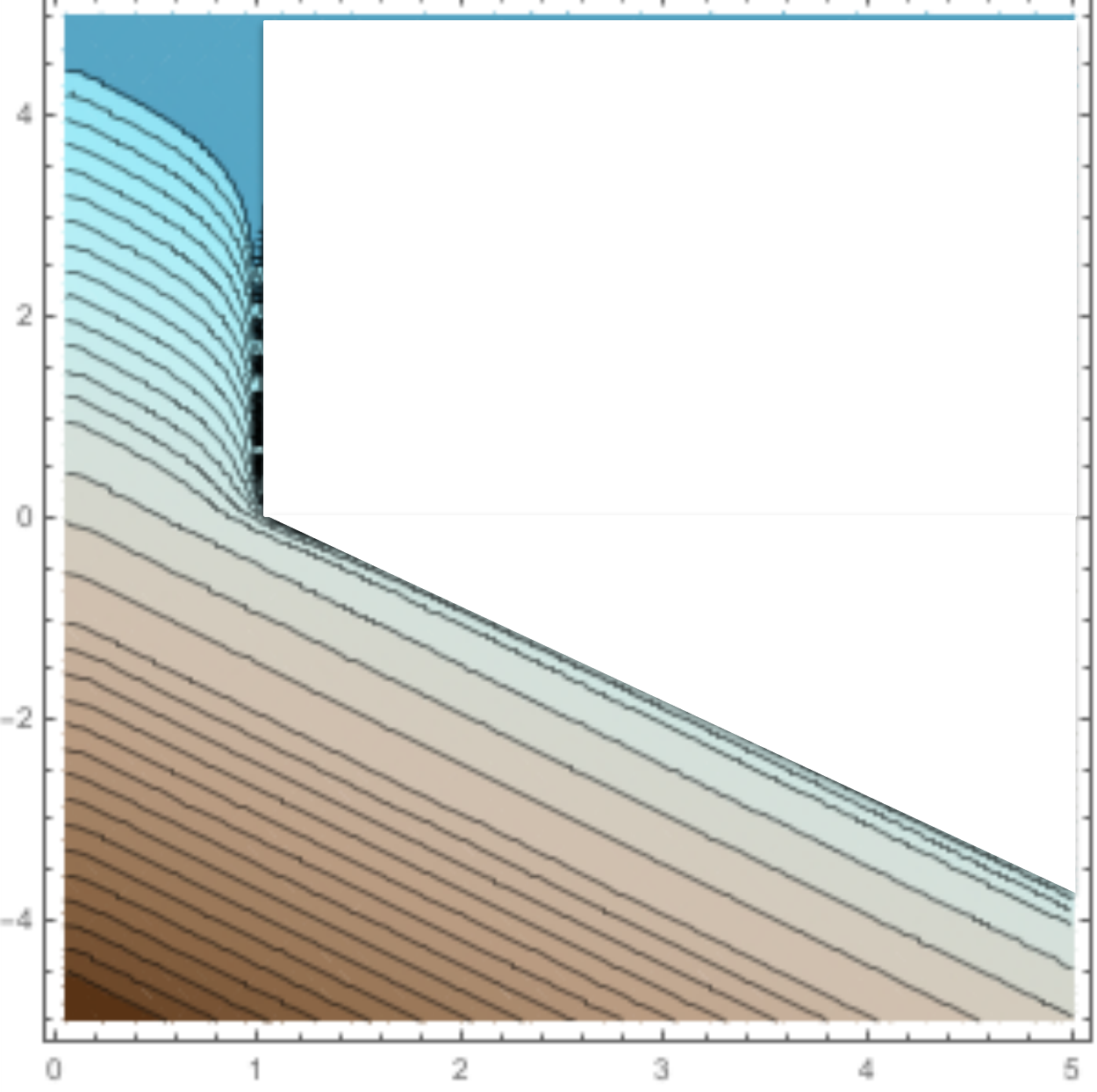}$\,\,$E$\,\,\,\,\,\,\,\,$$\,\,\,\,\,\,\,\,$$\,\,\,\,\,\,\,\,$
\includegraphics[width=5.5cm]{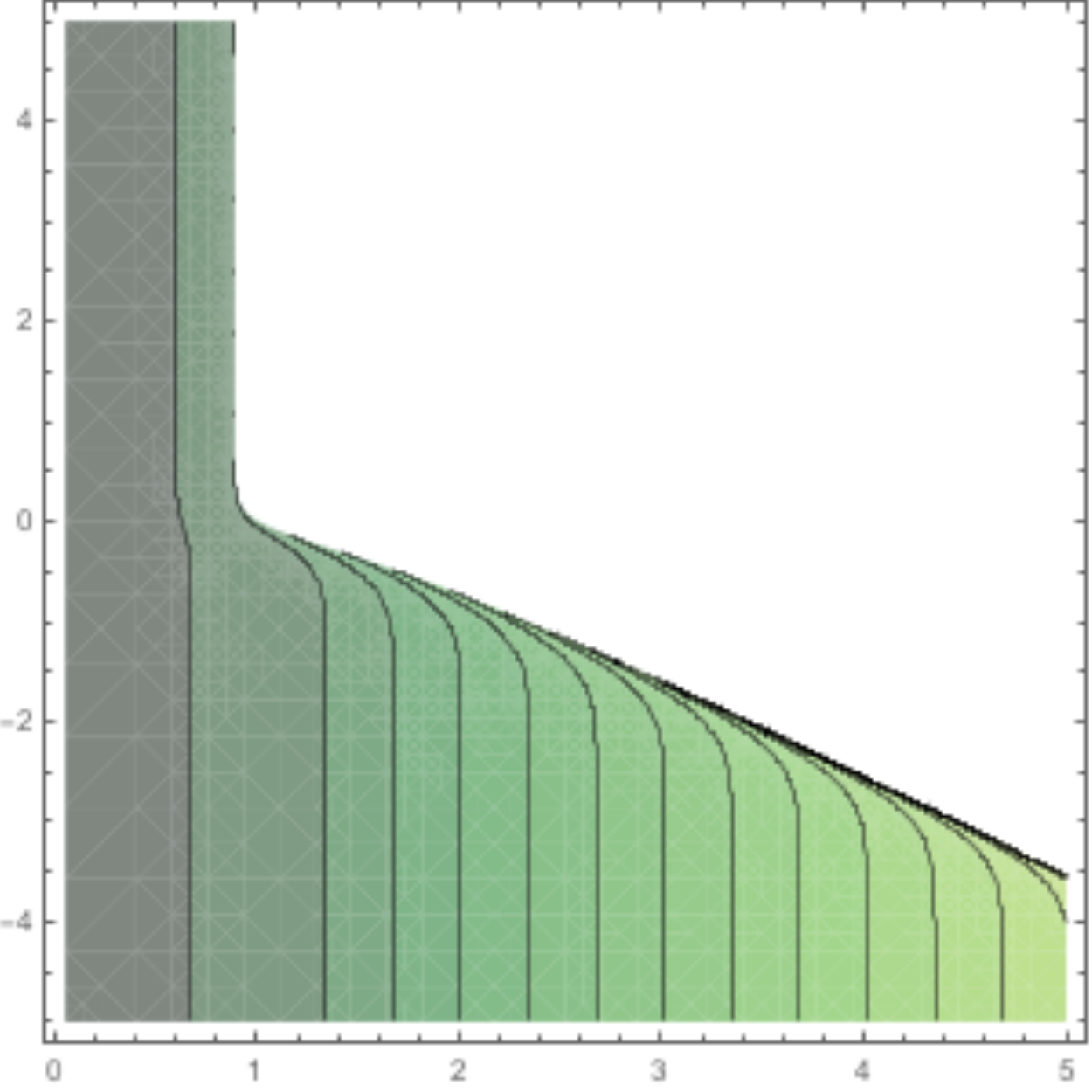}$\,\,$F
  \caption{Left panel:  contour plots for the boundary time as a function of initial conditions
  $z_*$ and $v_*$  for eqs. (\ref{5.3a})-(\ref{5.3b}) for $\nu=2,3,4$ ({\bf A,C,E, respectively}). Right panel:  contour plots for $z(l_{sing})$ as a function of  $z_*$ and $v_*$  for eqs. (\ref{5.3a})-(\ref{5.3b}) for $\nu=2,3,4$ ({\bf B,D,F, respectively}). The regions of white colour correspond to irrelevant initial conditions.}
        \label{fig:sing1}
 \end{figure}
 
It is also interesting to find the behaviour of the function $f$ (\ref{f}) as a function of position on the constructed solutions to  eqs. (\ref{5.3a})-(\ref{5.3b}).   In Fig.\ref{fig:z-mN}.A  we present the behavior of $f(x)$  near to 1 in the region of the singular point.

  \begin{figure}[h!]
  \centering
 \includegraphics[width=5cm]{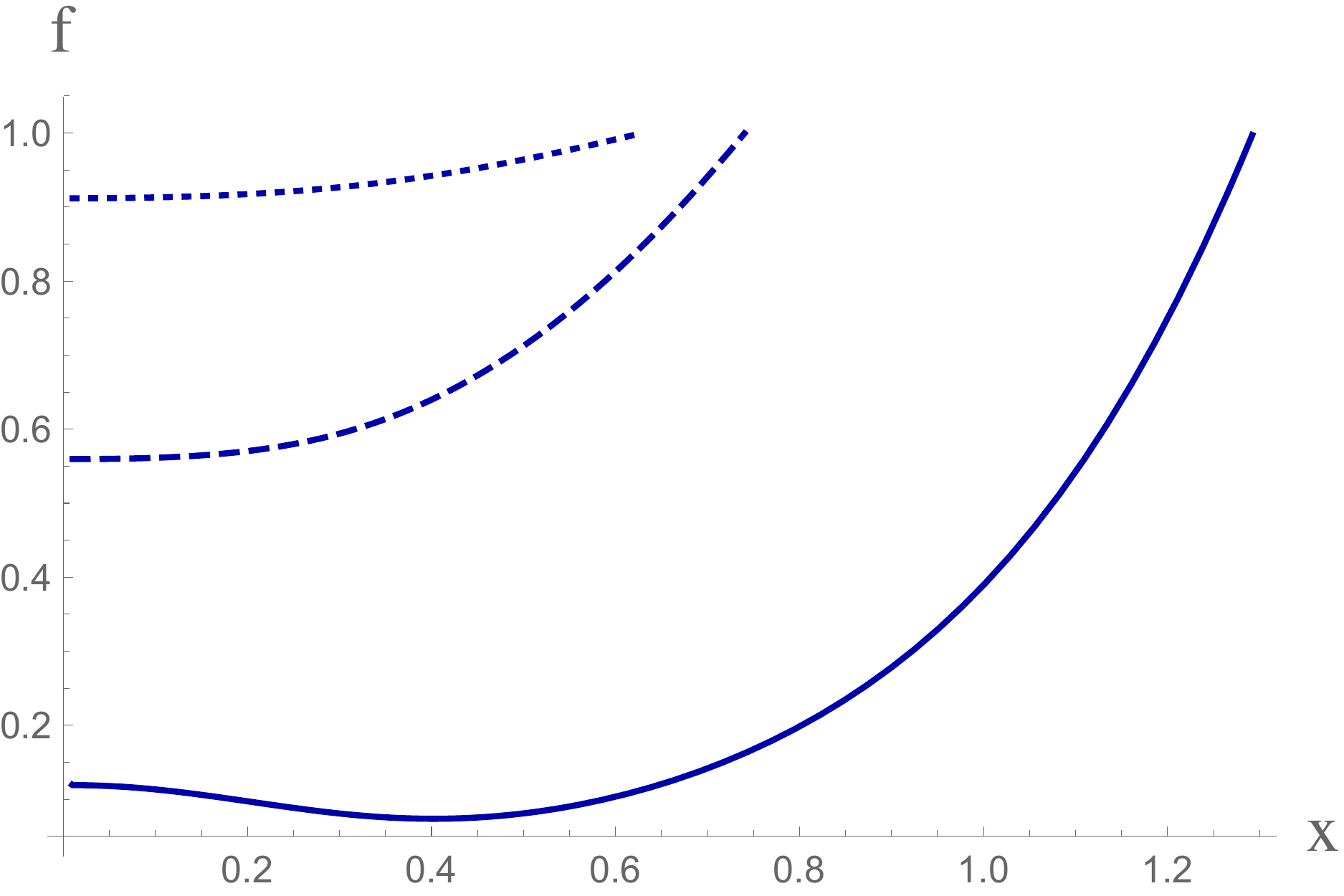}$\,\,\,\,\,\,$A$\,\,\,\,\,\,$$\,\,\,\,\,\,$$\,\,\,\,\,\,$$\,\,\,\,\,\,$
  \includegraphics[width=5.5cm]{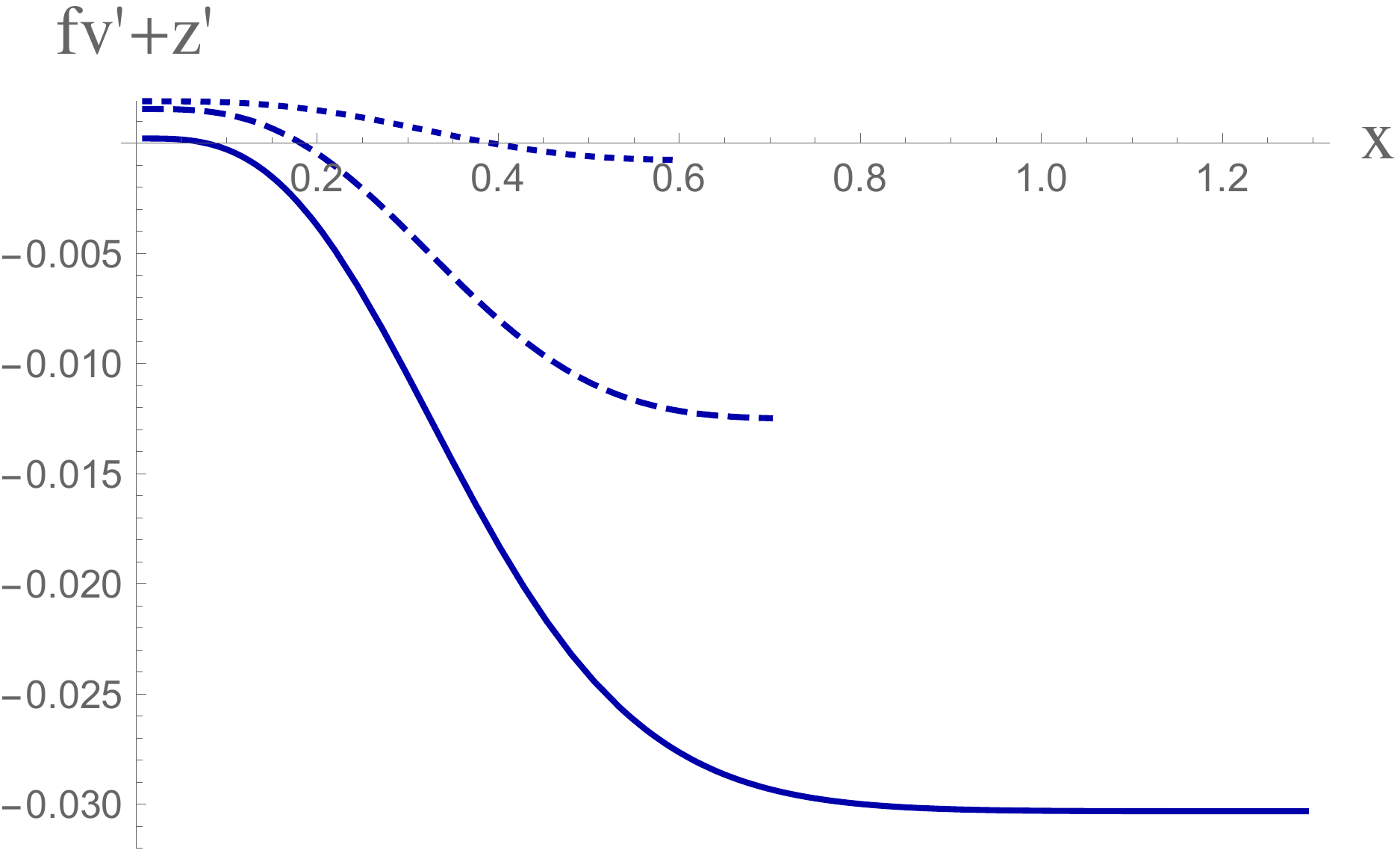}B
          \caption{{\bf A:} The dependence of $f(x)$ on $x$ on the solutions
        to eqs. (\ref{5.3a})-(\ref{5.3b}).
 {\bf B:} The dependence  of the quantity $fv' +z'$ on  the solutions  $z(x),v(x)$ to equations (\ref{5.3a})-(\ref{5.3b}).
   For   both plots $\nu =2$ and  different  masses: $m=1$, $m=0.5$ and $m=0.1$ shown by solid, dashed and dotted lines, respectively.}
        \label{fig:z-mN}
  \end{figure}

   In Fig.\ref{fig:b-fr-n} we check the asymptotic behaviour of
$\mathfrak{b}(z)$ defined by (\ref{b-fr}) on the solution $z(x)$ to equations (\ref{5.3a})-(\ref{5.3b}) for $\nu =2$. For these solutions $z_{*}$ is taken to be $1$. 
We see that for $x\to \ell$, i.e. near the end of the profile, $\mathfrak{b}(z)\to 1$.
\begin{figure}[h!]
 \includegraphics[width=5.5cm]{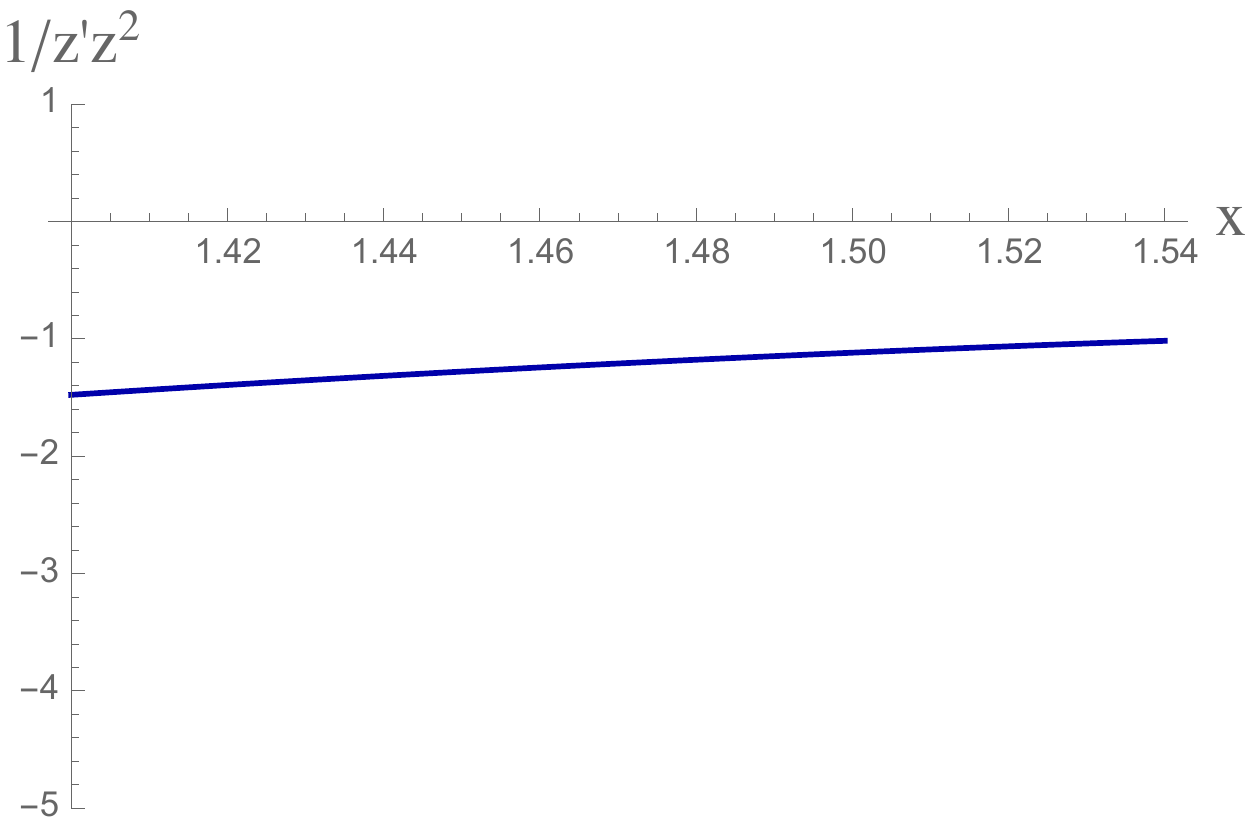}$\,\,\,\,$A$\,\,\,\,\,\,$$\,\,\,\,\,\,$$\,\,\,\,\,\,$$\,\,\,\,\,\,$$\,\,\,\,\,\,$$\,\,\,\,\,\,$
 \includegraphics[width=5.5cm]{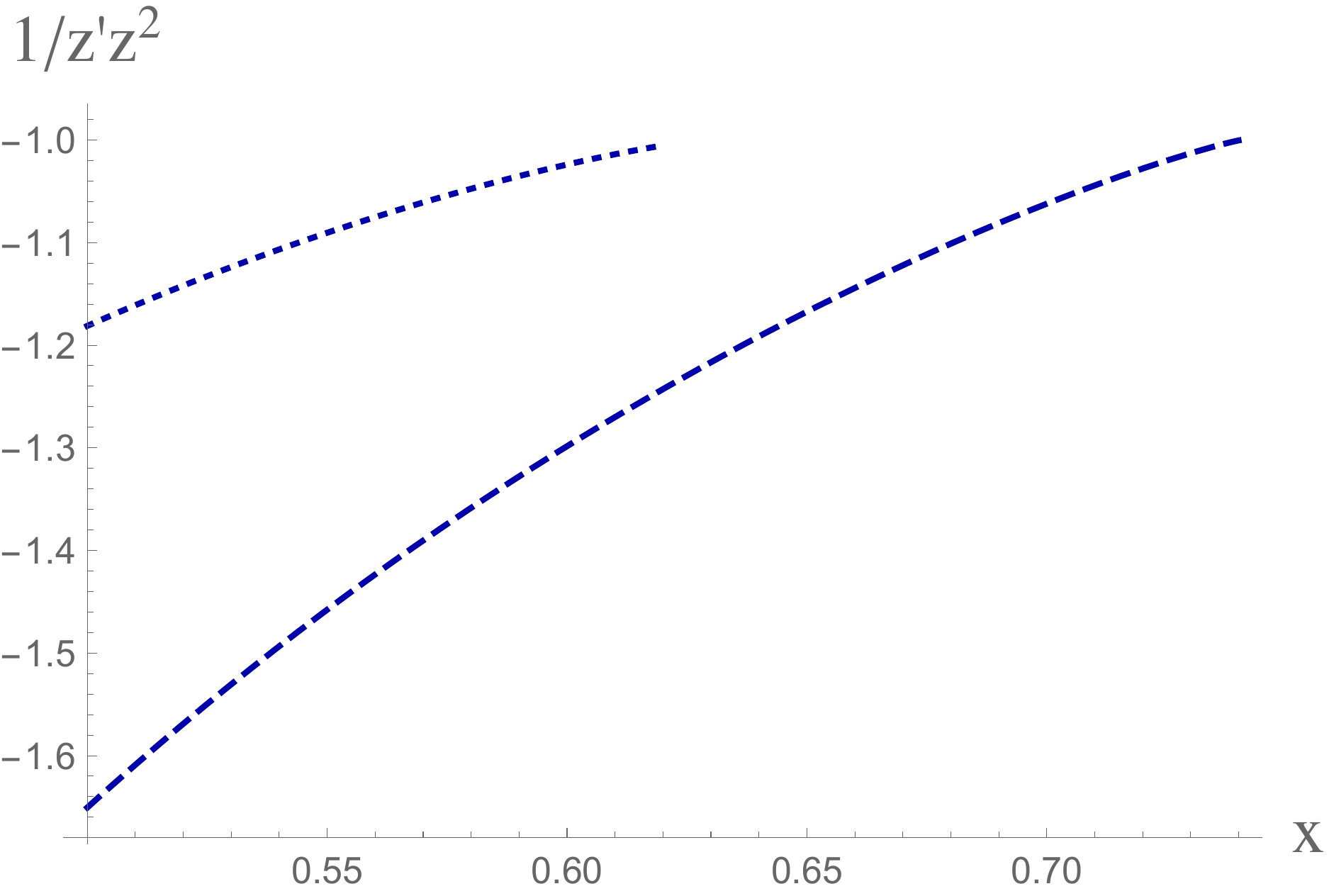}$\,\,\,\,\,\,$B
        \caption{{\bf A:} The dependence  of the quantity (\ref{b-fr}) on the solution $z(x)$ to equations (\ref{5.3a})-(\ref{5.3b}) for  $\nu =2$   and  $m=1$. {\bf B:} The dependence  of the quantity (\ref{b-fr}) on the solution  $z(x)$ to equations (\ref{5.3a})-(\ref{5.3b}) for different masses: $m=0.5$ and $m=0.1$ shown by dashed and dotted lines, respectively, $\nu =2$. In both cases $z_{*} = 1$ and $\ell=0.63$, $\ell=0.75$ for {\bf B} and {\bf C}, respectively.}
        \label{fig:b-fr-n}
  \end{figure}
 \newpage
\subsection{Equations (\ref{EOM-1-2}), (\ref{EOM-2-2})}

In Fig.\ref{fig:zx2} we show the profiles of the solutions to equations (\ref{EOM-1-2})-(\ref{EOM-2-2})
for $z_*=1$ and  different values of $\alpha$ and different $v_*$. We see that the profile for
$\alpha=0.05$ is sharper, as can be obviously expected.

 \begin{figure}[h!]
  \centering
     \includegraphics[scale=0.3]{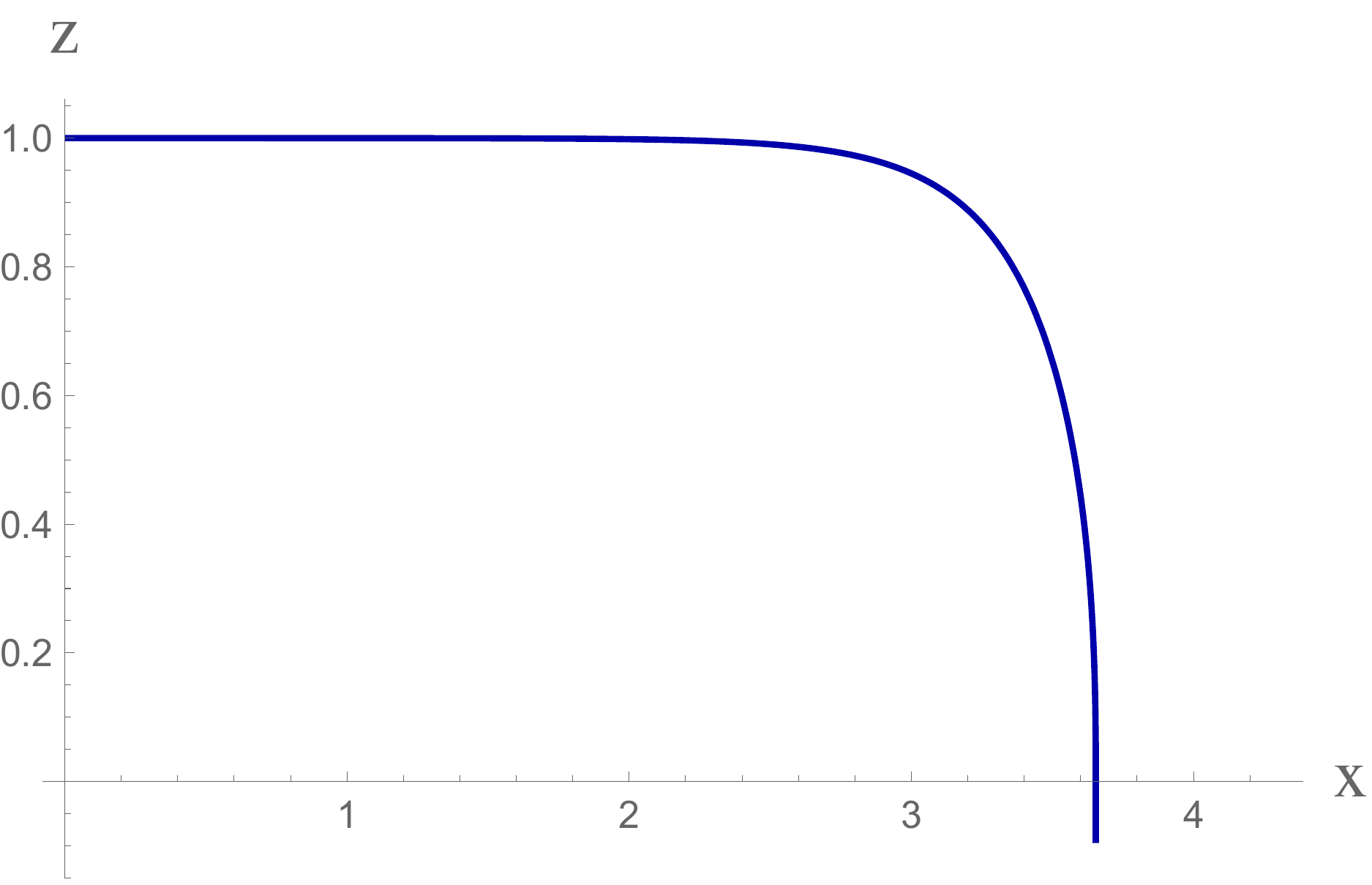}$\,\,\,$A.     $\,\,\,\,\,\,$ $\,\,\,\,\,\,$
      \includegraphics[scale=0.3]{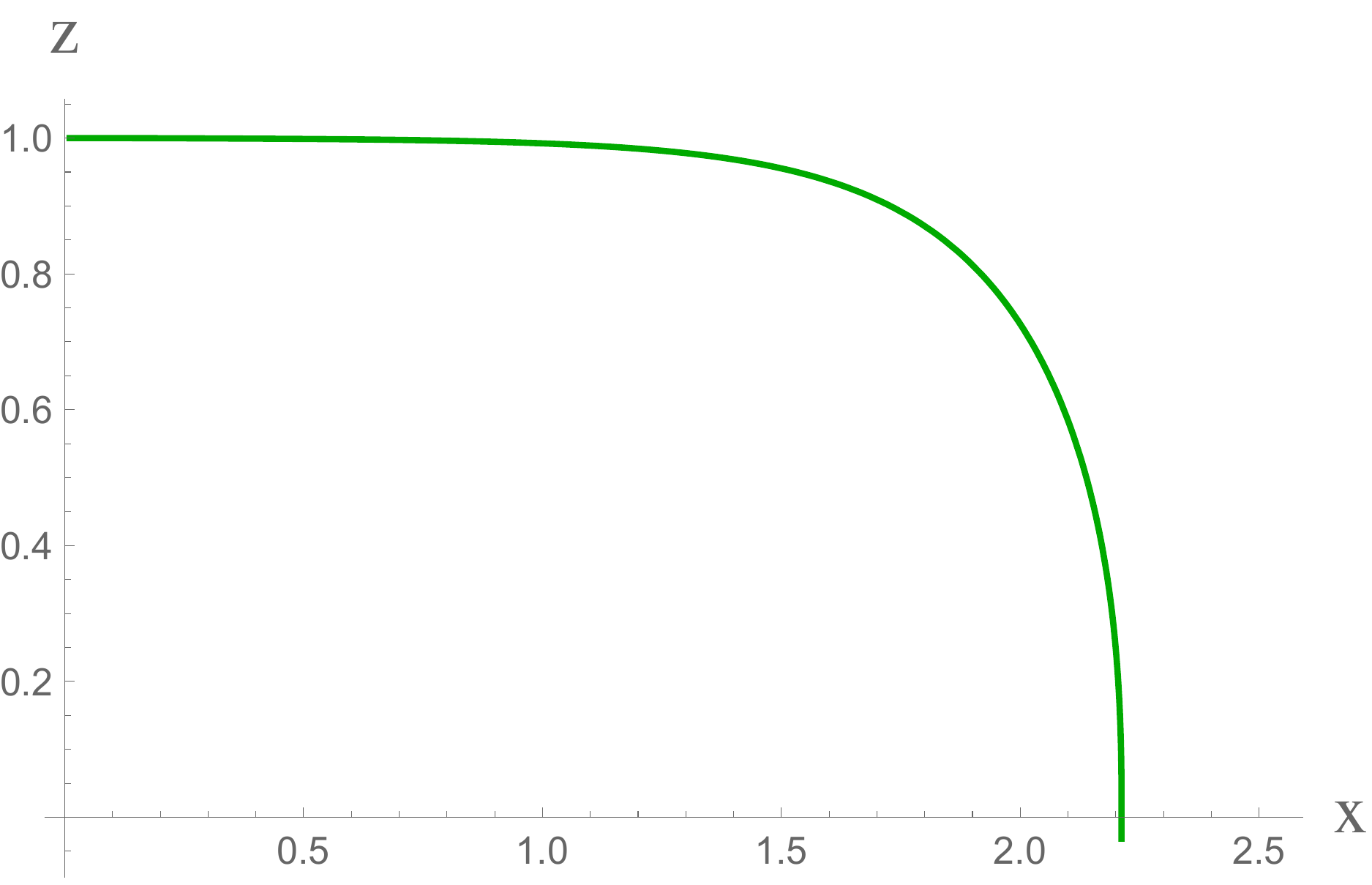}$\,\,\,$B
            \includegraphics[scale=0.3]{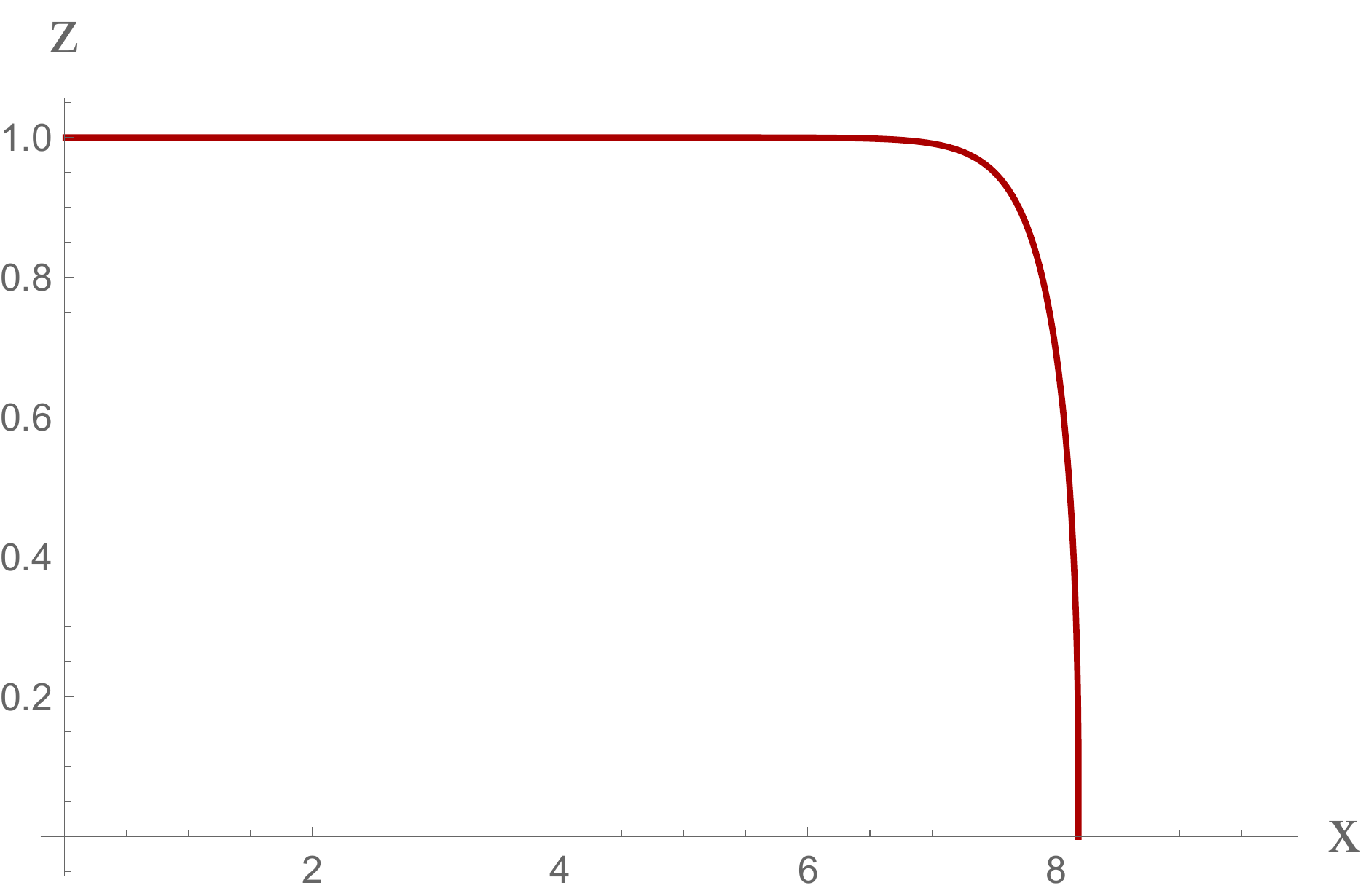}$\,\,\,$C$\,\,\,\,\,\,$$\,\,\,\,\,\,$
       \includegraphics[scale=0.3]{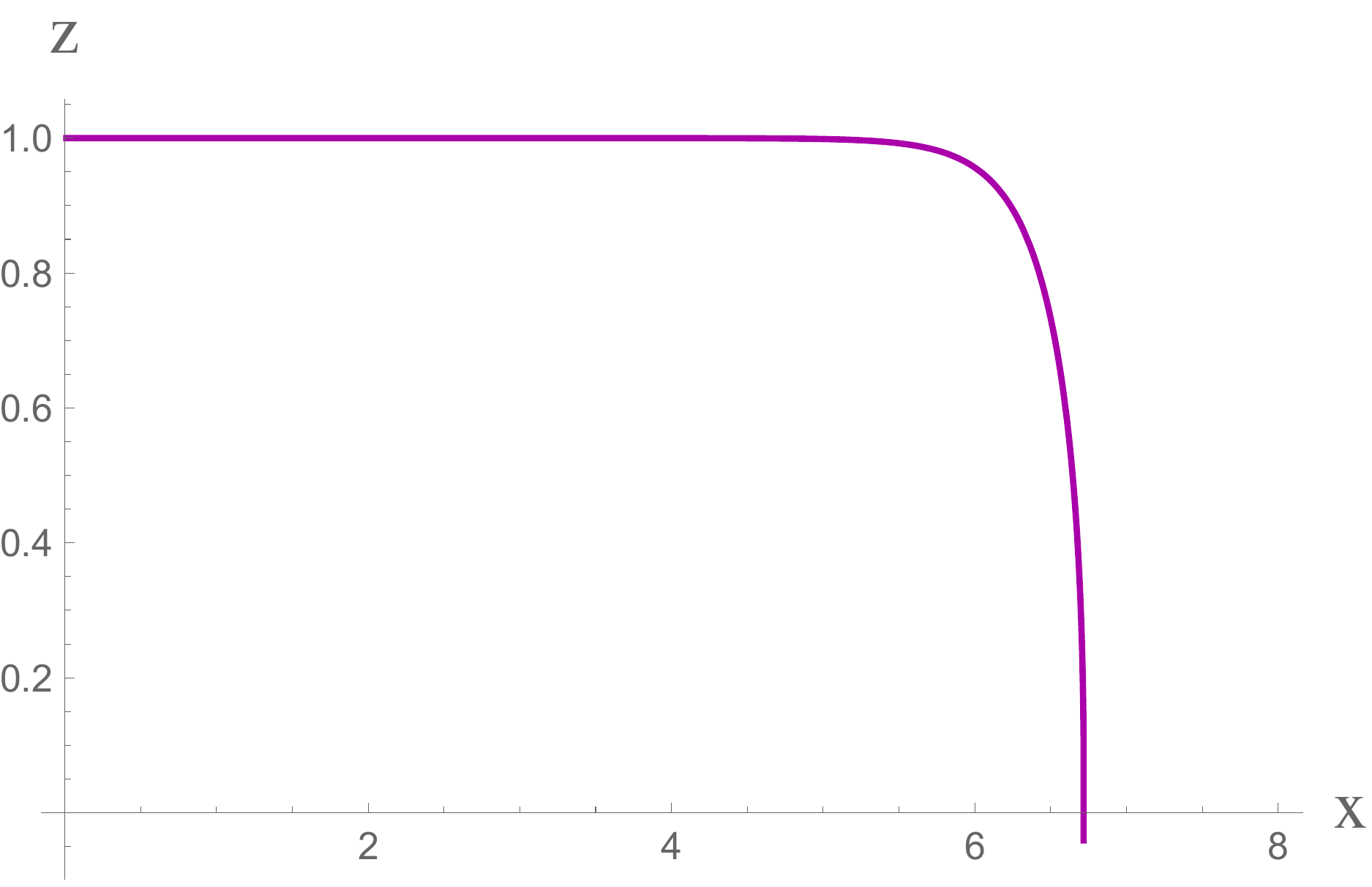}$\,\,\,$D
      \caption{Forms of the profiles of the solutions to eqs. (\ref{EOM-1-2})-(\ref{EOM-2-2}). {\bf A:} $v_*=1$, {\bf B:} $v_*=0.5$.  For both cases $z_*=1$ and $\alpha=0.2$.
      {\bf C:} $v_*=1$, {\bf D:} $v_*=0.5$.  In both cases $z_*=1$ and $\alpha=0.05$.}
        \label{fig:zx2}
 \end{figure}

     \begin{figure}[h!]
\centering
     \includegraphics[scale=0.32]{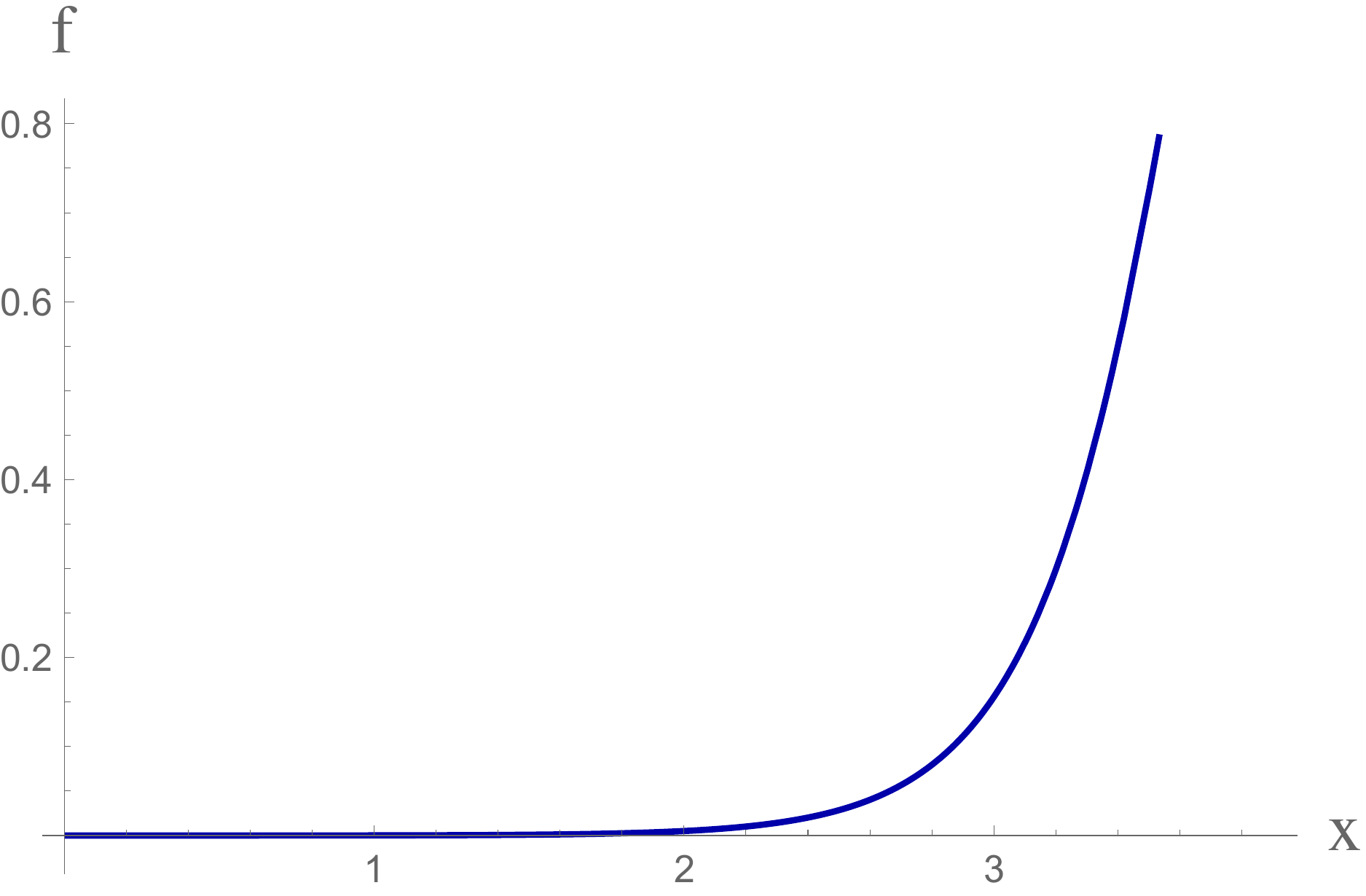}$\,\,\,\,\,\,$A
      \includegraphics[scale=0.32]{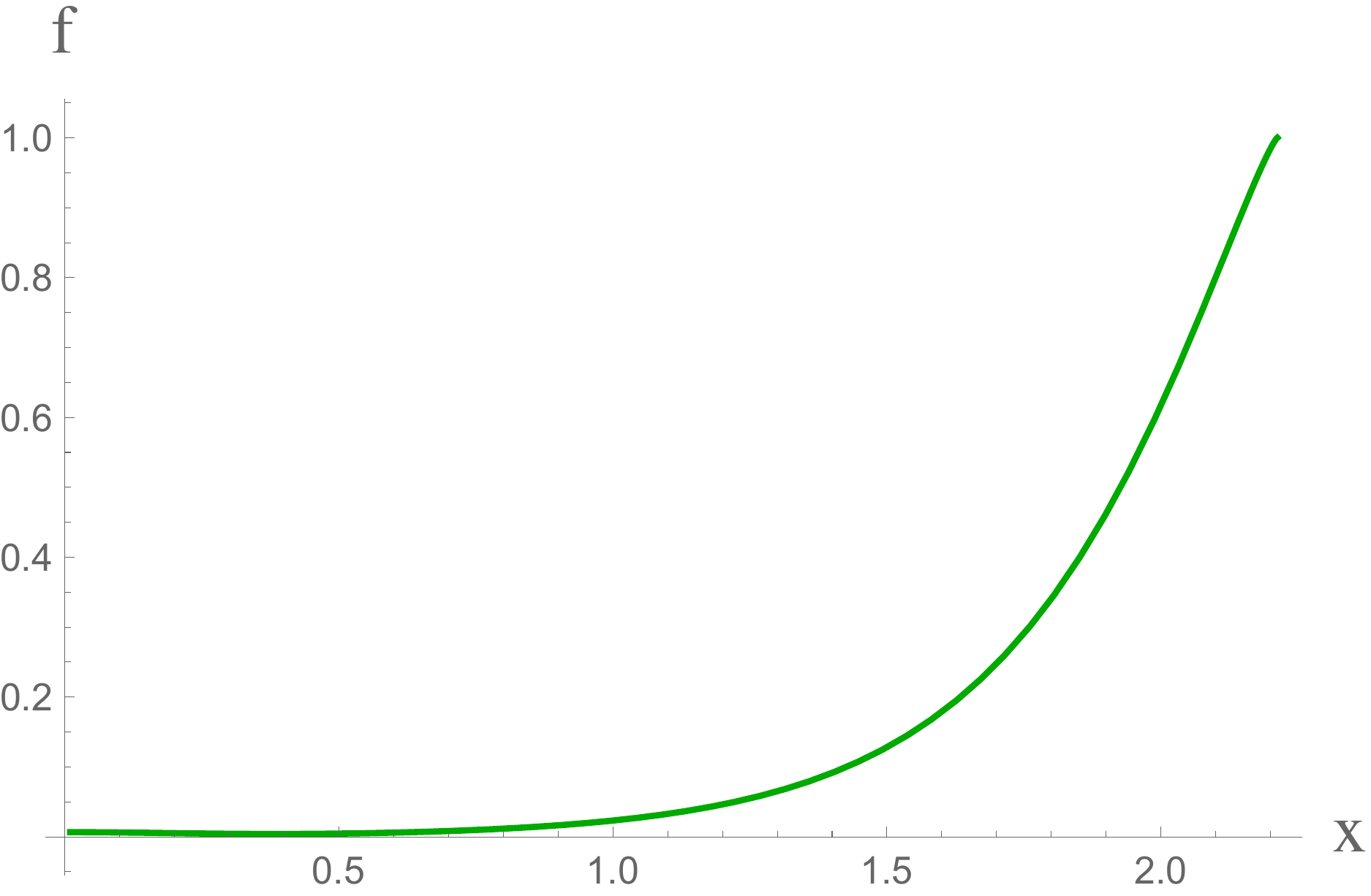}$\,\,\,\,\,\,$B
         \caption{
         The dependence $f=f(x)$ on the solutions to  eqs. (\ref{EOM-1-2})-(\ref{EOM-2-2}). {\bf A:} $v_*=1$, {\bf B:} $v_*=0.5$. For both cases $z_*=1$.}
        \label{fig:fx}
 \end{figure}

As it has been mentioned in Sect.~\ref{Sect:4.2.2} on the region, where we can guarantee, that  the quantity
 \be
 {\cal Q} = \frac{f(z)\dot{v}_{y} + \dot{z}_{y}}{2z^{2 - 2/\nu}}
 \ee
  is conserved on the solution, we can say  the solution can be approximated by the static solution in this region. 
  We also present the dependence of the "quasi" conserved quantity ${\cal Q}$ on $x$ for $\nu=2$  in Fig.\ref{fig:QC} .
  
  \begin{figure}[t]
\centering
     \includegraphics[scale=0.32]{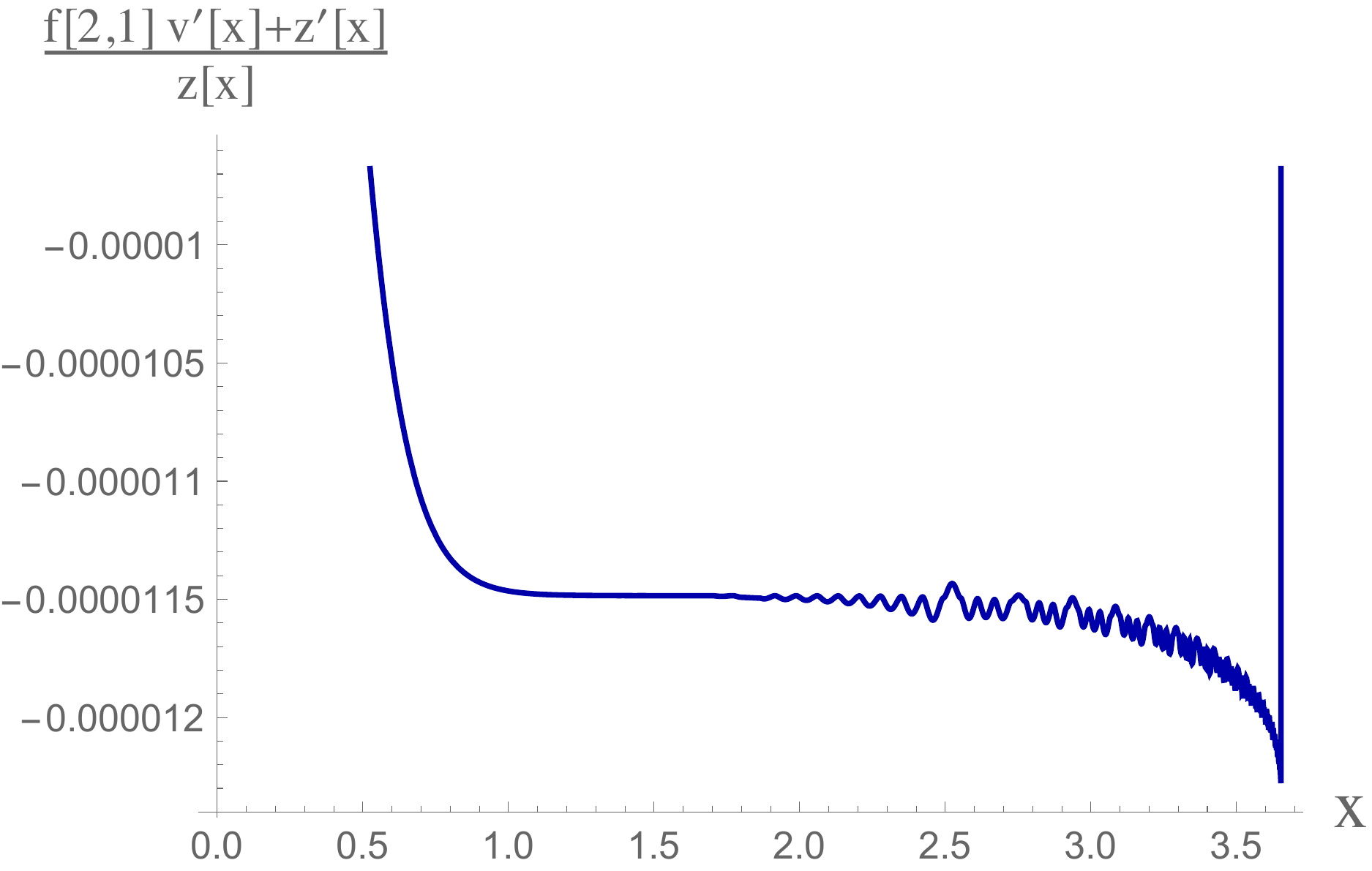}$\,\,\,\,\,\,$A$\,\,\,\,\,\,$$\,\,\,\,\,\,$$\,\,\,\,\,\,$
      \includegraphics[scale=0.32]{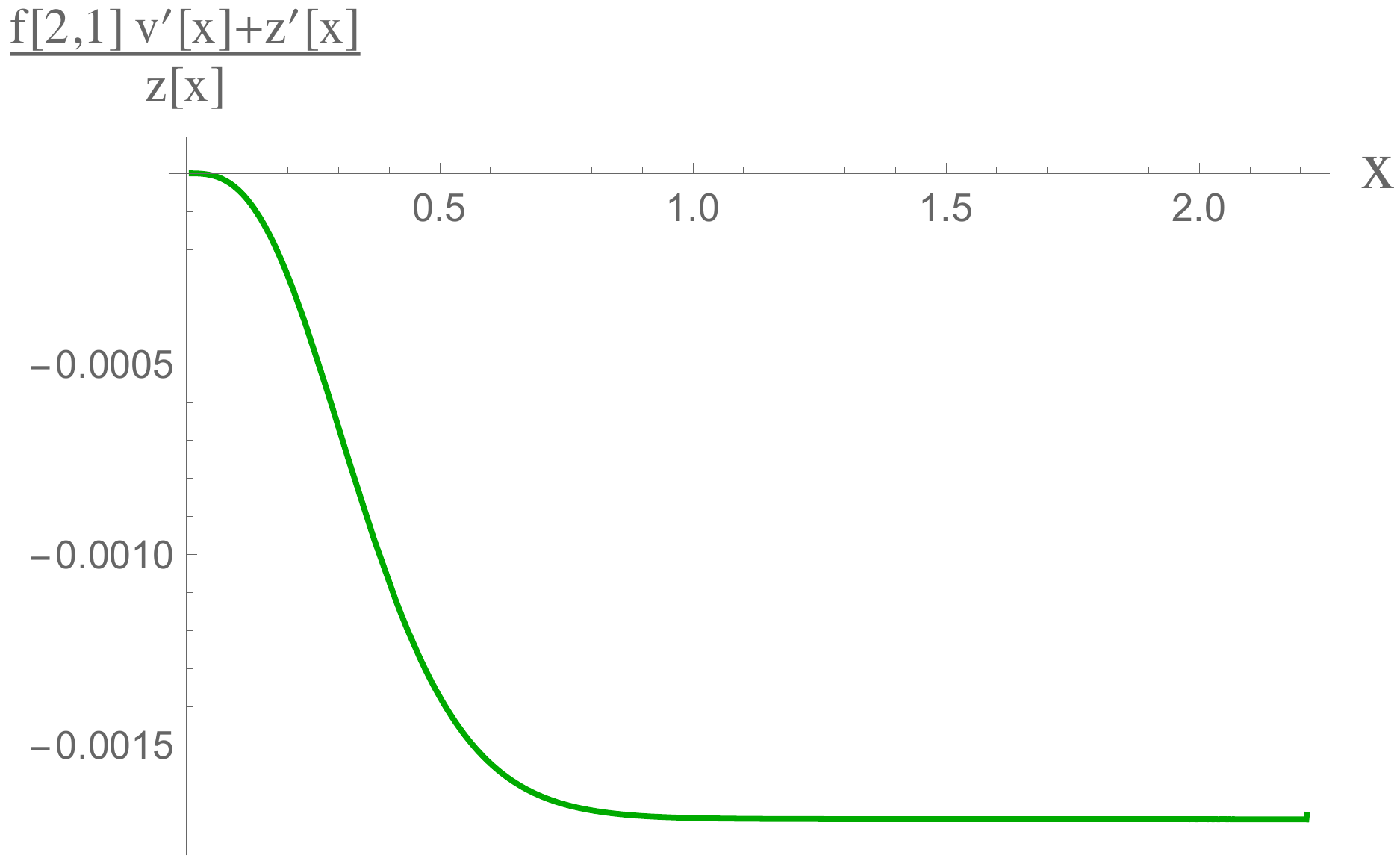}$\,\,\,\,\,\,$B
 \caption{Behaviour of a "quasi" conserved quantity ${\cal Q}$ for $\nu=2$. {\bf A:} $v_*=1$, {\bf B:} $v_*=0.5$. For both cases $z_*=1$}
        \label{fig:QC}
 \end{figure}

 In Fig.\ref{fig:asymp} we check the asymptotic behaviour of $\mathfrak{b}(z)$ defined by (\ref{b-got-2}).
      \begin{figure}[h!]
\centering
     \includegraphics[scale=0.3]{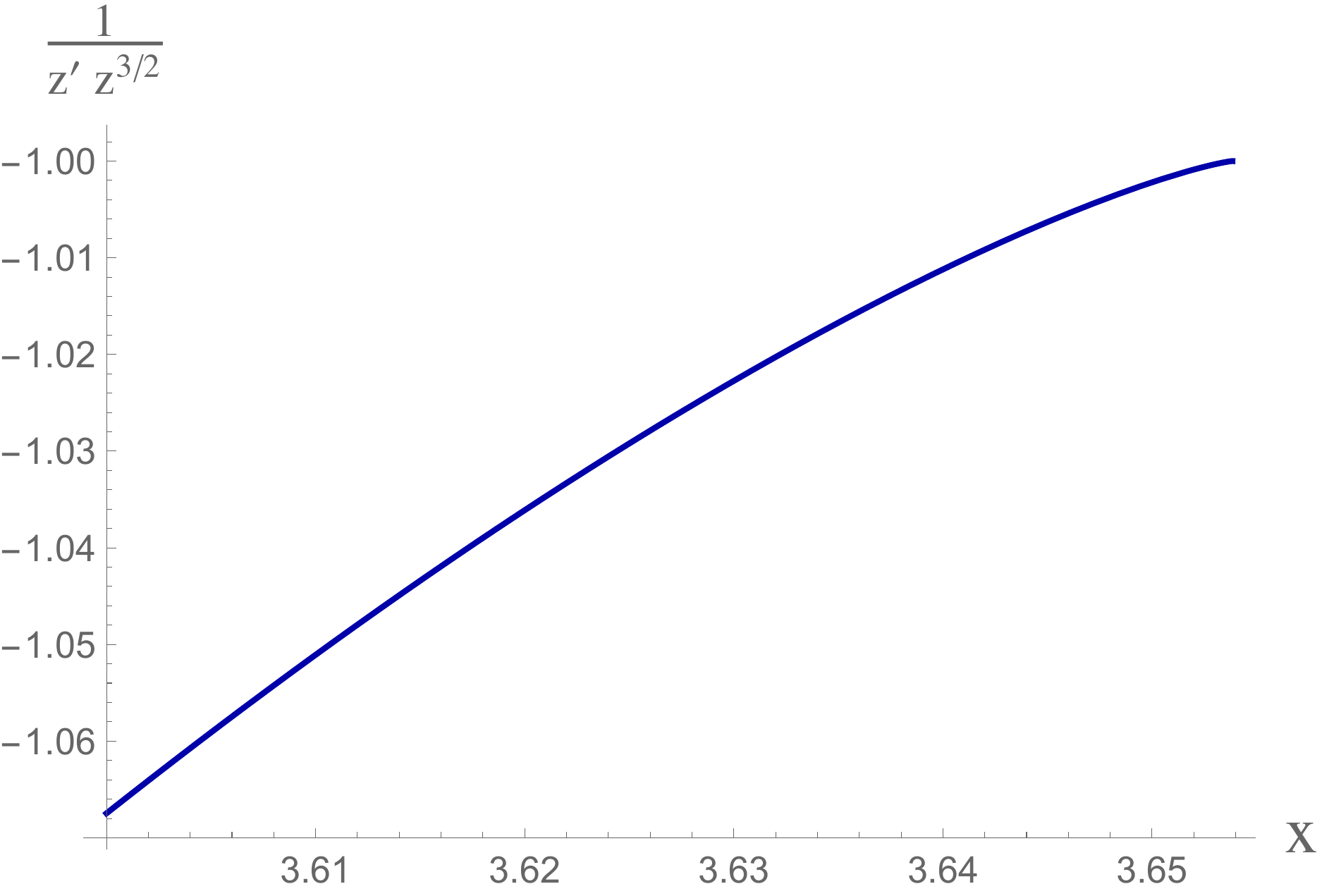}$\,\,\,\,\,\,$A$\,\,\,\,\,\,\,\,\,\,\,\,$
      \includegraphics[scale=0.3]{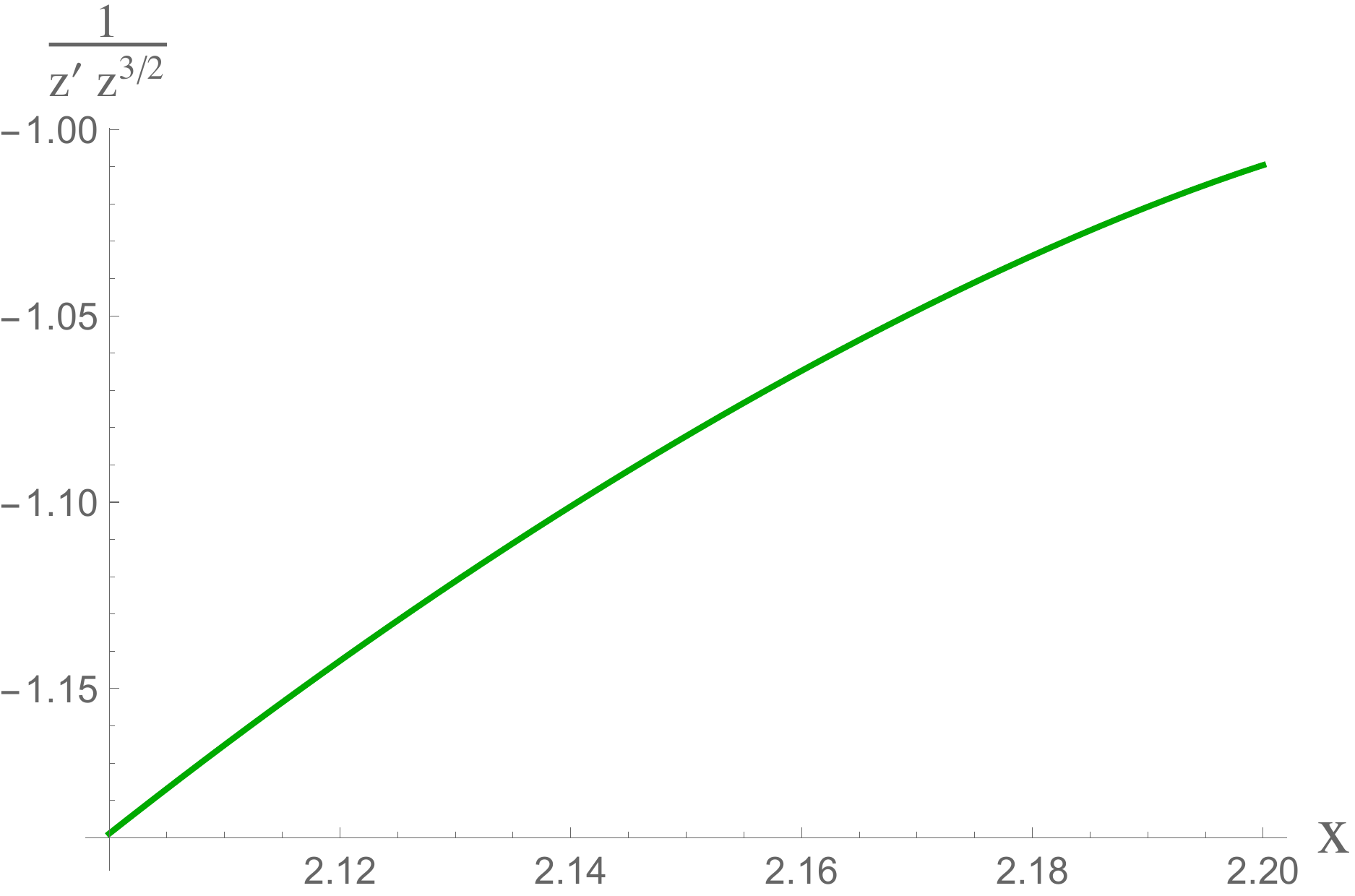}$\,\,\,\,\,\,$B\\
       \includegraphics[scale=0.3]{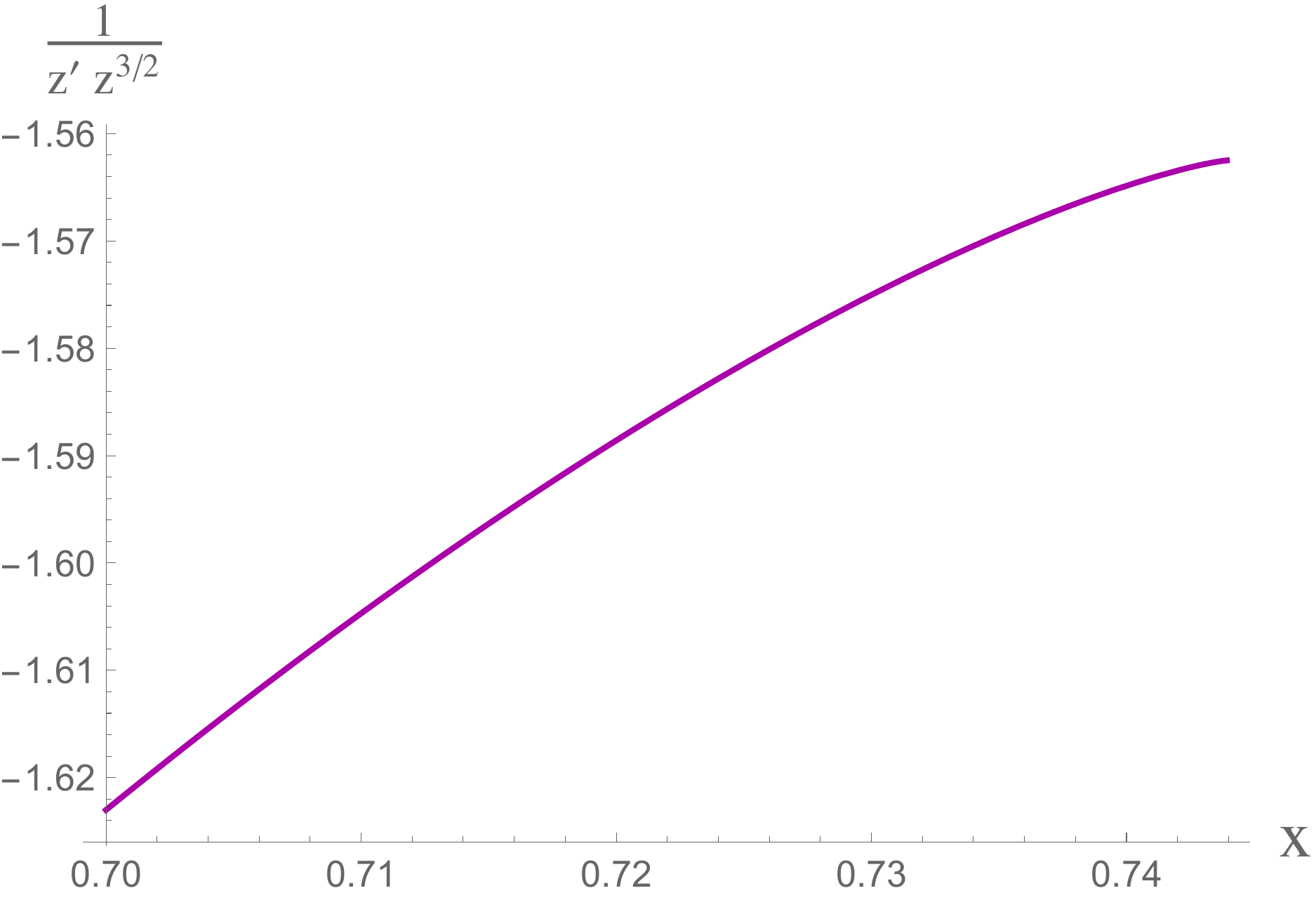}C $\,\,\,\,\,\,\,\,\,\,\,\,\,\,\,\,\,\,$
       \includegraphics[scale=0.3]{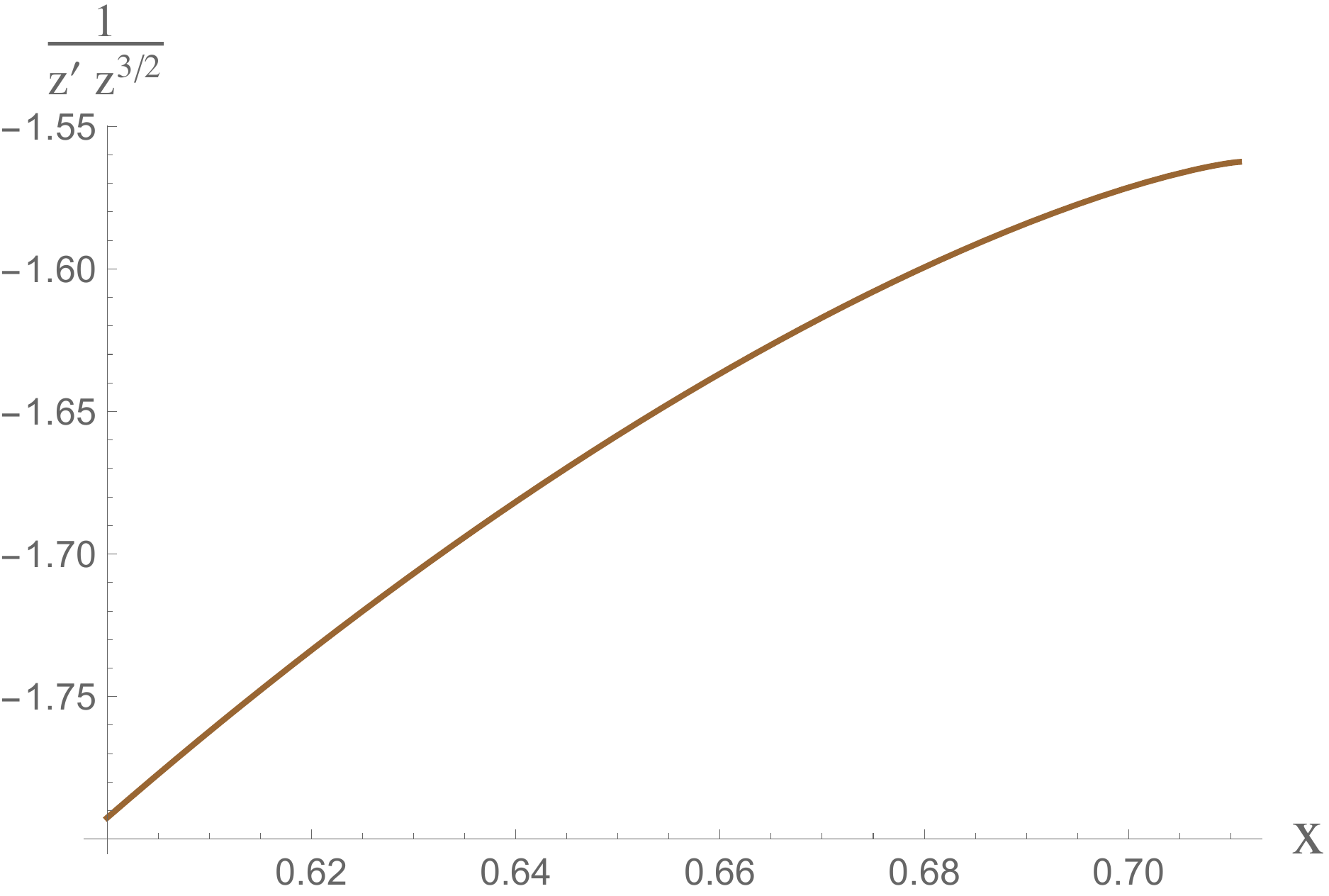}\,\,\,\,\,\,$D$\\
     \includegraphics[scale=0.3]{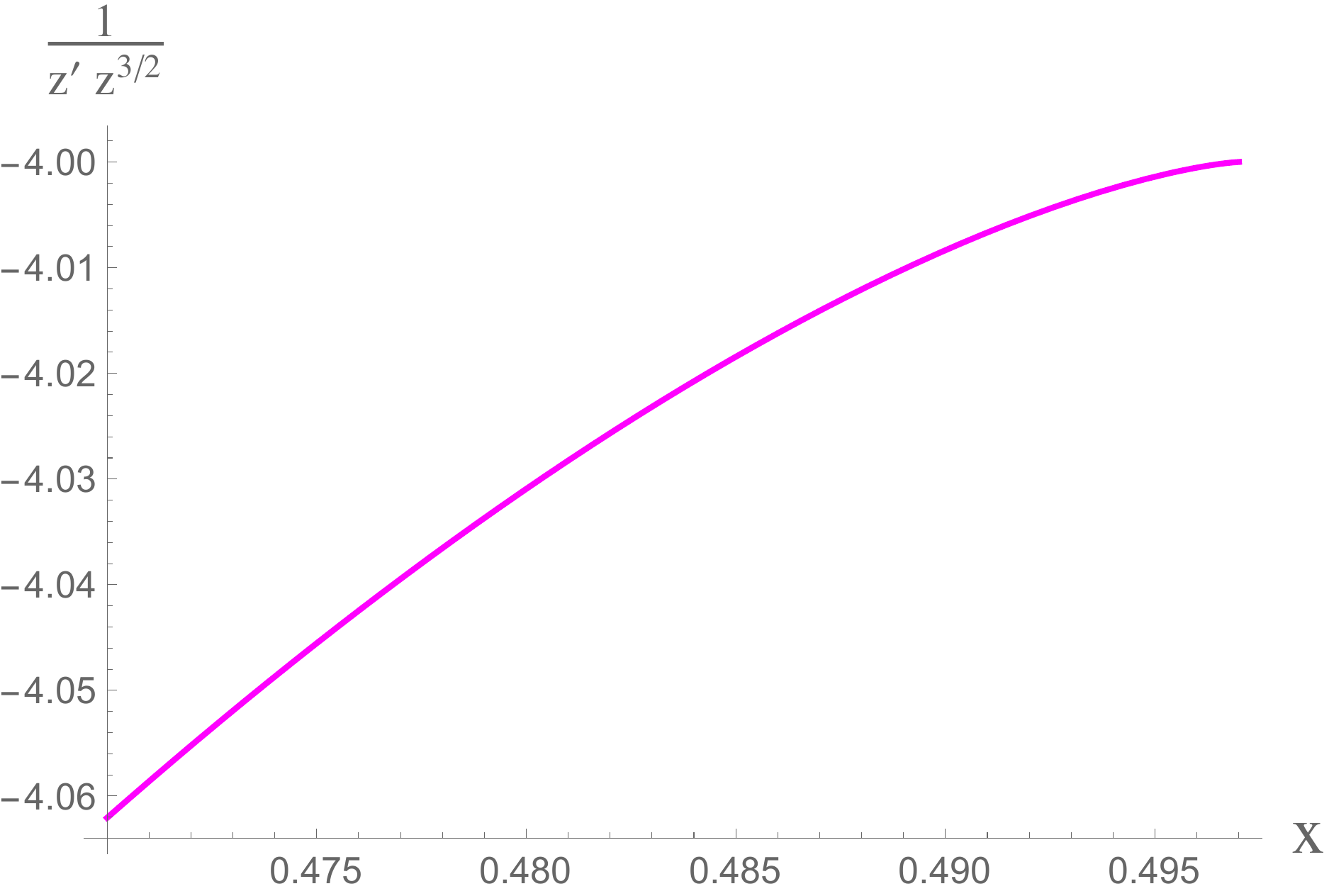} $\,\,\,\,\,\,$E$\,\,\,\,\,\,\,\,\,\,\,\,$
     \includegraphics[scale=0.3]{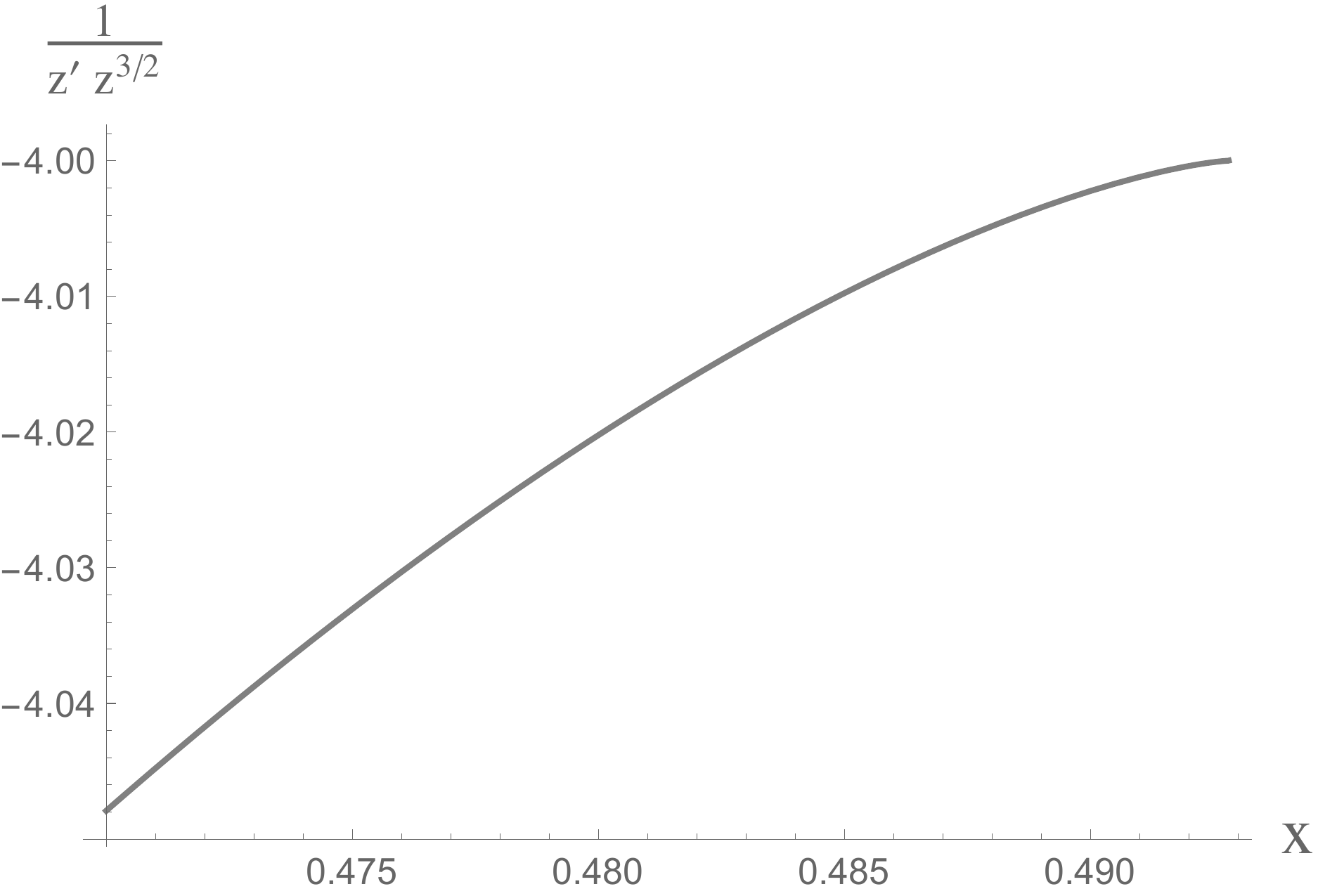}$\,\,\,\,\,\,$F
      \caption{Check of asymptotic behaviour of $\mathfrak{b}(z)$ near $z=0$. {\bf A:} $v_*=1$, {\bf B:} $v_*=0.5$. In both cases $z_*=1$
     {\bf C:} $v_*=1$, {\bf D:} $v_*=0.1$. In both cases $z_*=0.8$; {\bf E:} $v_*=1$, {\bf F:} $v_*=0.1$. In both cases $z_*=0.5$
     The asymptotic is $1/z_*^2$, i.e. for $z_*=0.8$  it is 1.5625 and for $z_*=0.5$ it is 4.}
        \label{fig:asymp}
 \end{figure}

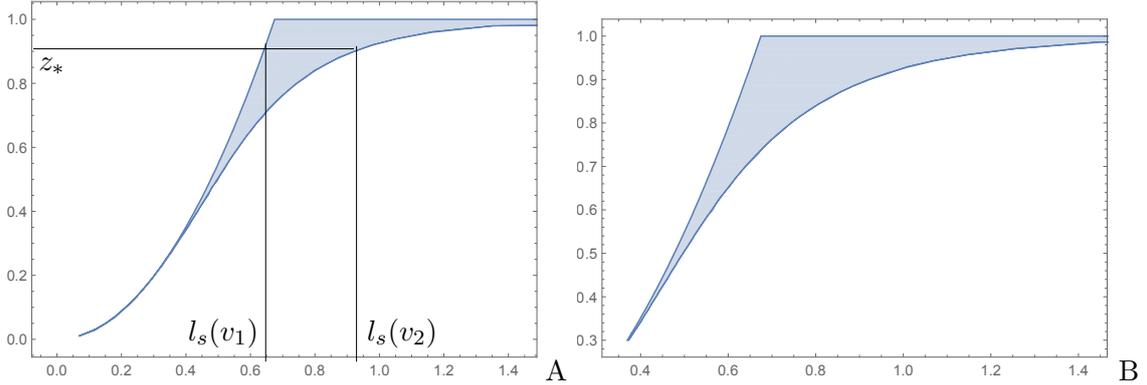
\begin{figure}[h!]
\centering
\begin{picture}(198,170)
\put(0,0){\includegraphics[width=7cm]{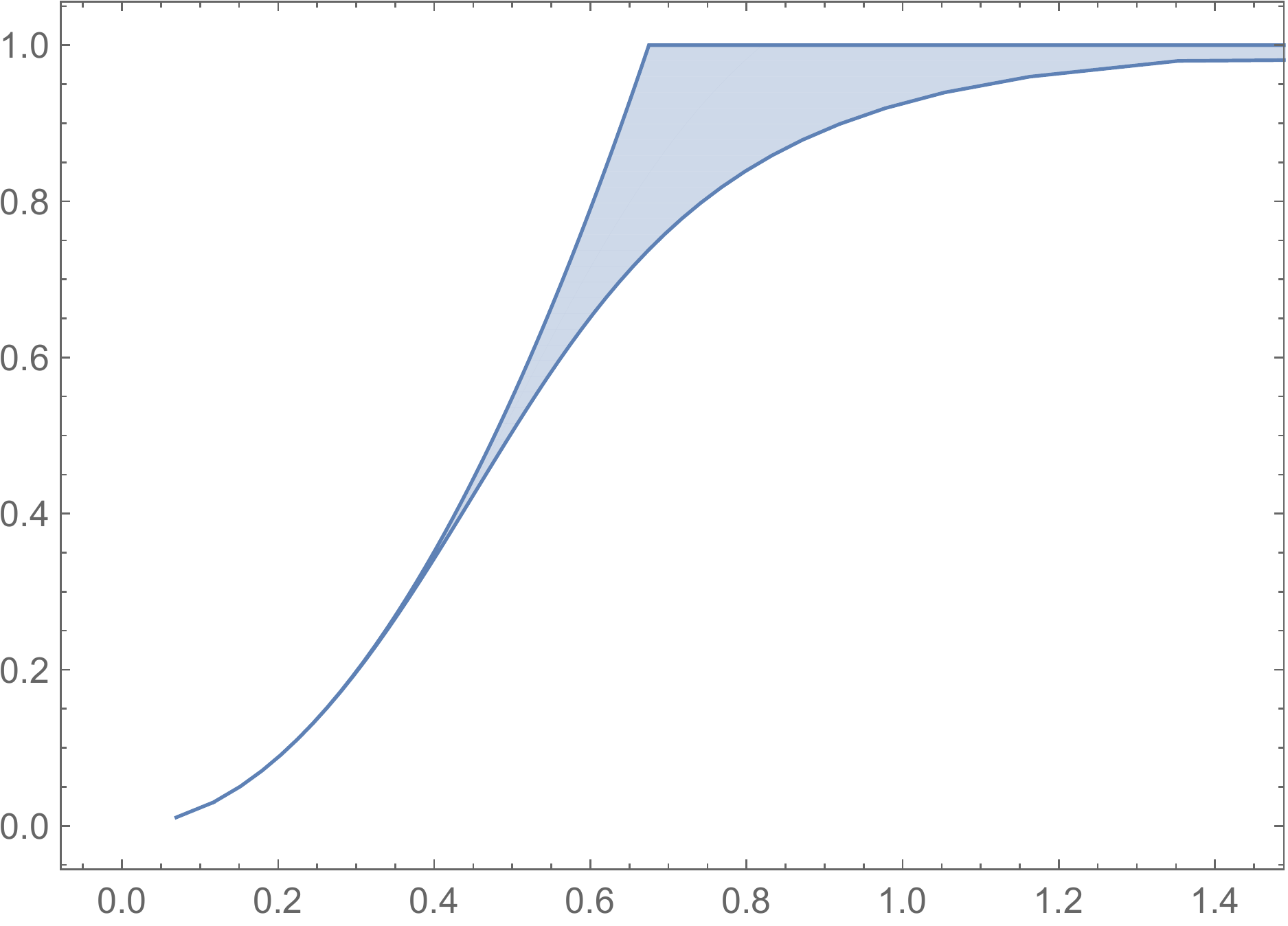}}
\linethickness{0.1mm}
\put(10,125){\line(1,0){120}}\linethickness{0.1mm}
\put(97,7){\line(0,1){120}}
\put(131,7){\line(0,1){119}}\put(12,117){$z_*$}
\put(68,15){$l_{s}(v_ {1})$}
\put(135,15){$l_{s}(v_{2})$}
\end{picture}$\,\,$A
\includegraphics[width=7cm]{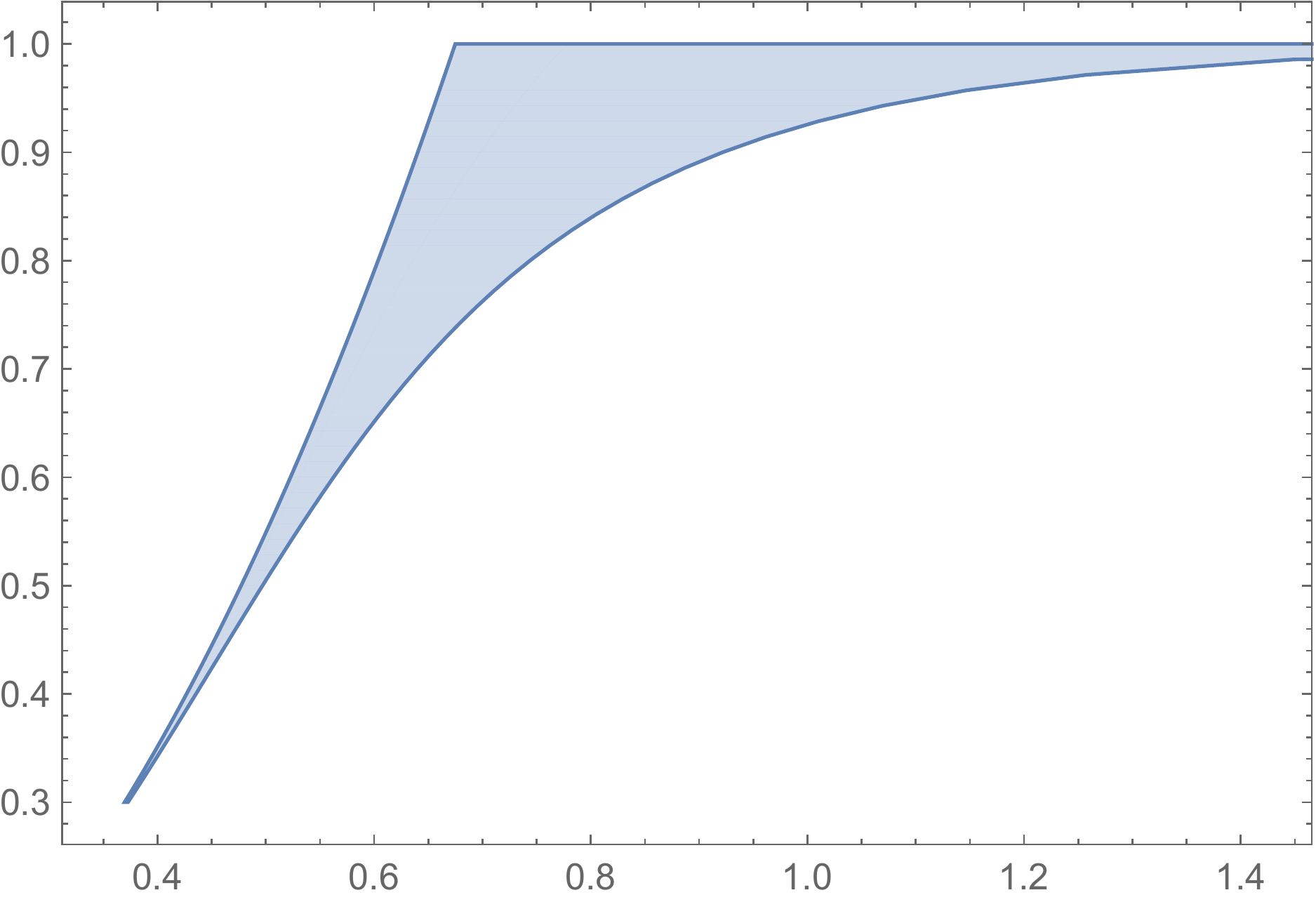}$\,\,$B\\
  \caption{{\bf A:}  Positions of  the singular point  for given $z_* \leq 1$ and given $v_*$  belonging to the variety $v_1\leq v_*\leq v_2$,
  $l_{s}(v_{1})\leq l_{s}(v_{1})\leq l_{s}(v_{2})$. In this plot $v_1=-5,\,v_2=25$.  $\nu =2$. {\bf B:} The same for $\alpha=0.05$.}
        \label{fig:sing-area}
 \end{figure}

We present the dependence of $l_{sing}$ on the turning point $z_{*}$ and  $v_*$ in Fig.\ref{fig:sing-area}  and Fig.\ref{fig:sing-aream}.
Here we again observe that one can get different positions of $l_{s}$ in the range from $l_{s}(v_{1})$ to $l_{s}(v_{2})$ varying $v_{*}$ and fixing $z_{*}$.

\begin{figure}[h!]
\centering
\includegraphics[width=5.5cm]{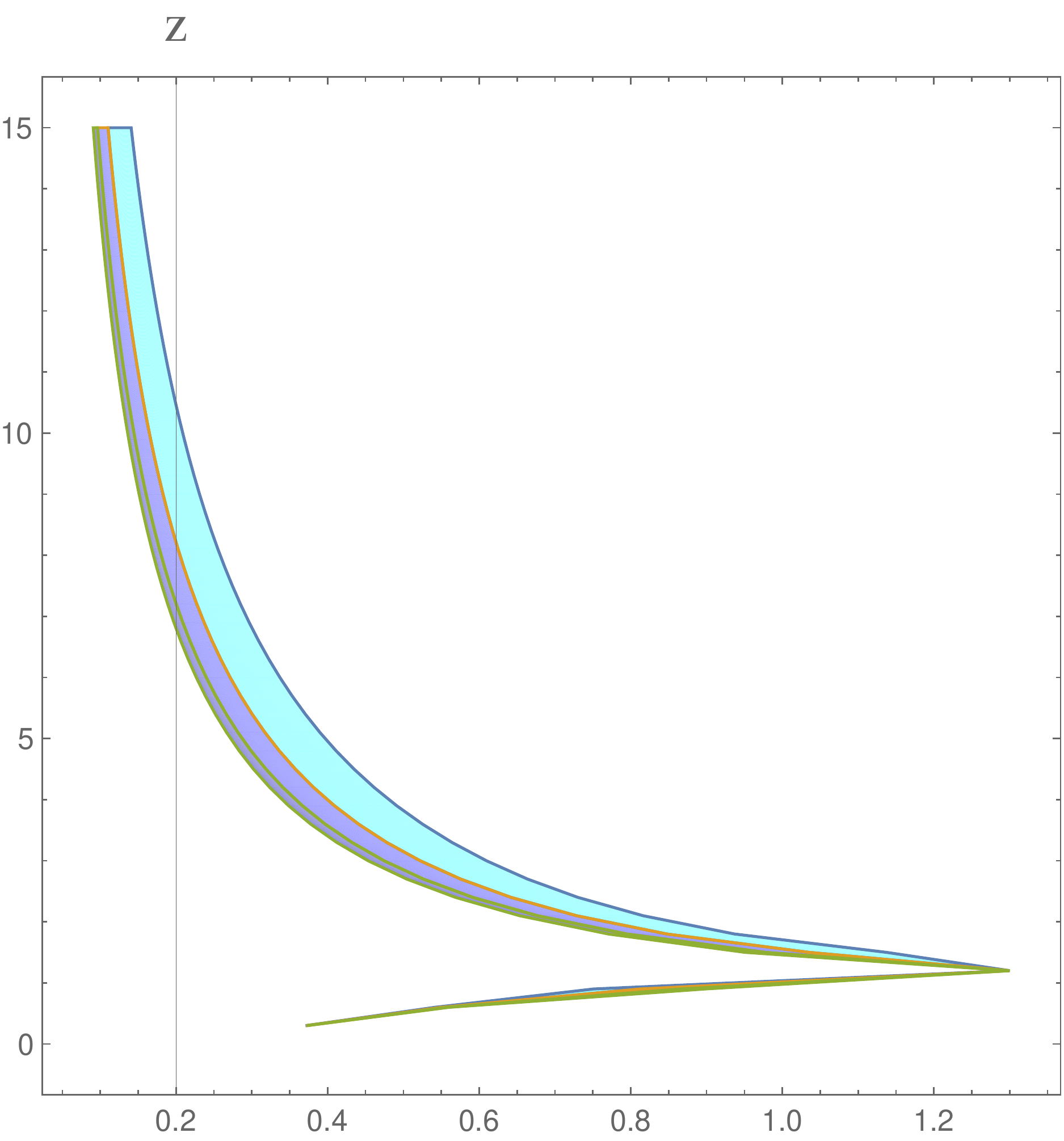}$\,\,$C$\,\,\,\,$$\,\,\,\,$$\,\,\,\,$$\,\,\,\,$$\,\,\,\,$$\,\,\,\,$$\,\,\,\,$$\,\,\,\,$
\includegraphics[width=5.5cm]{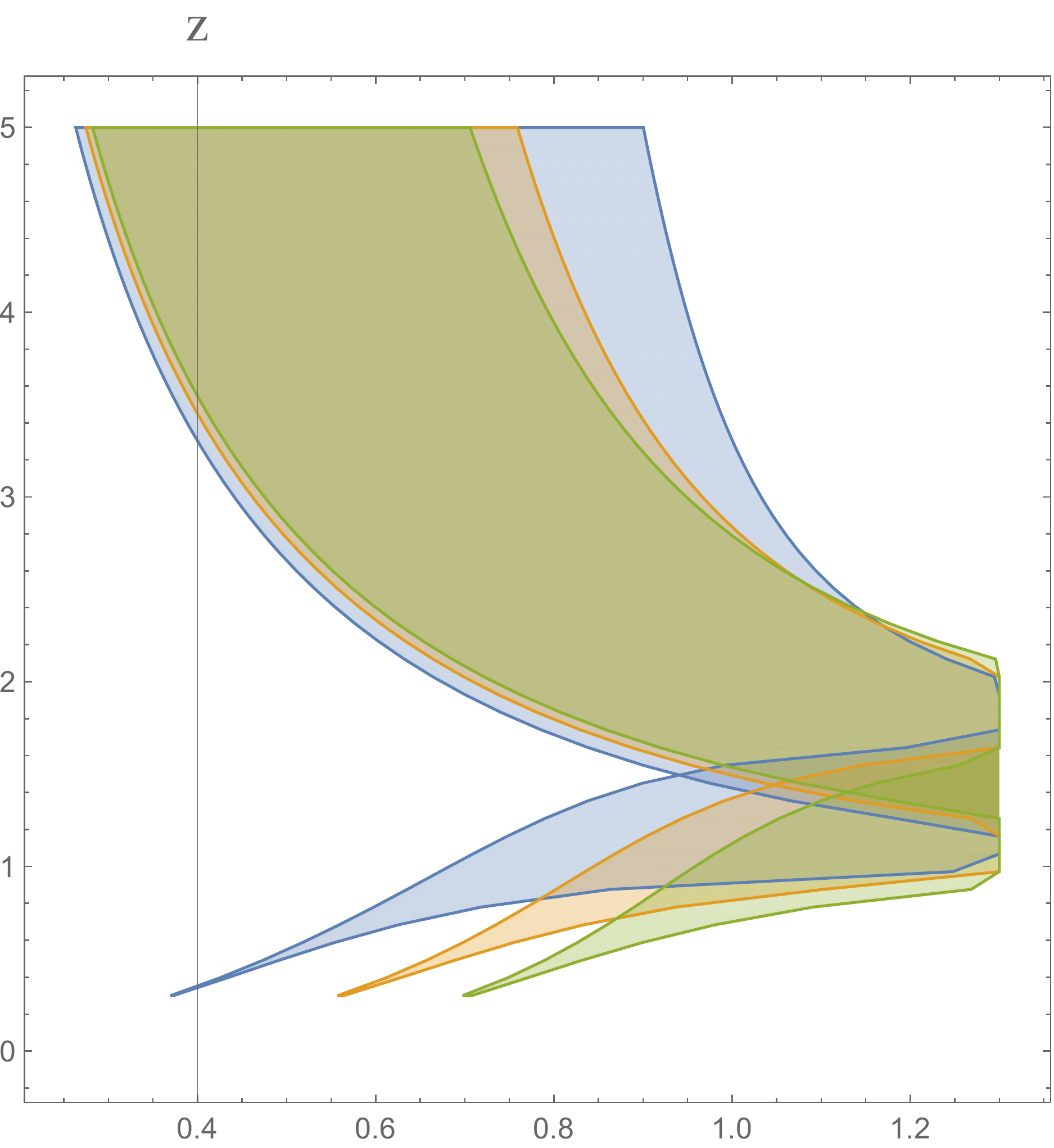}$\,\,$D
  \caption{{\bf A:} Positions of the singular point  for given $z_* $,  $0< z_*<15$ and given $v_*$  belonging to the variety $v_1\leq v_*\leq v_2$, $v_1=-5,\,v_2=25$.  $\nu =2$. {\bf B:} The same for $0<z_* <5$ and $\nu=2,3,4$.}
        \label{fig:sing-aream}
 \end{figure}

In the three left panels of Fig.\ref{fig:sing-areah} we show contour plots for the boundary time
depending on the initial conditions  $z_*$, $v_*$ for $\nu=2,3,4$. In the three right panels of Fig.\ref{fig:sing-areah} we present contour plots for
  $z(l_{sing})$ as a function of initial conditions $z_*$, $v_*$ for $\nu=2,3,4$. As in the previous case, regions of white colour correspond to the irrelevant initial conditions.
  
\begin{figure}[h!]
\centering
\includegraphics[width=5.5cm]{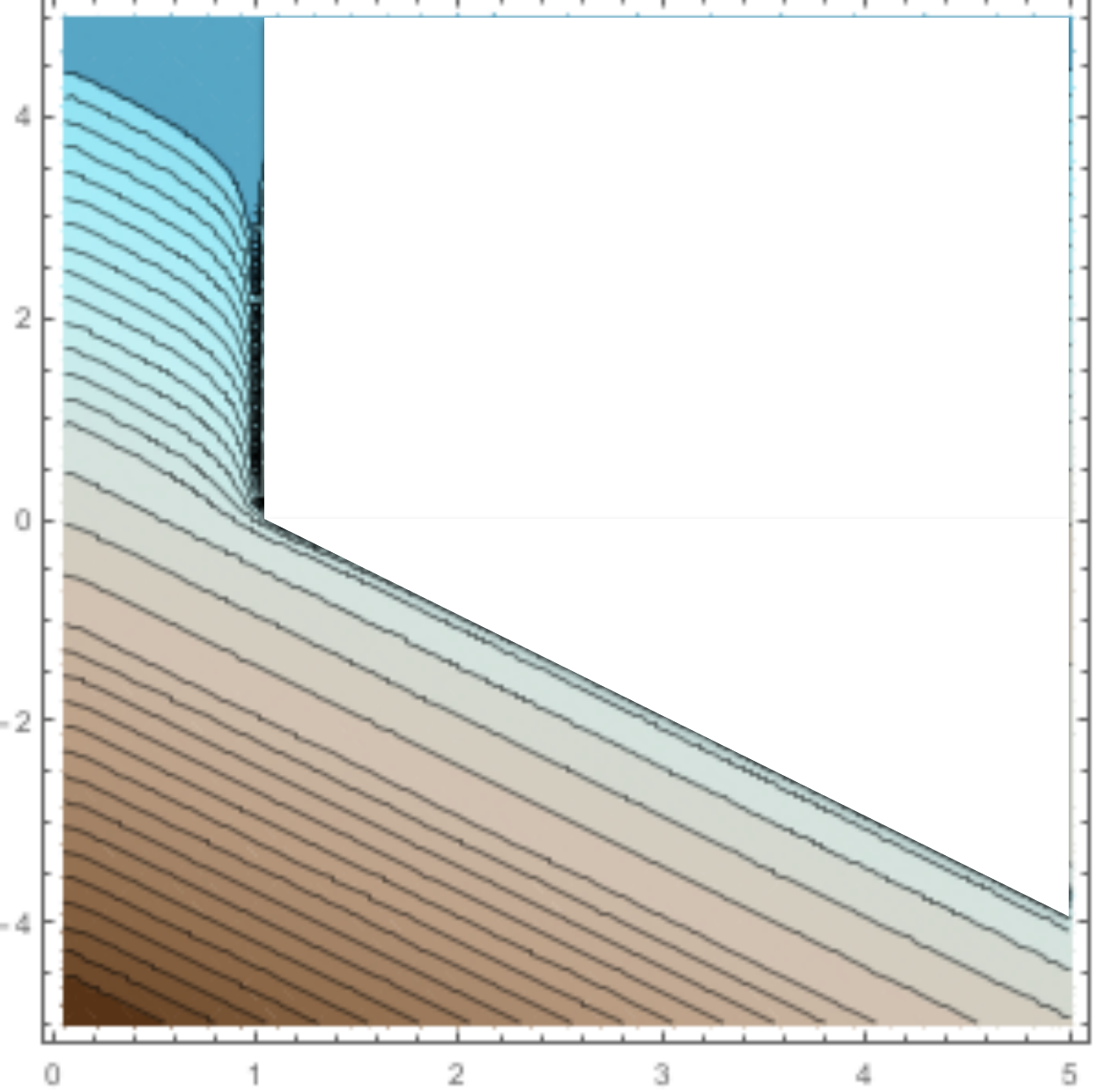}$\,\,$A$\,\,\,\,\,\,\,\,$$\,\,\,\,\,\,\,\,$$\,\,\,\,\,\,\,\,$
\includegraphics[width=5.5cm]{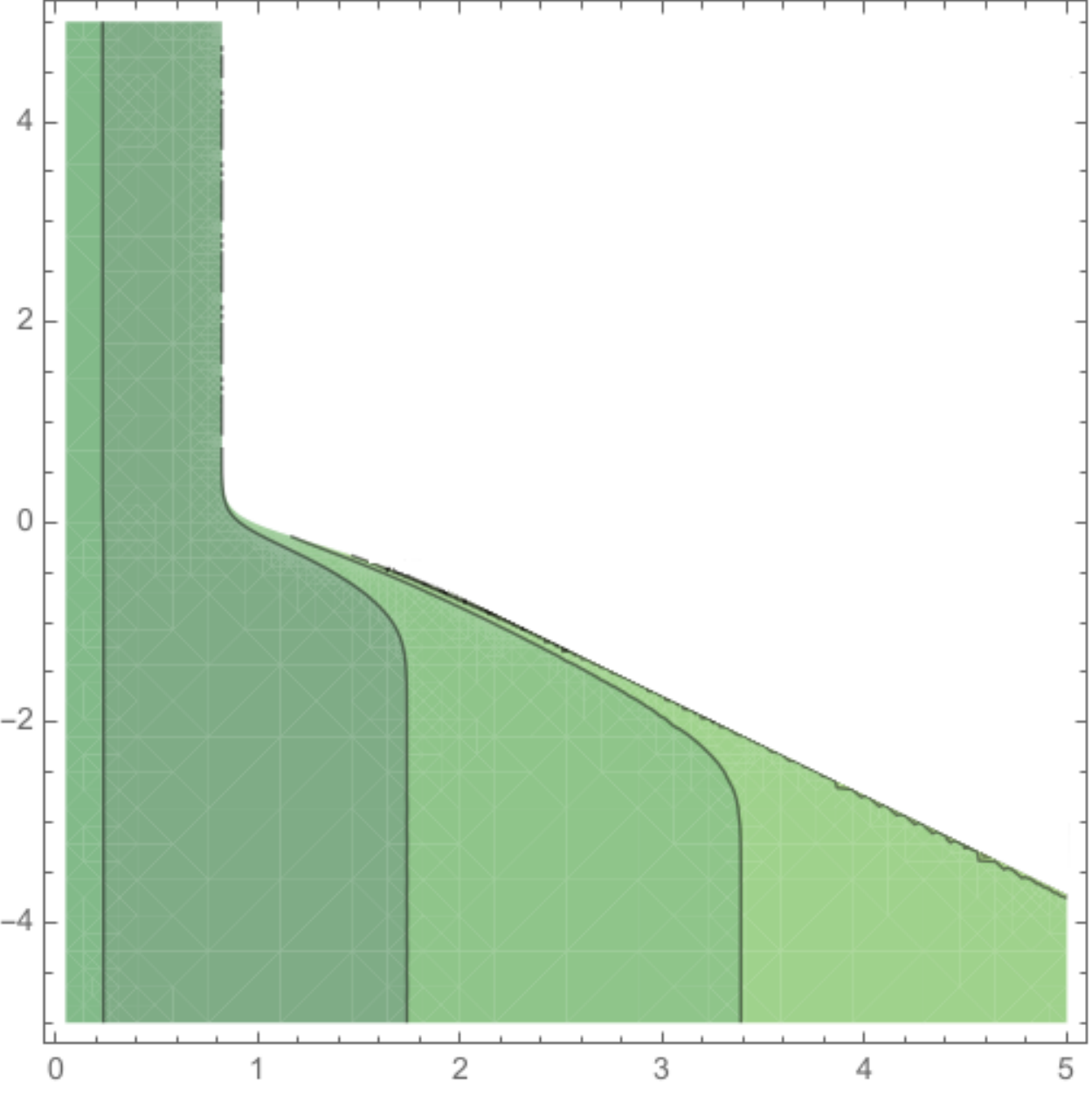}$\,\,$B
\includegraphics[width=5.5cm]{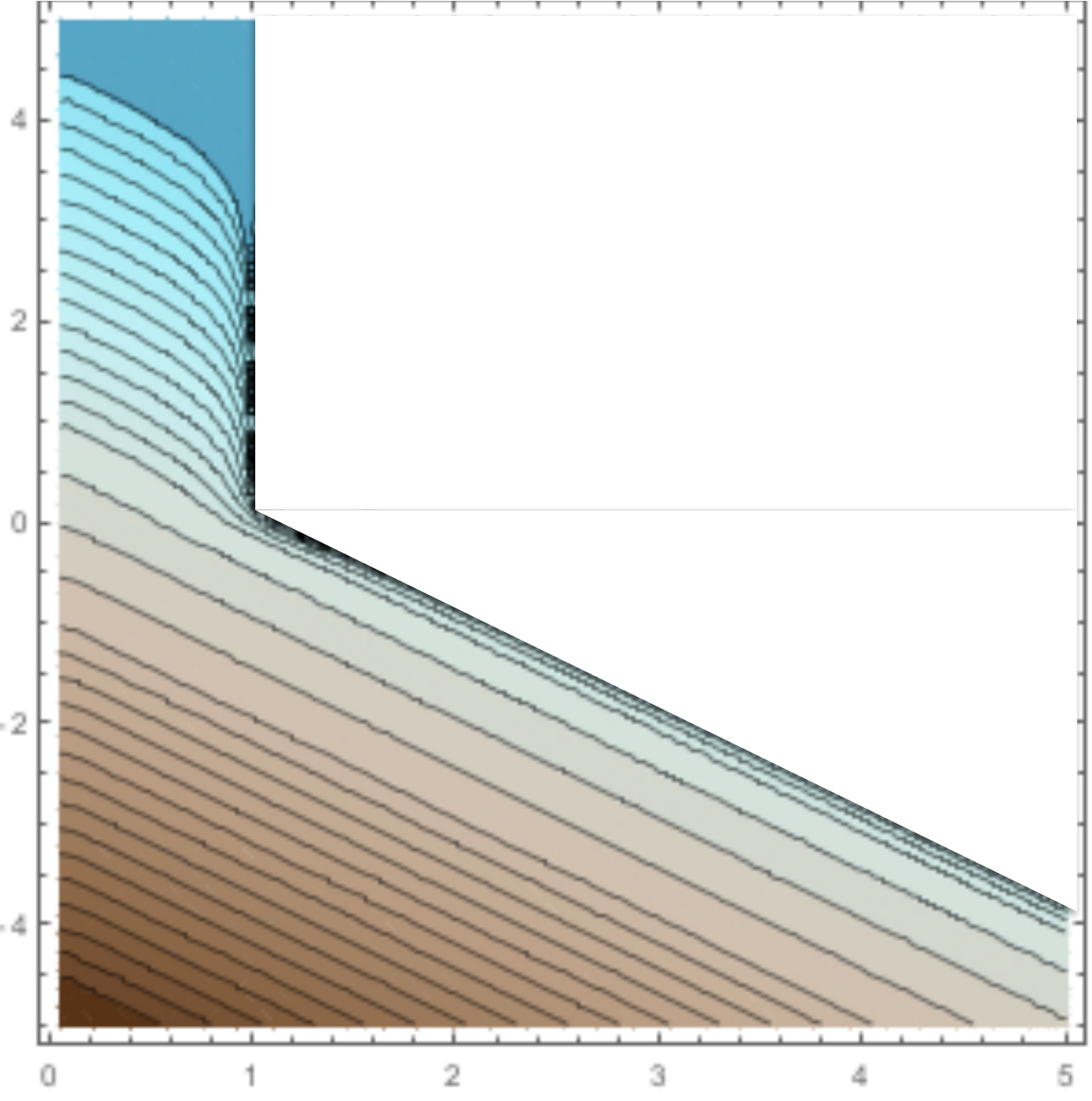}$\,\,$C$\,\,\,\,\,\,\,\,$$\,\,\,\,\,\,\,\,$$\,\,\,\,\,\,\,\,$
\includegraphics[width=5.5cm]{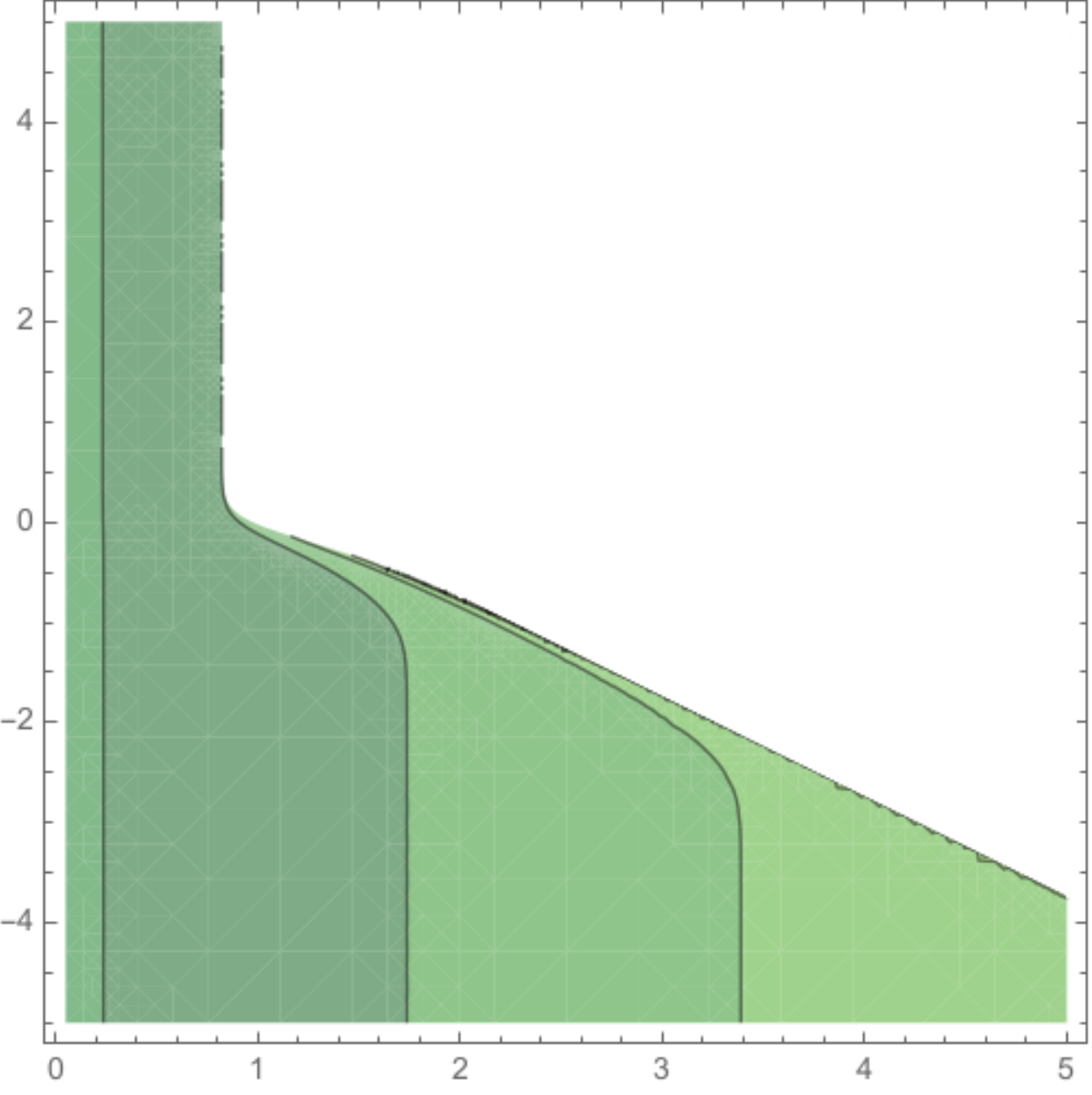}$\,\,$D
\includegraphics[width=5.5cm]{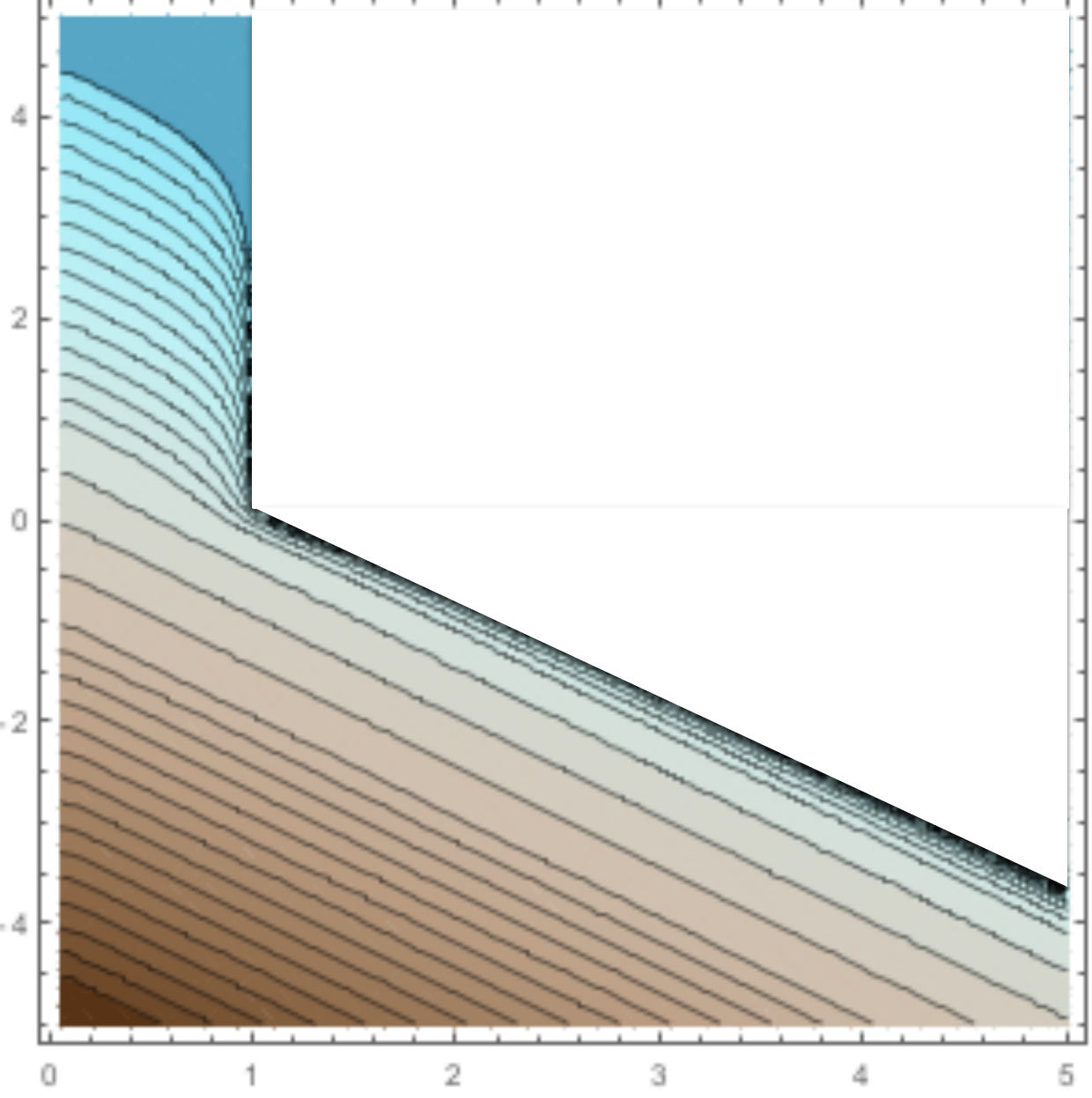}$\,\,$E$\,\,\,\,\,\,\,\,$$\,\,\,\,\,\,\,\,$$\,\,\,\,\,\,\,\,$
\includegraphics[width=5.5cm]{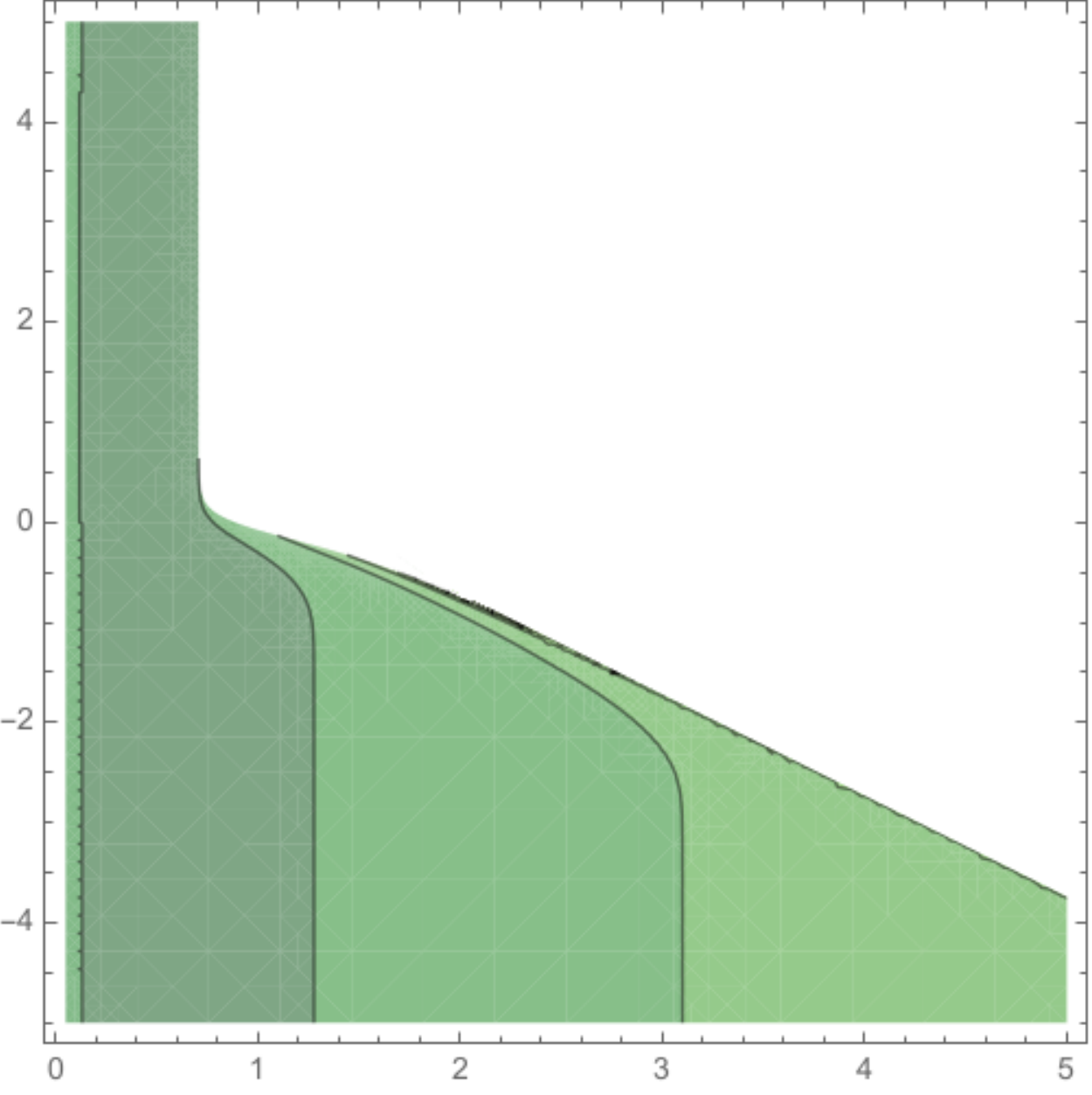}$\,\,$F
 \caption{Left panel:  contour plots for the boundary time as a function of initial conditions
  $z_*$ and $v_*$  for eqs.(\ref{EOM-1-2})-(\ref{EOM-2-2}) for $\nu=2,3,4$ ({\bf A,C,E, respectively}). Right panel:  contour plots for $z(l_{sing})$ as a function of  $z_*$ and $v_*$  for eqs. (\ref{EOM-1-2})-(\ref{EOM-2-2}) for $\nu=2,3,4$ ({\bf B,D,F, respectively}). The regions of white colour correspond to irrelevant initial conditions.}
        \label{fig:sing-areah}
 \end{figure}
\newpage
$$\,$$

\end{document}